\documentclass[12pt,twoside]{article}
\usepackage[dvips]{graphicx,color}
\usepackage{rotating}

\usepackage{amssymb}

\def\mdmatm{\Delta m^2_{32}}
\def\dmatm{$\mdmatm$}
\def\mdmsol{\Delta m^2_{21}}
\def\dmsol{$\mdmsol$}
\def\numunue{$\nu_\mu \rightarrow \nu_e$}
\def\anumunue{$\bar\nu_\mu \rightarrow \bar\nu_e$}
\def\meV{e\mbox{V}}
\def\eV{$\meV$}

\def\GeV{G\eV}
\def\cerenkov{Cherenkov}
\def\linac{LINAC}

\begin{document}

\begin{titlepage}
  \enlargethispage{20pt}
\begin{minipage}[t]{.49\textwidth}
October 28, 2002
\end{minipage}%
\begin{minipage}[t]{0.49\textwidth}
\flushright{BNL-69395}

\end{minipage}
  \begin{center}    
    \noindent{\large \bf

      Report of the BNL Neutrino Working Group:  \\
      Very Long Baseline Neutrino Oscillation 
      Experiment for Precise Determination of Oscillation 
      Parameters and Search for \numunue{} Appearance 
      and CP Violation. 

    }
  \end{center}

  \begin{center}

Coordinators: M.~Diwan, W.~Marciano, W.~Weng

Contributors and Participants

D.~Beavis, M.~Brennan, 
Mu-Chu Chen, 
R.~Fernow, 
J.~Gallardo, R.~Hahn,
S.~Kahn, H.~Kirk, 
D.~Lowenstein, H.~Ludewig,
W.~Morse,  
 R.~Palmer, Z.~Parsa, D.~Raparia, 
T.~Roser, A.~Ruggiero, 
J.~Sandberg, N.P.~Samios, C.~Scarlett, Y.~Semertzidis, 
N.~Simos,  N.~Tsoupas, B.~Viren, P.~Yamin, M.~Yeh\\
{\sl
  Brookhaven National Laboratory
  Box 5000, Upton, NY 11973-5000 }
\smallskip

W.~Frati, J.~R.~Klein, K.~Lande, A.~K.~Mann, 
R.~Van Berg and P.~Wildenhain \\
{\sl 
  University of Pennsylvania
  Philadelphia, PA 19104-6396 }

\smallskip

R.~Corey\\
{\sl South Dakota School of Mines and Technology
  Rapid City, S.D. 57701}

\smallskip

D.~B.~Cline, K.~Lee, B.~Lisowski, P.~F.~Smith \\
{\sl Department of Physics and Astronomy, University of California, Los Angeles, CA 90095 USA}

\smallskip

I.~Mocioiu, R.~Shrock \\ 
{\sl C.N.~Yang Institute for Theoretical Physics, 
  State University of New York, Stony Brook, NY 11974 USA}

\smallskip 

C.~Lu, K.T.~McDonald \\
{\sl Joseph Henry Laboratories, Princeton University, 
  Princeton, NJ 08544 USA}

\smallskip

Renato Potenza \\
{\sl Istituto Nazionale di Fisica Nucleare,
  Dipartimento de Fisica e Astronomia,
  Universita di Catania,
  64, Via S. Sofia,
  I-95123 Catania,
  Italy}

\end{center}


\end{titlepage}

\pagestyle{empty}

This document contains figures in color. The figures should be viewed in   color.

This work was performed under 
the auspices of the U.S. Department of Energy, 
Contract No. DE-ACO2-98CH10886.

\newpage 

\pagestyle{plain}

\pagenumbering{roman}

\tableofcontents

\newpage 

\pagenumbering{arabic}

\section{Executive Summary}

On Dec. 1, 2001, Associate Laboratory Director Tom Kirk appointed a
BNL based neutrino physics study group.  Its charge was to examine
future forefront neutrino oscillation experiments that could be
carried out using traditional $\nu_\mu (\bar \nu_\mu)$ beams of
exceptional intensity (super beams) from an upgraded AGS.  The study,
as reported in this document, addressed detector distances, sizes and
technologies as well as novel ideas for cost effective beam lines and AGS
upgrade paths.  Most important, it focused on the physics discovery
and study potential in its assessment of various options.

Given the success of solar and atmospheric neutrino studies in discovering
neutrino oscillations and measuring some mixing and mass parameters, it
became clear that the next generation accelerator based neutrino oscillation
program must be very ambitious.  In addition to improving measurements of
already approximately known 
$\Delta m^2_{ij} = m^2_i - m^2_j$
and the large mixing angles $\theta_{23}$ and
$\theta_{12}$, the next major effort should be capable of determining the as yet
unknown mixing angle $\theta_{13}$, the mass hierarchy of neutrinos and the 
phase  $\delta_{CP}$.
  Together these will provide a
measure of CP violation in the lepton sector via the Jarlskog invariant
$$J_{CP} = {1\over 8} \sin 2 \theta_{12} \sin 2 \theta_{23} 
\sin 2 \theta_{13}\cos \theta_{13} \sin \delta $$ \\
Indeed, CP violation is properly viewed as the Holy Grail of neutrino
oscillations, since it may be closely connected with the matter-antimatter
asymmetry of the universe.

In order to cover a significant region of the allowed $\theta_{13}$ parameter
space ($\sin^2 2 \theta_{13} \le 0.2$, $0\le \delta \le 2\pi$),
to allow for the determination of the mass ordering to the three neutrinos and 
the possible observation of CP violation
a very large detector of approximately 500 kton, a long baseline ($\ge$2000 km) and
an intense proton source of 1 megawatt are necessary.  For that reason, our
studies concentrated primarily on a water \cerenkov{} detector where the
required technology is mature and capable of achieving the required
large tonnage.  The technical performance of the water 
detectors has also been fully demonstrated in the relevant event energy 
ranges.
Similarly, a relatively simple cost effective AGS
upgrade that primarily increases the repetition rate was examined.  Such a
large water \cerenkov{} detector could also be used to search for proton decay,
supernova neutrinos, $n\bar n$ oscillations, etc.  It could also be used to
significantly improve measurements of atmospheric neutrino oscillations.
Indeed, an extremely attractive picture that emerged from our studies was a
very large multi-physics water \cerenkov{} detector with outstanding discovery
potential in many frontier areas of physics as well as a robust guaranteed
program of detailed studies and precise measurements.

In this report, we describe our vision of the very long baseline neutrino
oscillation experimental component of that program.  It assumes that a 500
kton or larger water \cerenkov{} detector will be built somewhere in the USA
perhaps as a major component of a National Underground Lab and its distance
from BNL will be considerable, e.g.. BNL-Homestake (2540 km) or BNL-WIPP
(2900 km).  To have a sufficient number of detected neutrino events at that
distance, a 1 MW AGS proton source (currently the AGS has 0.14 MW of power)
is envisioned with targetry focusing and a decay tunnel capable of providing
an intense wide band neutrino beam (at 0 degree production) with good support in the
$0.5 \le E_\nu \le 7$ GeV energy range.

The experimental specifications described above were originally chosen with
the idea of measuring the CP violating parameter 
$\delta$  via $\nu_\mu \to \nu_e$ 
oscillations.  However, during the course of our studies, it became clear
that such an effort has a much richer and more diverse physics program.
Indeed, in the scenario we have studied in detail (BNL-Homestake), two
measurements, $\nu_\mu$ 
 disappearance oscillations detected via muon events and
$\nu_\mu \to \nu_e$ 
 appearance oscillations via electron events together provide a wealth
of information.

During the initial research program, 
a run of $5 \times 10^7$ sec (probably distributed over 5 years),
the $\nu_\mu$  disappearance study will resolve several oscillation maxima and
minima (thus firmly establishing oscillations) and measure $\Delta m^2_{32}$ to 
1\% or better
and $\sin^2 2 \theta_{23}$ to 1\% or better, 
significant improvements over existing or planned
measurements.  In the $\nu_\mu \to \nu_e$ 
 appearance mode, the $\nu_e + n \to e^- + p$  quasi-elastic events
over the 0.5 GeV range will allow the following investigations to 
be completed:

\begin{enumerate}
\item  Search for and measurement of $\sin^2 \theta_{13}$ 
  to below 0.005 via matter enhanced
  oscillations.
\item Determine the sign of $\Delta m^2_{31}$, 
  i.e. whether $m_3$ is the
  largest or smallest of the 3 neutrino masses, also via matter enhancement or
  suppression effects in the 3-7 GeV region.
\item Measure $\sin \delta$ 
  (and $\cos \delta$) to
  about $\pm 25\%$ thus determining $J_{cp}$ and the $\delta$
  quadrant.
\item Measure $\Delta m^2_{21}$ 
  and $\theta_{12}$
  from the $\nu_\mu \to \nu_e$ 
  oscillations of low energy $0.5-1.0$ GeV
  neutrinos with about the same sensitivity as Kamland, but in an appearance
  rather than disappearance mode.

\end{enumerate}

The above program is extremely rich, covering essentially all the parameters
of 3 generation neutrino mixing as currently envisioned.  It is also robust,
offering important measurements even if some parameters 
whose values we have assumed in our calculations
change significantly.  
Together with the search for proton decay and
study of cosmic neutrinos, our accelerator based long baseline neutrino
oscillation program represents a major step forward in the advancement of
science.  Beyond the first research period,
one could envision further accelerator and beam
upgrades, antineutrino runs, or additional beams from other accelerator
facilities.  Indeed, the large detector that forms the centerpiece of this
effort should be expected to function  for half a century or more
expanding our knowledge of all the
above noted research areas.

This report will show that the bold program envisioned above is technically
feasible and economically attractive.  
We show that the
existence of the AGS machine at BNL with its straightforward and economical
upgrade to the needed 1 MW power level, taken together with the needed
very long baseline available for at least two appropriate detector sites,
makes
 this approach to a practical facility the best one for the next-generation
U.S. neutrino physics program.  The identified physics goals are compelling and
not covered by less ambitious alternatives.  Nevertheless, its realization
will require strong commitment and vision.  The high payoff is worth the
effort.

  \newpage

\cleardoublepage

  \section{Introduction}

  \vspace{1ex}

  Brookhaven National Laboratory and collaborators
started a neutrino 
  working group to identify new opportunities in the field of 
  neutrino oscillations and explore 
  how our laboratory facilities can be used to explore this field 
  of research. The memo to the working group and the charge are
  included in Appendix I.

  This report is the result of the deliberations of the working group. 
  Previously, we
  wrote a  letter of intent to build a new high intensity neutrino beam
  at BNL \cite{bnlloi}.  
  A new intense proton beam will be used to produce
  a conventional horn focussed neutrino beam directed at a detector 
located in either the Homestake mine in Lead, South Dakota
 at 2540 km or the Waste Isolation Pilot
Plant (WIPP) in Carlsbad, NM at 2880 km \cite{homestk, wipp}.
  As a continuation of the study that produced the letter of intent, 
  this report examines several items in more detail.
  We  mainly concentrate on the use of water \cerenkov{} detectors
  because of their size, resolution, and background rejection 
capability, and cost. We examine the prospects of building such a
detector in the Homestake mine.

  The accelerator upgrade will be carried out in phases. 
  We expect the first phase to yield a 0.4 MW proton beam and the 
  second phase to result in a 1.0 MW beam. The details of this 
  upgrade will be reported in a companion report. In this report 
  we assume accelerator intensity of 1 MW for calculating 
  event rates and spectra. We also assume a total experimental duration 
  of 5 years with running time of $10^7$ seconds per year.

  We examine the target station and 
  the horn produced neutrino beam with focus on two topics: 
  target and horn design for a 1 MW beam and the broad band spectrum of 
  neutrinos from a 28 GeV proton beam.

  \section{Neutrino Oscillations}

  The strongest evidence for neutrino
  oscillations comes from astrophysical observations of
  atmospheric neutrinos with $\mdmatm = (1.6 - 4.0) \times 10^{-3}
  ~\meV^2$ and maximal mixing \cite{sk},
  and from 
  solar neutrinos with $\mdmsol = (3
  -10) \times 10^{-5} ~\meV^2$ assuming the LMA solution~\cite{sno}.  
  The observation by the LSND
  experiment~\cite{lsnd} will soon be re-tested at Fermilab by the
  mini-Boone \cite{boon} experiment. Therefore we will not discuss it 
  further in this document.
  There are several accelerator based experiments (K2K,
  MINOS, and CNGS)  \cite{k2k, j2k, numi-off, minos, cngs} currently  
   in the construction phase or taking data 
   to confirm the atmospheric neutrino signatures for
  oscillations. 
   There is now a consensus that there are four main goals
  in the field of neutrino oscillations 
   that should be addressed soon with accelerator neutrino
  beams:

  \begin{enumerate}
  \item Precise determination of \dmatm{} and $\sin^2 2 \theta_{23}$ 
and definitive observation of oscillatory behavior.
  \item Detection of \numunue{} in the appearance mode.  If the measured
    $\Delta m^2$ for this measurement is near \dmatm{} then this
    appearance signal will show that $\left|U_{e3}\right|^2 (=
    \sin^2\theta_{13})$ from the neutrino mixing matrix in the standard
    parameterization is non-zero.
  \item Detection of the matter enhancement effect in \numunue{} in the
    appearance mode. This effect will also allow us to measure the sign
    of \dmatm{}, i.e. which neutrino is heavier.
  \item Detection of CP violation in neutrino physics.  The neutrino
    CP-violation in Standard Model neutrino physics comes from the phase
    multiplying $\sin \theta_{13}$ in the mixing matrix. This phase
    causes an asymmetry in the oscillation rates
    \numunue{} versus \anumunue{}.

  \end{enumerate}

  In this report we describe how all of these goals can
  be achieved under reasonable assumptions for the various parameters
  using the new intense AGS based beam and the  very long
  baseline of BNL to Homestake laboratory of 2540 km. 

  In Section 3 we estimate the event rates, backgrounds and oscillation 
signals. This section highlights the physics measurements achievable with the detector 
being proposed, focusing on its sensitivity to various oscillation parameters.

  In Section 4 of this report we briefly describe the 
  accelerator upgrade path 
  to achieve a  proton source with intensity greater than 1 MW.

  In Section 5 we examine the conventional neutrino beam 
  spectrum and the target-horn station.

  In Sections 6 we summarize the report and give a breakdown of the expected
costs.

\section{Very Long Baseline Experiment} 

\vspace{1ex}

\begin{figure}[htbp]
  \begin{center}
  \includegraphics*[width=\textwidth]{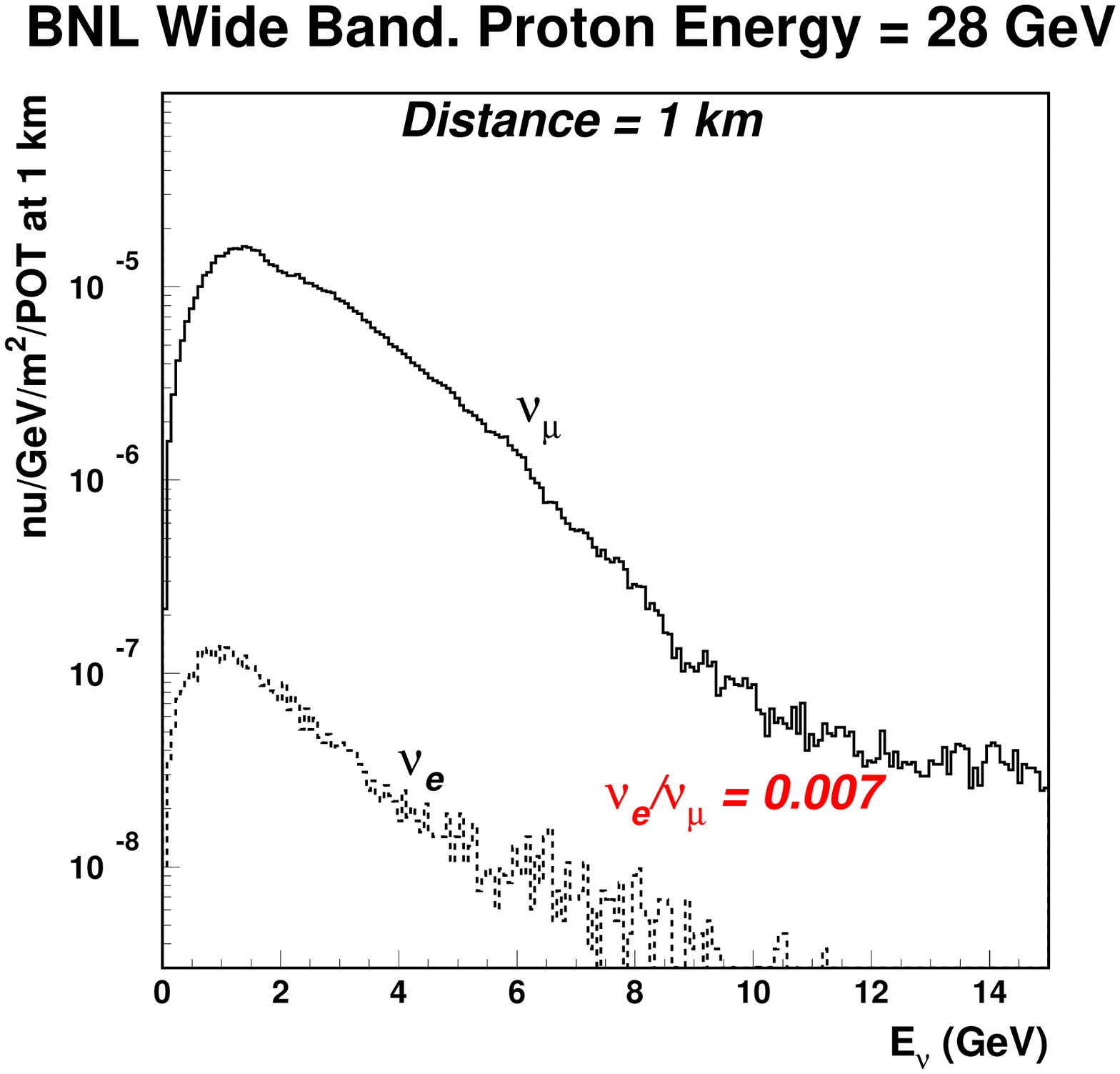}

  \caption[BNL wide band neutrino beam spectrum]{BNL wide band spectrum with the new graphite 
target and horn design. This spectrum is at 0 degrees with respect to 
the proton beam on target and the normalization is at 1 km from the 
target. 
}
   \label{wspec}  
  \end{center}
\end{figure}

We calculate the event rate without oscillations assuming a
1.0 MW proton beam power with 28 \GeV{} protons ($1.1 \times 10^{14}$
protons per pulse), a 0.5 MT fiducial mass 
water \cerenkov{} detector and 5 years of running.  Because
BNL's Alternating Gradient Synchrotron (AGS) can run in a parasitic 
mode to the Relativistic Heavy Ion Collider (RHIC), we expect to get 
beam for
as much as $1.8\times 10^7$ sec per year.  However, we conservatively
assume only $1.0\times 10^7$ sec of AGS running 
per year here.  Using these
parameters, the $0^\circ$ flux from Figure \ref{wspec} 
 and the relevant cross sections, we calculate that the
number of quasi-elastic
 charged current muon neutrino events in a
detector located at 2540 km will be $\sim 12000$ 
in five years running.
Table \ref{evcount}  shows the number of different kinds of 
events we expect in the absence of oscillations. 
The large statistics combined with the long baseline
make many of the following important measurements possible.

\begin{table}

\caption[Expected event rates]{Number of events of different types for the very long
baseline experiment. The parameters are 1 MW of beam, 0.5 MT of
fiducial mass, and 5 years of running with $10^{7}$ seconds of 
live time each year.
 CC, NC, QE, stands for charged current, neutral current, 
and quasielastic, respectively. 
The $\nu_e$ interaction rate is from the electron neutrino 
contamination in the beam.} \smallskip 
\begin{center}
\begin{tabular}{|l|r|}
\hline 
Reaction & Number  \\
\hline 
CC $\nu_\mu + N \to \mu^- + X$ & 51800  \\
NC $\nu_\mu + N \to \nu_\mu + X$ & 16908 \\ 
CC $\nu_e + N \to e^- + X$  & 380  \\
\hline 
QE $\nu_\mu + n \to \mu^- + p$ & 11767 \\
QE $\nu_e + n \to e^- + p$ & 84  \\
\hline 
CC $\nu_\mu + N \to \mu^- + \pi^+ + N$ & 14574 \\
NC $ \nu_\mu  + N \to \nu_\mu + N + \pi^0$ & 3178 \\
NC $ \nu_\mu + O^{16}  \to \nu_\mu + O^{16} + \pi^0$ & 574 \\
\hline 
CC $\nu_\tau + N \to \tau^- + X$ &  319 \\
(if all $\nu_\mu \to \nu_\tau$) &   \\
\hline 
\end{tabular}
\end{center} 

\label{evcount}
\end{table}

\newpage

\subsection{$\nu_\mu$ disappearance}

\begin{figure}[htbp]
  \begin{center}
  \includegraphics*[width=\textwidth]{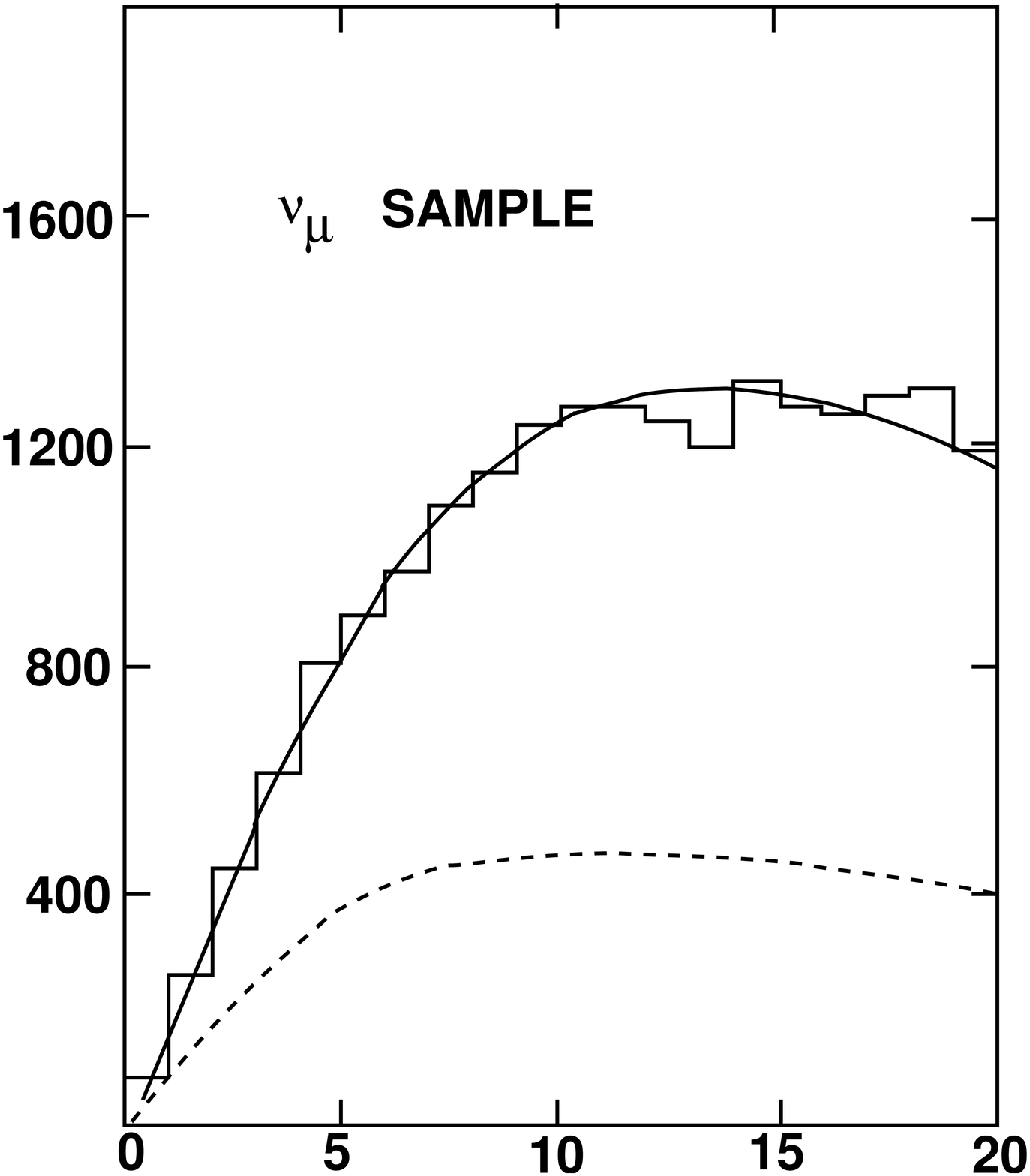}
  \caption[Neutrino produced muon angle distribution, data and Monte Carlo.]
{Angular distribution of muons from the process 
 $\nu_\mu n \rightarrow \mu^- p$ (top curve) and 
background from  
 $\nu_\mu N \rightarrow \mu^- N' \pi$ (bottom curve).
The histogram is data from AGS experiment
E734 (year 1986) and the lines are Monte Carlo.}
  \label{fig:e734mu}
  
  \end{center}
\end{figure}

\begin{figure}
  \begin{center}
    \includegraphics*[width=\textwidth]{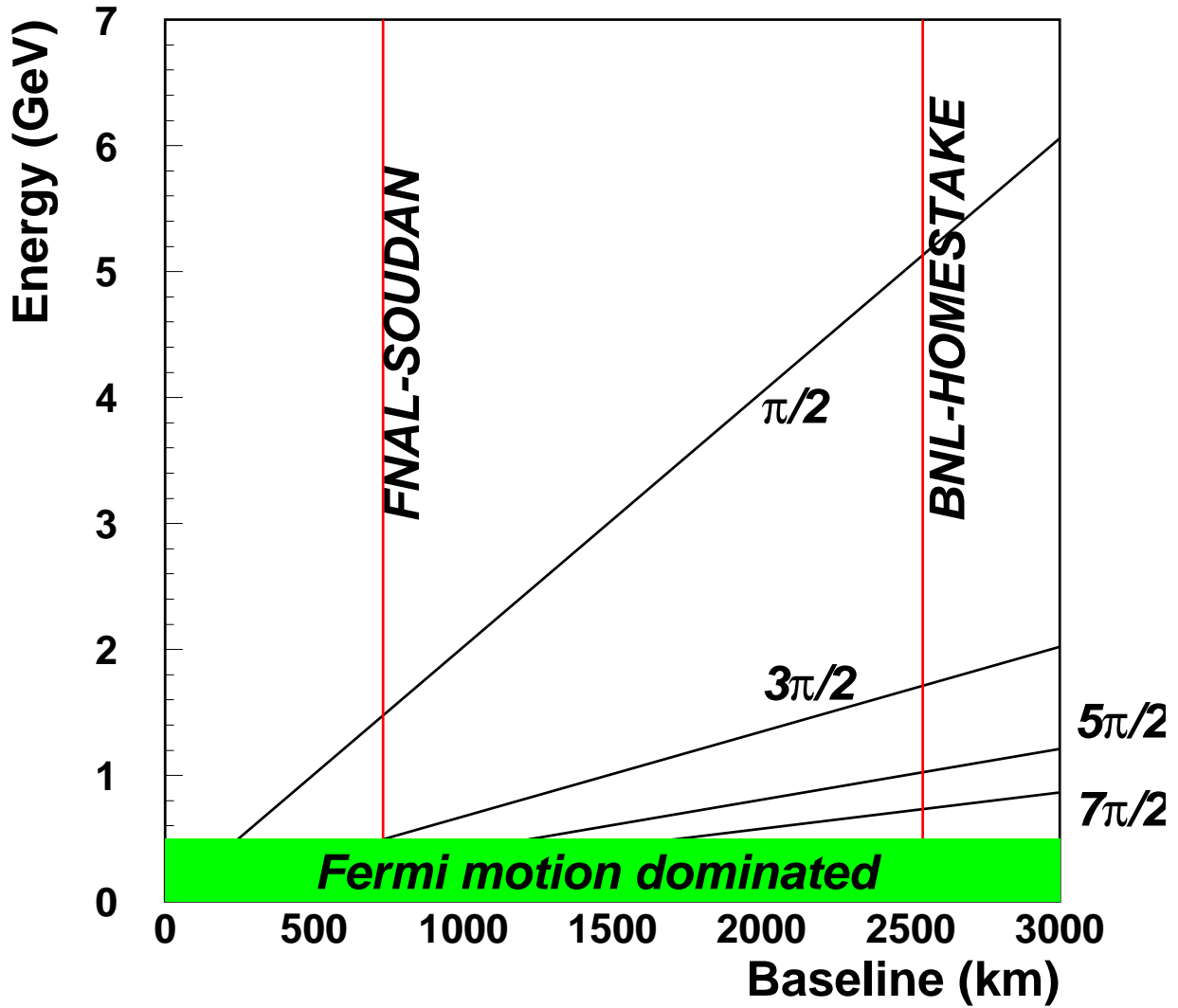} 
    \caption[Oscillation nodes {\it vs.} distance.]
{Nodes of neutrino oscillations 
for disappearance (Not affected by matter effects) as a 
function of oscillation length and
      energy for $\mdmatm = 0.0025~\meV^2$. 
 The distances from FNAL to Soudan  (the distance from BNL to Morton 
salt works is approximately the same\cite{imb}) 
and from BNL to Homestake are shown by the vertical lines. 
}
    \label{nodes}  
  \end{center}
\end{figure}

\begin{figure}
  \begin{center}
    \includegraphics*[width=\textwidth]{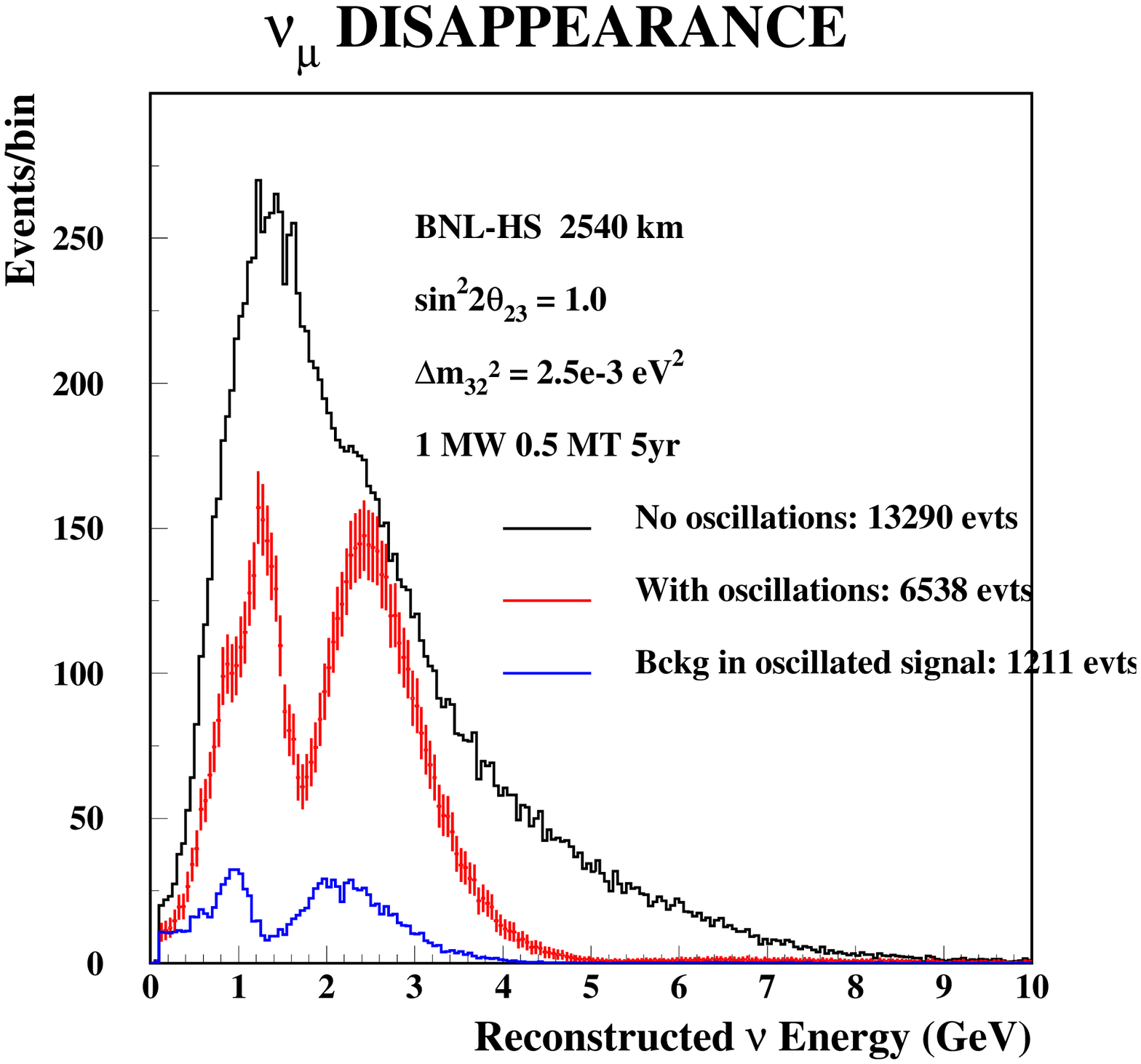}
    \caption[Expected $\nu_\mu$ disappearance spectra, $\Delta m^2_{32} = 0.0025$]
{Spectrum of detected  events in a 0.5 MT detector at
      2540 km from BNL including quasielastic signal and CC-single pion 
background. 
 We have assumed 1.0 MW of beam power and 5
      years of running.  The top histogram is without oscillations;
the middle error bars are with oscillations and the bottom histogram is
the contribution of the background to the oscillated signal only.  
 This plot is for $\mdmatm =
      0.0025~\meV^2$.
      The error bars correspond to the statistical error expected in
      the bin. A 10 \% detector energy resolution is assumed.
        At low energies the Fermi movement, which is included in 
simulation, will dominate the resolution.}      
    \label{wcnodesa}  
  \end{center}
\end{figure}
\begin{figure}
  \begin{center}
    \includegraphics*[width=\textwidth]{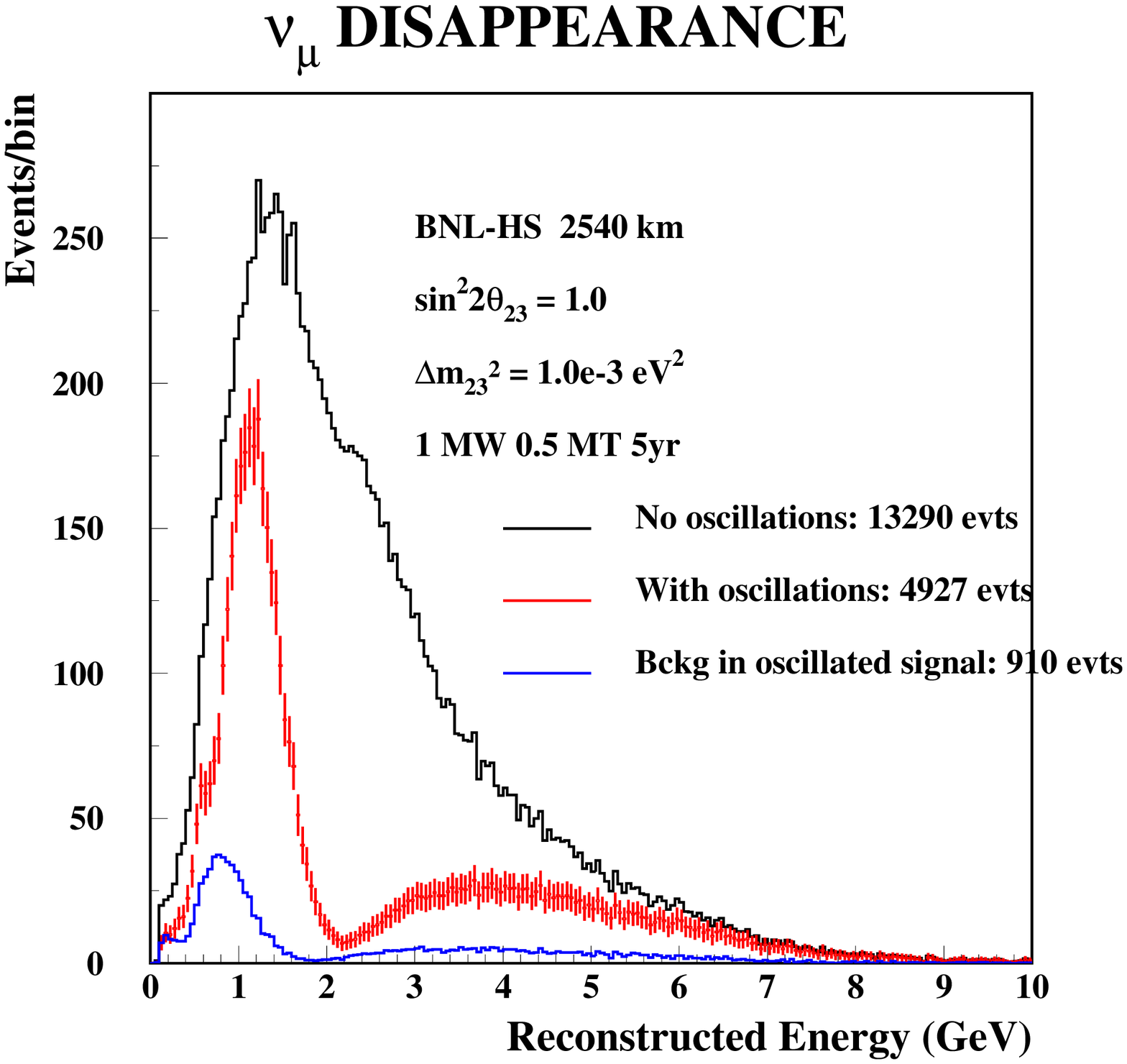}
    \caption[Expected $\nu_\mu$ disappearance spectra, $\Delta m^2_{32} = 0.001$]
{Spectrum of detected  events in a 0.5 MT detector at
      2540 km from BNL including quasielastic signal and CC-single pion 
background. 
 We have assumed 1.0 MW of beam power and 5
      years of running.  The top histogram is without oscillations;
the middle error bars are with oscillations and the bottom histogram is
the contribution of the background to the oscillated signal only.  
 This plot is for $\mdmatm =
      0.001~\meV^2$.
      The error bars correspond to the statistical error expected in
      the bin. A 10 \% detector energy resolution is assumed.
        At low energies the Fermi movement, which is included in 
simulation, will dominate the resolution.}      
    \label{wcnodesb}  
  \end{center}
\end{figure}

The angular distribution of the muons from the quasi-elastic process
$\nu_{\mu} + n \rightarrow \mu^- + p$ produced by the $0^o$ beam in
Figure~\ref{bnlspec} was measured 
in experiment E734 (1986) at BNL.  It 
is shown again in Figure~\ref{fig:e734mu} along with the principal
background, $\nu_{\mu} + N \rightarrow \mu^- + N + \pi$ \cite{e734d}.
A variety of strategies is possible to reduce this background further
in a water \cerenkov{} detector.
 Knowing the direction of
an incident $\nu_{\mu}$ accurately and measuring the angle and energy 
 of the
observed muon allows the energy of the $\nu_{\mu}$ to be calculated,
up to Fermi momentum effects. 
This method is used by the currently running K2K experiment
\cite{k2k}.  The known capability of large water \cerenkov{} detectors
indicates that at energies lower than 1 \GeV{} the $\nu_\mu$ energy
resolution will be dominated by Fermi motion and 
nuclear effects\cite{kasuga}.
  The contribution to the resolution from water \cerenkov{} track
reconstruction depends on the photo-multiplier tube coverage.  With
coverage greater than $\sim$ 10\%, we expect that the
reconstruction resolution should be more than 
adequate for our purposes \cite{e889}. In the
following discussion 
we assume a 10\% resolution on the $\nu_\mu$ energy. 
This is consistent with the resolution projected for 
10\% coverage from the K2K experience \cite{sharkey}.

The range of $\mdmatm \sim 1.24{E_\nu\mbox{[\GeV{}]} \over
  L\mbox{[km]}}$ covered by the proposed experiment using the beam in
Figure \ref{wspec} extends to the low value of about 
$5 \times 10^{-4} ~\meV^2$.  The lower end of this extensive range of values is
considerably below the corresponding values for other  long
baseline terrestrial experiments~\cite{minos,cngs}.  If the value of
\dmatm{} turns out to be towards the lower end ($\sim 10^{-3}$) of its
current range, or if the value of \dmsol{} turns out to be towards its
high end ($\sim$ $10^{-4}~\meV^2$), then  large and very
interesting interference effects in the very long baseline experiment
will be possible. 

Extra-long neutrino flight paths open the possibility of observing
multiple nodes (minimum intensity points) of the neutrino oscillation
probability in the disappearance experiment.  Observation of one such
pattern will for the first time directly demonstrate the oscillatory
nature of the flavor changing phenomenon.  The nodes occur at
distances $L_n = 1.24 (2n-1) E_{\nu}/\mdmatm$, $n= 1,2,3, $ \ldots.
In Figure \ref{nodes}, as an example, we show the flight path $L$ versus
$E_{\nu}$ relationship of the nodes for $\Delta m^2 = 0.003 
  ~eV^2$, a value close to the value measured in atmospheric
neutrino experiments \cite{sk}.  An advantage of having a very
long baseline is that the multiple node pattern is detectable over a
broad range of $\Delta m^2$.  For \dmatm{} as small as 0.001 $e\mbox{V}^2$,
the oscillation effects will be very large.

The two single charged pion reactions $\nu_\mu + p \rightarrow \mu^- + p + \pi^+$ and
$\nu_{\mu} + n \rightarrow \mu^- + n + \pi^+$ produce a signal which is
 somewhat larger  than  the quasi-elastic total
in Table \ref{evcount}.  
For these  events, 
 if both the muon and the pion produce more than 50 photoelectrons
each, the event can be easily identified as a two ring event in a water \cerenkov{} detector 
and rejected. 50 photoelectrons 
corresponds to about 170 MeV/c (250 MeV/c) for muons (pions) for a detector 
with 10\% photo-multiplier coverage. 
An additional cut to require the muon to be within 60$^o$ of the neutrino
direction reduces the background further. 
With such a cut, 
we find that 18\% of the events will show one ring (principally the $\mu^-$). 
The detection of two muon decays, one from the $\mu^-$ the other from 
the decay chain $\pi \to \mu \to e$,   could be used 
to further suppress this background by approximately a factor of 2.
More importantly, background events can be tagged by the two muon decays 
to determine the shape of the background from the data itself. This will 
greatly increase the confidence in the systematic error due to this  background. 
The reaction $\nu_{\mu} + n \rightarrow \mu^- + p + \pi^0$ (the
only allowed CC-$\pi^0$ reaction) is $\sim$15\% of the total quasi-elastic rate. The
momentum distribution of $\mu^-$ and $\pi^0$ are essentially the same as those
for CC-charged pion production. Only 0.5\% of the CC-$\pi^0$ events
will look like quasi-elastic muon events because at least one 
of the  gamma rays from the 
$\pi^0$ decay is usually visible.
Thus this background is negligible
in the quasi-elastic sample.

 The expected plot of signal and background is shown in Figures
\ref{wcnodesa} and \ref{wcnodesb}.
They show the disappearance of
muon type neutrinos as a function of neutrino energy measured  
 in quasi-elastic events. The 
background, which will be mainly charged current, will also oscillate, but 
the reconstructed neutrino energy will be systematically lower for the 
background. Nevertheless, the main effect will be to slightly 
broaden the large dips due to disappearing muon neutrinos. 

\begin{figure}
  \begin{center}
    \includegraphics*[width=0.8\textwidth]{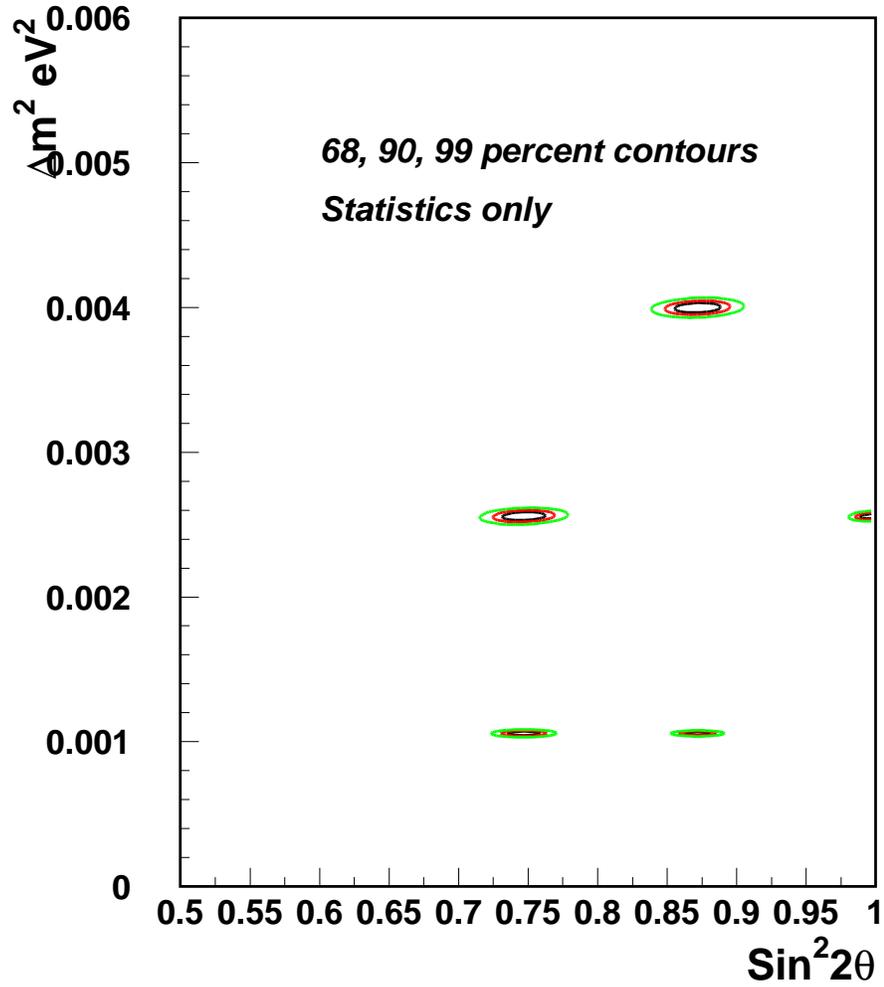}
    \caption[Statistical uncertainty for $\Delta m^2_{32}$ and $\sin^22\theta_{23}$]
{ Statistical resolution at 68\%, 90\% and 
99\% confidence level on $\Delta m_{32}^2$ and $\sin^2 2\theta_{23}$
for the 2540 baseline experiment; assuming 1 MW, 0.5 MT, and 5 years
of exposure. }
    \label{cntr1}  
  \end{center}
\end{figure}

In Figure \ref{cntr1} we show the statistical 
precision expected on the measurement of $\Delta m^2_{32}$ and
$\sin^2 2 \theta_{23}$
for several different points in the parameter space.  
It is clear that since the signal and the statistics are large, the 
systematic error in fitting the spectrum 
 will dominate the final error. 
We  list  various effect that must be considered for the 
measurement with brief comments about each.  

\begin{itemize}

\item The determination of $\Delta m^2$ has a statistical uncertainty of 
approximately $\pm 0.7\%$ at $\Delta m^2 = 0.0025 ~eV^2$ with maximum mixing.
It is about $\pm 1.0\% $ when $\sin^2 2 \theta_{23} = 0.75$. 
Clearly, the knowledge of the energy scale will be very important in 
measuring this  number.  If the energy scale uncertainty is 
$\delta E/E$ then the final error will be given by 
$$({\sigma(\Delta m^2)\over \Delta m^2})^2 = ({\sigma_{stat} \over \Delta m^2})^2
 + ({\delta E \over E})^2$$
Therefore, it will be very important to understand the energy calibration of 
the detector to about 1 \% for  muon energy of $\sim 1$ GeV. 
One solution could  be 
a magnetic spectrometer to measure the momentum of cosmic ray  muons 
entering the detector. This consideration could  affect the depth at which 
this detector should be mounted. Another option could be a 
linear accelerator that could provide protons or electrons at a rate
of few Hz at $\sim 100$ MeV.

\item Even if the overall energy scale is known well,
the energy calibration could vary non-linearly over the entire spectrum. 
The worst effects of these fluctuations
will be  where the spectrum has the 
maximum slope. This effect will cause additional 
smearing of  the spectrum and reduce the resolution on $\Delta m^2$. 
We assume a 5\% uncertainty of the energy calibration over the 
entire range. 

It should be pointed out that the oscillation minima should be at 
energies that are in precisely known ratios of integers: 3, 5, 3/5, etc. 
This could be used to determine the relative 
energy scale precisely. On the other hand these ratios could be important 
to determine the presence of new physics
(non-sinusoidal depletion of muon neutrinos) in the oscillations. 

\item The model of Fermi motion and reconstruction resolution 
will affect both the shape of the signal and the background 
used in the fit. The consequences of this  effect are probably the same 
as the previous one in terms of  the resolution of fitted 
parameters.

It was pointed out earlier that some of the 
the CC-$\pi^+$ background could be 
tagged by two muon decays. This sample of events can be 
used in separate fits to put more constraints on the 
detector  simulations. 
 The large number of
charged current 
 events ($\sim 52000$) 
that are not quasielastic could also be used in the same 
manner.

\item The statistical uncertainty in the determination of 
$\sin^2 2 \theta_{23}$ is $\pm 0.016$ at $\sin^2 2 \theta_{23}=0.75$ and 
$\Delta m^2_{32} = 0.0025 ~eV^2$. This determination is somewhat better at 
smaller $\Delta m^2$. At maximum mixing, Figure \ref{cntr1} shows that we 
can determine $\sin^2 2 \theta_{23} > 0.99$ at 90\% confidence level. 

We expect this error to be even smaller if proper background subtraction is 
performed on the data. 
Normally the determination of this quantity depends on  the 
systematic error for the normalization of the flux. However, 
in the case of very 
long baseline,
 the largest part of the sensitivity comes from the shape of the spectrum
or how deep the valleys are compared to the peaks (see Figure \ref{wcnodesa}).
Therefore, this determination is not affected  greatly by the systematic error 
for the overall normalization.
This is demonstrated as follows: 
 for $\Delta m^2_{32} = 0.003 ~eV^2$, 
even without background subtraction, the valleys at $\pi /2$ 
and $3\pi/2$ have only 2\% and 30\% of the  un-oscillated event rate
(see Figure \ref{wcnodesa}). 
If  we assume the flux normalization  error to be 5\%, which is consistent 
with what has been achieved by the K2K experiment\cite{sharkey}, 
then the expected error due to flux normalization 
on $\sin^2 2 \theta_{23}$ is 
$0.02\times 0.05 = 0.001$.

\item We note that within the parameter region of interest there should be 
very little correlation in the determination of $\Delta m^2_{32}$ and 
$\sin^2 2 \theta_{32}$. 

\end{itemize}

\begin{figure}
  \begin{center}
    \includegraphics*[width=0.8\textwidth]{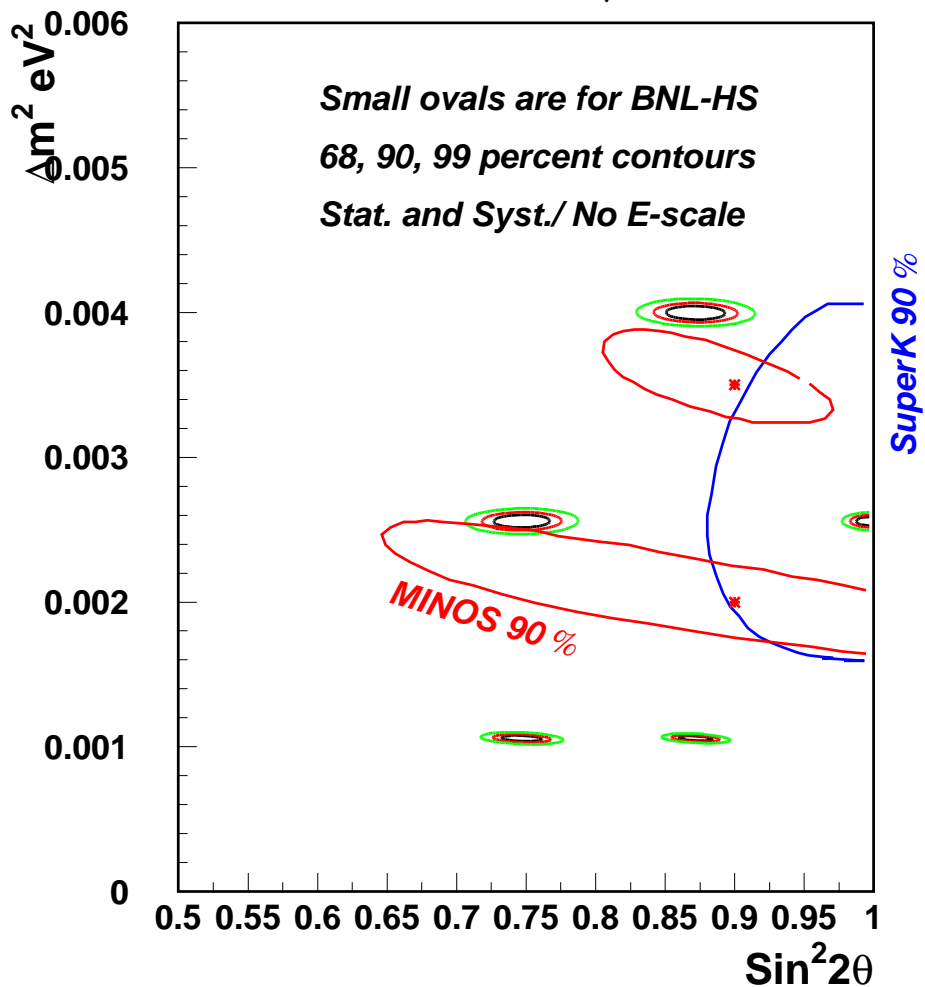}
    \caption[Statistical and systematic uncertainty for $\Delta m^2_{32}$ and $\sin^22\theta_{23}$, includes other's allowed regions.]
    {  Resolution including statistical and systematic effects 
at 68\%, 90\% and 
99\% confidence level on $\Delta m_{32}^2$ and $\sin^2 2\theta_{23}$
for the 2540 baseline experiment; assuming 1 MW, 0.5 MT, and 5 years
of exposure.  We have included a 5\% bin-to-bin systematic 
 uncertainty in the 
energy calibration as well as a 5\% systematic 
uncertainty in the normalization. The expected resolution 
from the MINOS experiment 
at Fermilab and the allowed region from SuperK is also 
indicated. 
}
    \label{cntr2}  
  \end{center}
\end{figure}


\begin{figure}
  \begin{center}
    \includegraphics*[width=0.8\textwidth]{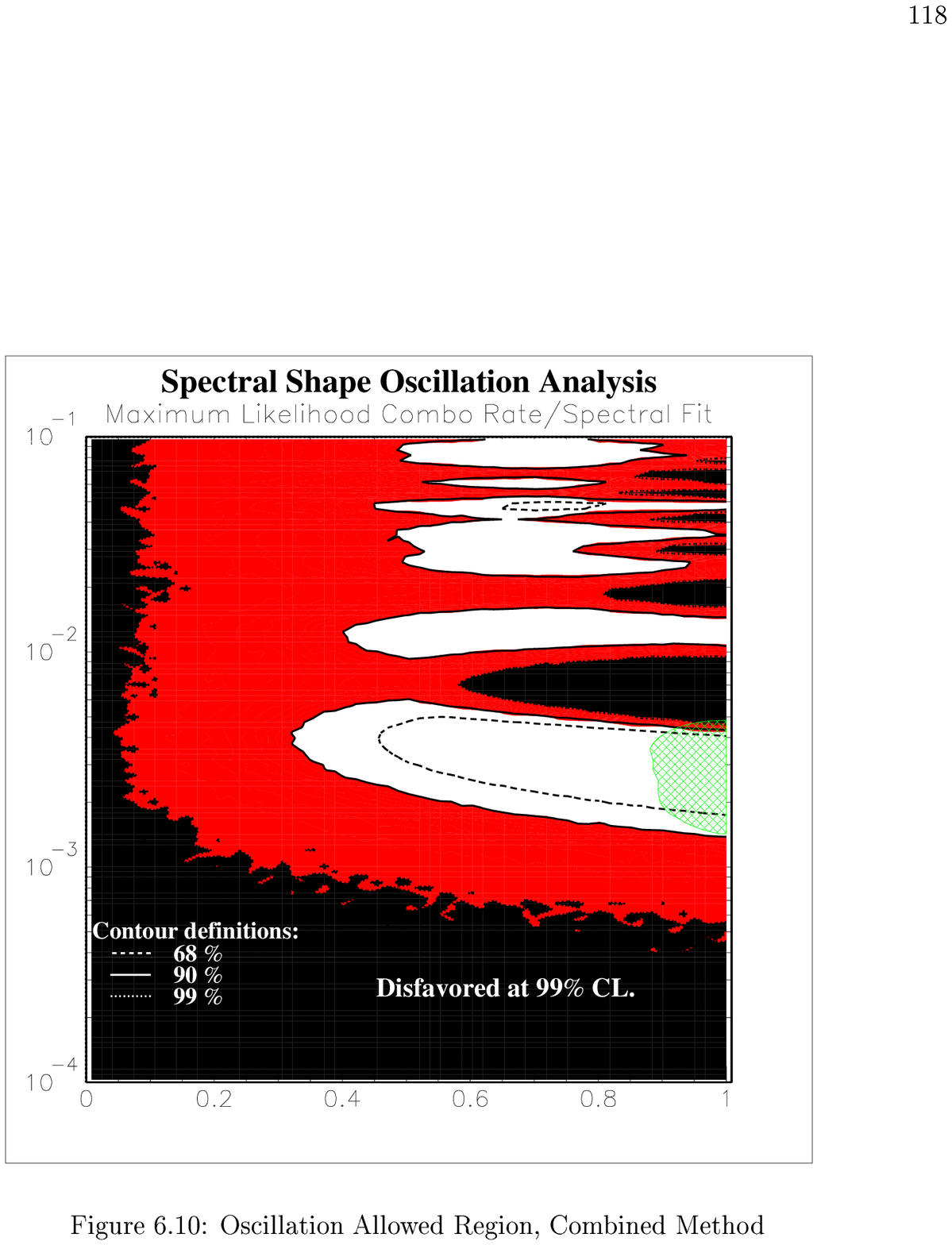}
    \caption[The allowed region from the K2K experiment.]{ The allowed region for $\Delta m^2_{32}$ and $\sin^22\theta_{23}$ from the K2K experiment. 
      From thesis by Eric Sharkey, SUNY at Stony Brook.}
    \label{k2kallowed}  
  \end{center}
\end{figure}

With the assumption on the systematic errors as above we obtain 
Figure \ref{cntr2}. The systematic errors introduce a small correlation in 
the $\Delta m^2_{32}$ vs.  
$\sin^2 2 \theta_{32}$ measurement. The error on the determination of 
$\Delta m^2_{32}$ at 0.0025 $eV^2$ increases to about $\pm 1.2\%$ 
at maximum mixing,
but there is only a small effect on the determination
 of  $\sin^2 2 \theta_{23}$.
As mentioned before, the energy scale uncertainty must be added in quadrature 
to the calculated uncertainty on $\Delta m^2_{32}$. 
The precision of this experiment can be compared 
with the precision expected from 
MINOS (Figure \ref{cntr2}) and the precision obtained so far from the K2K 
experiment (Figure \ref{k2kallowed}). It is expected that K2K will obtain twice
as much data; therefore we could naively estimate that the precision on the 
 parameter determination will improve as $1/\sqrt{2}$.  

Finally, we note that the flux normalization is usually 
obtained by placing a 
detector close to the neutrino source. For example, both K2K and MINOS
have large near detectors to determine the flux. Since
absolute  flux determination 
is not very important for parameter determination in our case, we argue that
the requirements on a  near detector need not be very severe for this 
measurement. It may not be necessary to build a near detector until 
sufficient statistics are obtained in the far detector to demand the
required systematic error reduction of a near detector.

\subsection{$\nu_\mu\to \nu_e$ appearance}

The oscillation of \numunue{} is 
discussed is several recent papers \cite{arafune, marciano, irina, irina2}. 
This oscillation in vacuum is  described fully by 
the following equation:  
\begin{eqnarray}
  \label{eq:one}
P(\nu_\mu\to\nu_e) & = & 4(s^2_{23}s^2_{13}c^2_{13} +J_{CP}\sin\Delta_{21})
\sin^2\frac{\Delta_{31}}{2} \nonumber \\
& & +2(s_{12}s_{23}s_{13}c_{12}c_{23}c^2_{13} \cos\delta -s^2_{12}s^2_{23}s^2_{13}c^2_{13}) \sin
\Delta_{31} \sin \Delta_{21} \label{eq9} \\
& & +4(s^2_{12}c^2_{12}c^2_{23}c^2_{13} +s^4_{12}s^2_{23}s^2_{13}c^2_{13} -2s^3_{12}s_{23}s_{13}c_{12}c_{23}c^2_{13}
\cos\delta -J_{CP} \sin\Delta_{31}) \sin^2\frac{\Delta_{21}}{2}
\nonumber \\
& & +8(s_{12}s_{23}s_{13}c_{12}c_{23}c^2_{13} \cos\delta - s^2_{12}s^2_{23}s^2_{13}c^2_{13}) \sin^2
\frac{\Delta_{31}}{2} \sin^2 \frac{\Delta_{21}}{2} \nonumber
\end{eqnarray}
where 
\begin{equation}
  \label{eq:blah}
  J_{CP} \equiv s_{12}s_{23}s_{13}c_{12}c_{23}c^2_{13}\sin\delta \label{eq6}
\end{equation}

\begin{figure}
  \begin{center}
    \includegraphics*[width=\textwidth]{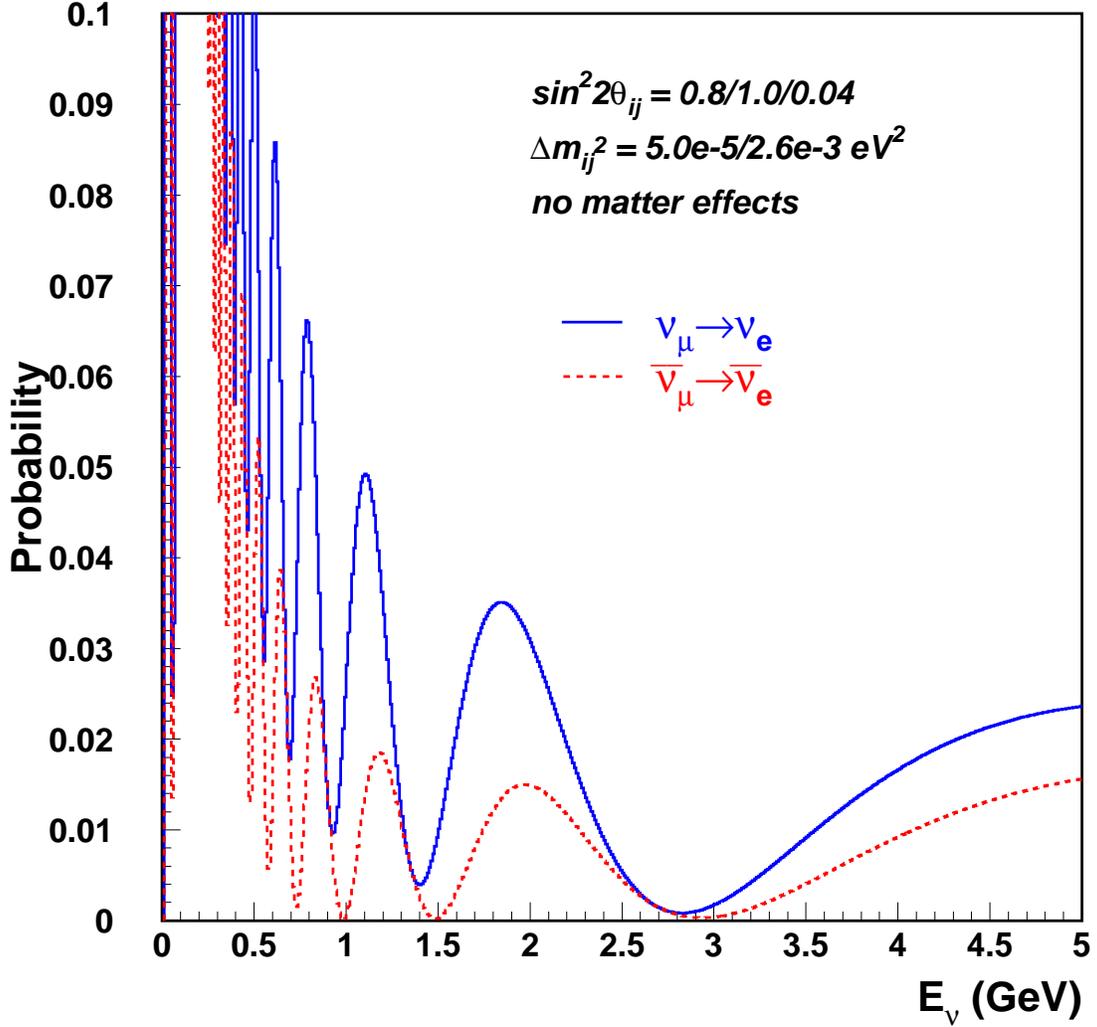} \\
    \caption[P($\nu_\mu \to \nu_e$), $\delta_{CP} = 45^\circ$.]{Probability of \numunue{}
      and \anumunue{}  oscillations at 2540 km in vacuum 
      assuming a $\delta_{CP}=+45^o$ CP violation phase. It can be seen that the 
      CP asymmetry between $\nu_\mu$ and $\bar\nu_\mu$ increases
      for lower energies because the CP asymmetry is proportional
      to $\mdmsol L /E$ which increases for lower energies. 
        The parameters listed in the figure are 
$\sin^2 2 \theta_{12}=0.8$, $\sin^2 2 \theta_{23}=1.0$, and 
$\sin^2 2 \theta_{13}=0.04$ and 
$\Delta m^2_{21}=5.0\times 10 ^{-5}~eV^2$, 
$\Delta m^2_{32}=0.0026~eV^2$.
    }
    \label{cpasym}
  \end{center}
\end{figure}

$J_{CP}$ is an invariant that quantifies CP violation in the neutrino
sector. The abbreviations $s_{ij} \equiv \sin\theta_{ij}$,
 $c_{ij} \equiv \cos\theta_{ij}$,
and $\Delta_{ij}
\equiv \Delta m^2_{ij} L / 2 E_{\nu}$ are used.
The formula for $P(\bar\nu_\mu\to\bar\nu_e)$ is the same as
above except that the $J_{CP}$ terms have opposite sign.   
The vacuum oscillations for a baseline of 2540 km are
illustrated in Figure~\ref{cpasym} as a function of energy for both muon
and anti-muon neutrinos.  The main feature of the oscillation is due
to the term linear in $\sin^2\frac{\Delta_{31}}{2}$. The oscillation
probability rises for lower energies due to the terms linear in
$\sin^2 \frac{\Delta_{21}}{2}$.  The interference terms involve CP
violation and they create an asymmetry between neutrinos and
anti-neutrinos.  The vacuum oscillation formula (Eq.\ref{eq:one})
  and Figure~\ref{cpasym}
show that the CP asymmetry also grows as $1/E$ in the 0.5-3.0 \GeV{}
region.  The parameters listed in the figure are 
$\sin^2 2 \theta_{12}=0.8$, $\sin^2 2 \theta_{23}=1.0$, and 
$\sin^2 2 \theta_{13}=0.04$ and 
$\Delta m^2_{21}=5.0\times 10 ^{-5}~eV^2$, 
$\Delta m^2_{32}=0.0026~eV^2$.
Similar notation for parameters will be followed in the rest of the document.
Because of this effect it is argued that the figure of merit
for measuring CP violation is independent of the baseline. For very
long baselines  the statistics for a given size detector at a
given energy are poorer by one over the square of the distance, but 
the CP asymmetry grows linearly in distance \cite{marciano}.
The background to the electron neutrino signal comes from
contamination in the beam ($\nu_e/\nu_\mu \sim 0.7\%$) and 
neutral current events. At small distances the 
systematic error on this background could limit the ability to extract 
the CP violating effect, but at large distance the background 
reduces as $1/(distance)^2$ and allows us to greater sensitivity to CP 
violating effects. We rely on this important observation in the rest of this 
section.

\begin{figure}
  \begin{center}
    \includegraphics*[width=\textwidth]{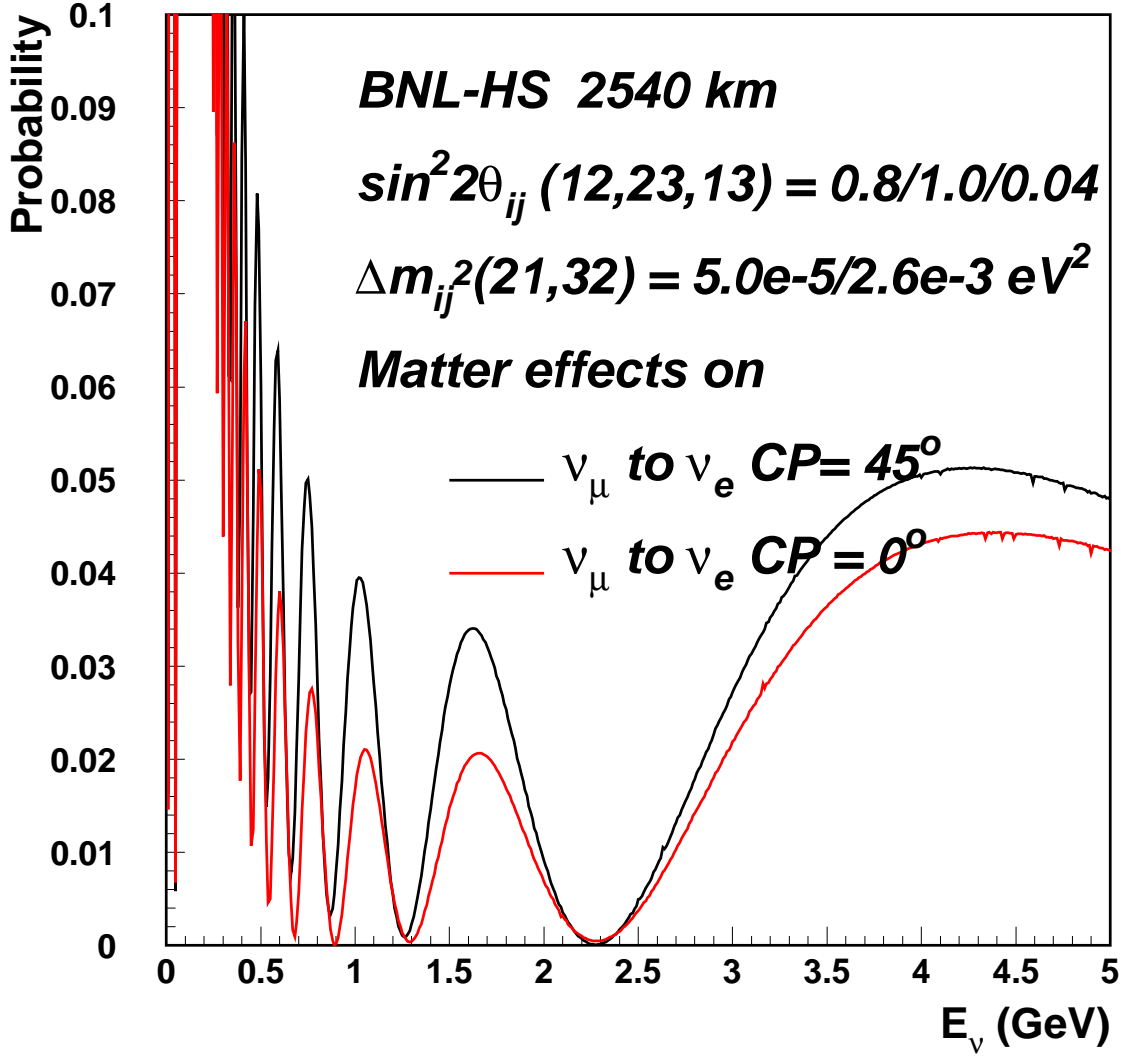}
    \caption[P($\nu_\mu \to \nu_e$), $\delta_{CP} = 0,45^\circ$, matter effects.]
    {Probability of $\nu_\mu$ oscillating into 
      $\nu_e$ after 2540 km.  The parameters assumed are listed in the
      figures.  The upper and lower  curves correspond to CP phase angle of 
$+45^o$ and $0^o$ respectively.  We point out that the effect of 
CP phase increases for lower energies. 
}
    \label{pnumunuea}
  \end{center}
\end{figure}
\begin{figure}
  \begin{center}
    \includegraphics*[width=\textwidth]{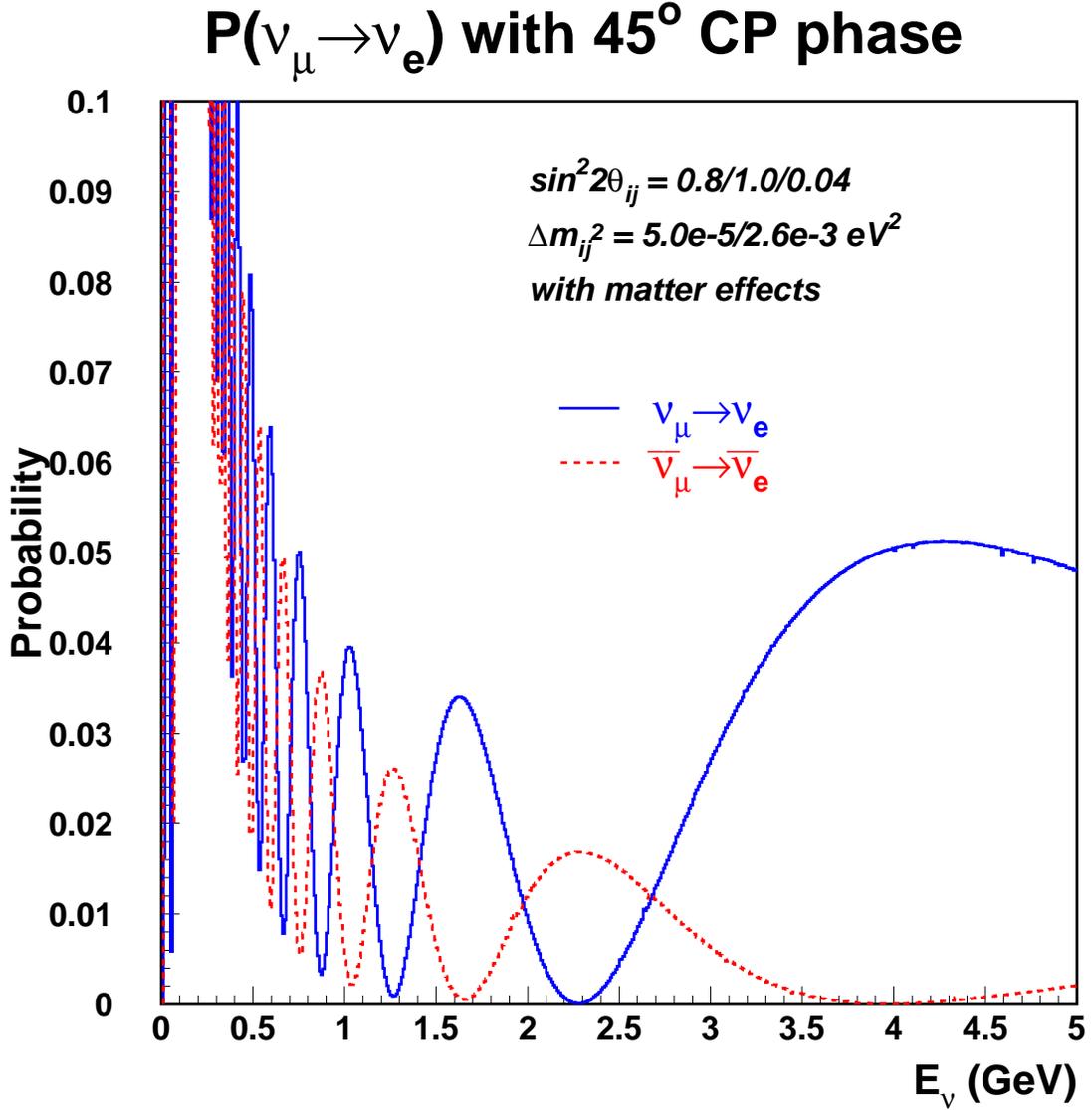}
    \caption[P($\nu_\mu \to \nu_e$) and P($\bar\nu_\mu \to \bar\nu_e$), $\delta_{CP} = 45^\circ$, matter effects.]
    {Probability of $\nu_\mu$ oscillating into 
      $\nu_e$ after 2540 km.  The parameters assumed are listed in the
      figures.  This plot assumes a CP violation phase of $+45^o$. 
The upper and lower curves are for neutrinos and anti-neutrinos, 
respectively. We see that for distance of 2540 the matter effects will be 
large and will lead to almost complete reversal of nodes and anti-nodes 
for neutrinos and anti-neutrinos. The probability for neutrinos with 
$\Delta m^2_{32} < 0$ will be similar to (but not exactly the same as)  anti-neutrinos.}
    \label{pnumunueb}
  \end{center}
\end{figure}

The vacuum oscillation formulation must be modified to include the
effect of matter \cite{irina}.  The \numunue{} probability in the
presence of matter is shown in Figures~\ref{pnumunuea} and
\ref{pnumunueb}. When compared to Figure~\ref{cpasym} we can see that
matter will enhance (suppress) neutrino (anti-neutrino) conversion at
high energies and will also lower (increase) the energy at which the
oscillation maximum occurs.  The effect is opposite (enhancement for
anti-neutrinos and suppression for neutrinos) if the sign of \dmatm{}
is negative.  The Figures \ref{cpasym} to \ref{pnumunueb} gives us hints about possible 
strategies in understanding neutrino oscillation parameters.  

In the low energy region from 0 to 1.0 GeV, the probability for 
$\nu_\mu \to \nu_e$ is dominated by the effects of $\Delta m^2_{21}$ 
if the solution to the solar neutrino deficit is the large mixing angle 
(LMA) solution. An excess of electron like events in this region 
would be sensitive to $\Delta m^2_{21}$ and $\sin^2 2 \theta_{12}$.  

In the intermediate energy region from 1.0 to 3.0 GeV, we see that 
the CP violating phase $\delta_{CP}$ has a large effect on the 
oscillation probability and the effects of matter 
are relatively small. Therefore this energy region could be used 
to measure the CP violating phase $\delta_{CP}$ from the 
observed spectrum of electron like events. 
 
The higher energy region with energy greater than 3.0 GeV
is clearly the region of discovery for $\nu_\mu \to \nu_e$ oscillations
as well as the sign of $\Delta m^2_{32}$. 
In the case of the normal 
mass hierarchy ($m_3 > m_2 > m_1$) the oscillation signal in the high
energy region for neutrinos will be enhanced by more than a factor of 2. 
Moreover, as we will discuss below, the backgrounds from 
both neutral currents and intrinsic $\nu_e$ will fall 
in this region. Therefore the appearance signal will have a 
distinctive shape to distinguish it from the background.  
In the case of ($ m_2 > m_1 > m_3$) the oscillation signal in the high
energy region will be almost completely suppressed. However, there will be 
a peak between 2 and 3 GeV. If $\sin^2 2 \theta_{13}$ is sufficiently 
large, this will be a clear signature for $\Delta m^2_{32} < 0$, 
a very important result in particle physics. 

Finally, matter enhancement of the oscillations has been postulated for a long time 
without experimental confirmation~\cite{wolfenstein}.  Detection of 
such an effect by measuring a large asymmetry between neutrino and 
anti-neutrino oscillations or by measuring the spectrum of electron 
neutrinos is a major goal for neutrino physics. This measurement will 
also yield the sign of \dmatm{}.

\subsection{Backgrounds}

While the $\nu_\mu$ disappearance result will be 
affected by systematic errors, the $\nu_\mu \to \nu_e$ appearance 
result will be affected mainly by the backgrounds.
The signal we are looking for consists of clean, single ring electron events in 
the detector. The signal will mainly result from the 
quasielastic reaction $\nu_e + n \to e^- + p$.  The main backgrounds 
will be from neutral current reactions and the intrinsic electron 
neutrinos in the beam. Most of the $\sim$ 17000 neutral current reactions 
from Table \ref{evcount} are either elastic scattering off nucleons or 
single pion production channels. Of these, the  channels that produce 
single $\pi^0$  will be the major source of backgrounds. We estimate that 
approximately 
2800 NC events will have multiple pions in the final state. 
Half of these will have at least one $\pi^0$. 
We expect that  these can be rejected much more effectively 
than the single $\pi^0$ production 
channels which will have $\sim$ 3700 events (see Table \ref{evcount}). 
This number includes the coherent production channel of 
$\nu_\mu + O^{16} \to \nu_\mu + O^{16} + \pi^0$. 
The charged current background channel,
 $\nu_\mu + n \to \mu^- + p + \pi^0$, 
in which the muon remains invisible was shown to be small for a similar beam spectrum
in the E889 proposal \cite{e889}.

\begin{figure}
    \includegraphics*[width=\textwidth]{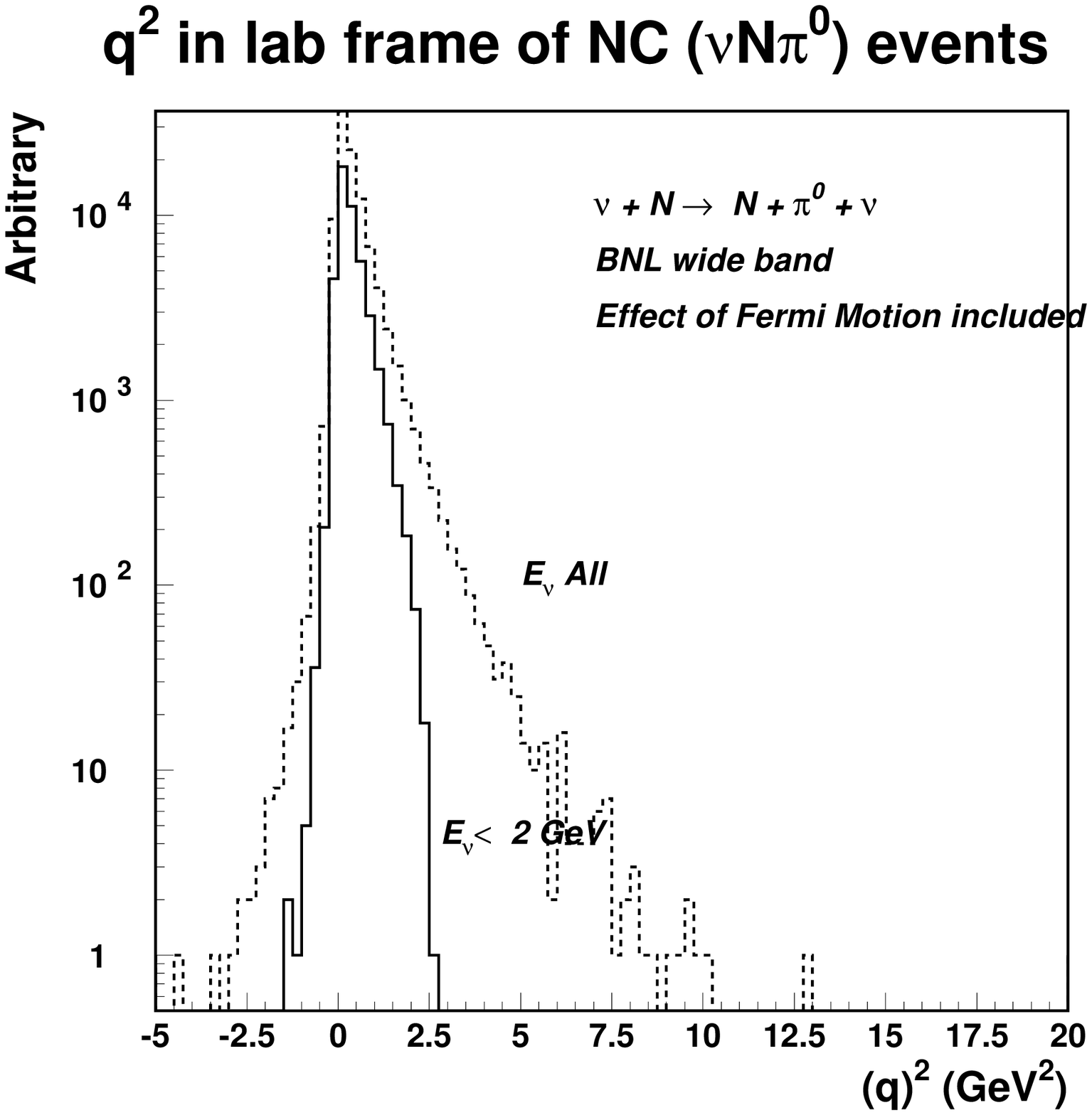}
\caption[The $q^2$ distribution for NC($\nu$N$\pi^0$) events]
{The $q^2$ distribution of $\nu_\mu + N \to \nu_\mu + N + \pi^0$ 
channels. Here $q^2 = ((p'_N + p'_\pi) - p_N)^2$. $p_N$ is the 
initial 4 momentum of the target nucleon (assumed to be at rest in 
the lab frame). $p'_N$ and $p'_\pi$ are the 4-momenta of the 
final state nucleon and pion, respectively. 
The peak of the distribution is 
independent of neutrino energy. The neutrino energy 
only determines the physical cutoff of the $q^2$ distribution.
The slightly negative behavior of the distribution is caused by 
the Fermi motion of the target nucleus which was assumed to be at rest in 
the above formula.} 
\label{ncq2}
\end{figure}

\begin{figure}
    \includegraphics*[width=\textwidth]{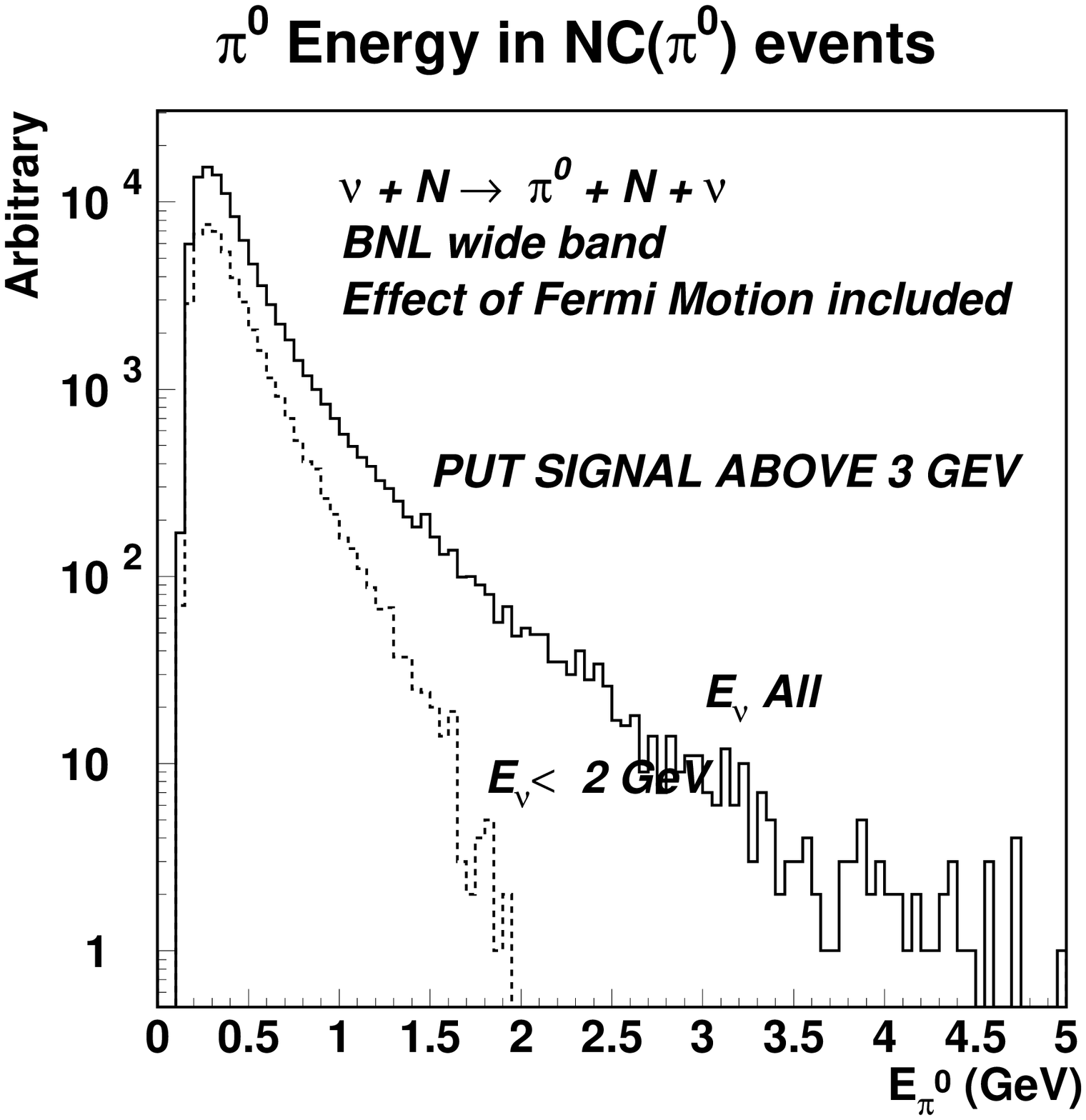}
\caption[NC($\pi^0$) energy spectrum]
{The $\pi^0$ energy 
 distribution of $\nu_\mu + N \to \nu_\mu + N + \pi^0$ 
channels with no cuts. 
 The peak of the distribution is 
independent of neutrino energy. The neutrino energy 
 determines the high energy cutoff of the distribution.
The distribution is about 3 orders of magnitude 
suppressed above 2.5 GeV where we expect the 
signal from  $\nu_\mu \to \nu_e$ appearance. 
}
\label{pi0e}
\end{figure}

For a baseline of 2540 km, the matter enhanced oscillation 
signal will be above 3 GeV. 
Our strategy for obtaining a unique, clear signal therefore 
depends on the observation that 
 neutral current background  will peak at low energies and fall 
rapidly as a function of observed energy. 
This is demonstrated in Figures \ref{ncq2} and \ref{pi0e} for 
the neutral current single pion production channel. 
In Figure  \ref{ncq2} we see that the $q^2$ distribution peaks at 
low values and is nearly independent of the  neutrino energy. 
The neutrino energy 
only determines the kinematic limit of the $q^2$ value.
This behavior leads most neutral current events to be at low energies.

Figure \ref{pi0e} shows the distribution of total $\pi^0$ energy for 
single pion production events with no detector cuts. We see that 
the distribution is about 3 orders of magnitude 
suppressed above 2.5 GeV where we expect the 
signal from  $\nu_\mu \to \nu_e$ appearance
(see Figure \ref{pnumunuea}). Therefore, we 
propose that even a modest rejection of neutral current background 
above 2.5 GeV is sufficient to provide us with good sensitivity 
for $\nu_\mu \to \nu_e$ appearance. 

\begin{figure}
  \begin{center}
    \includegraphics*[width=0.49\textwidth]{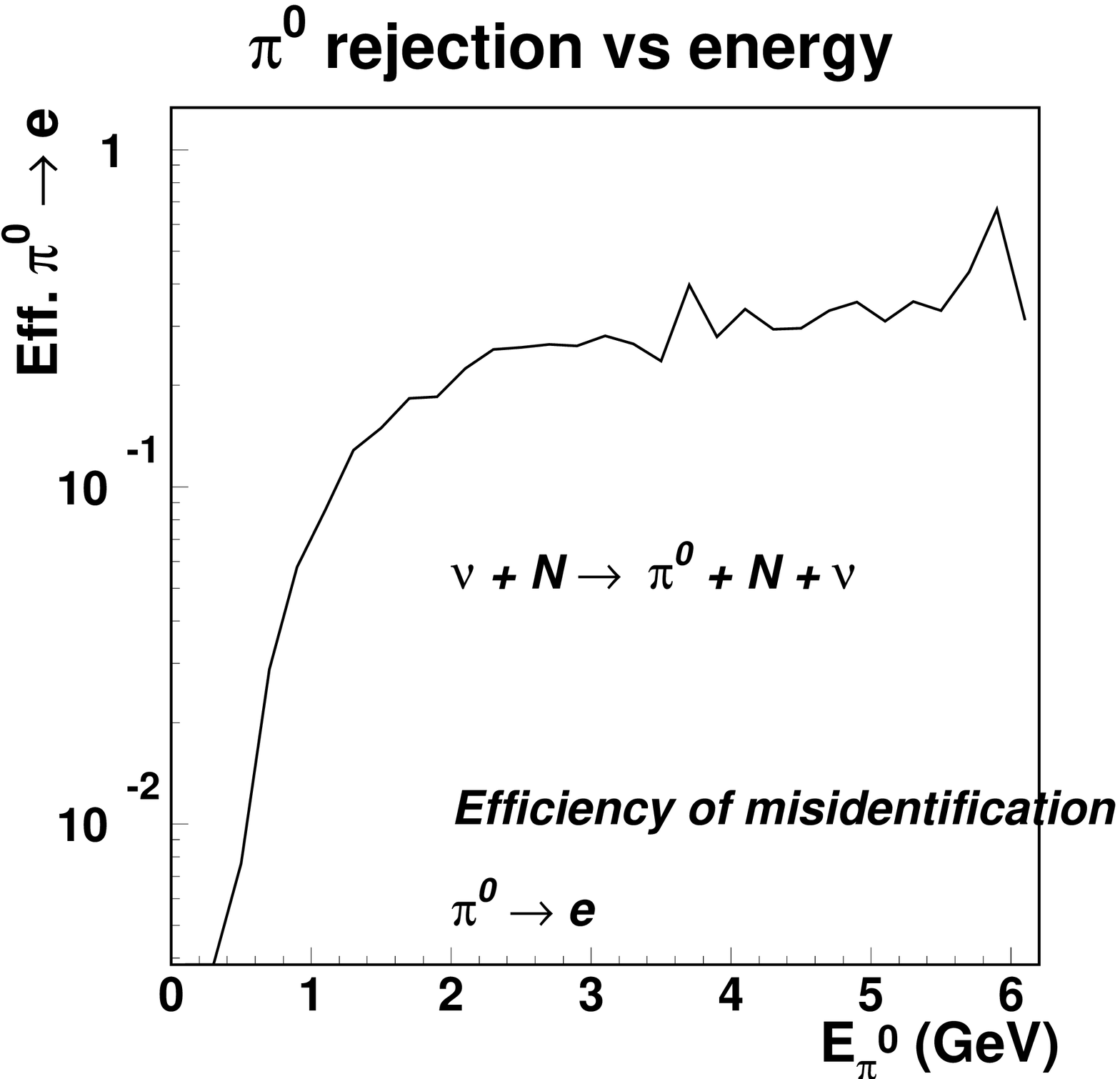}
    \includegraphics*[width=0.49\textwidth]{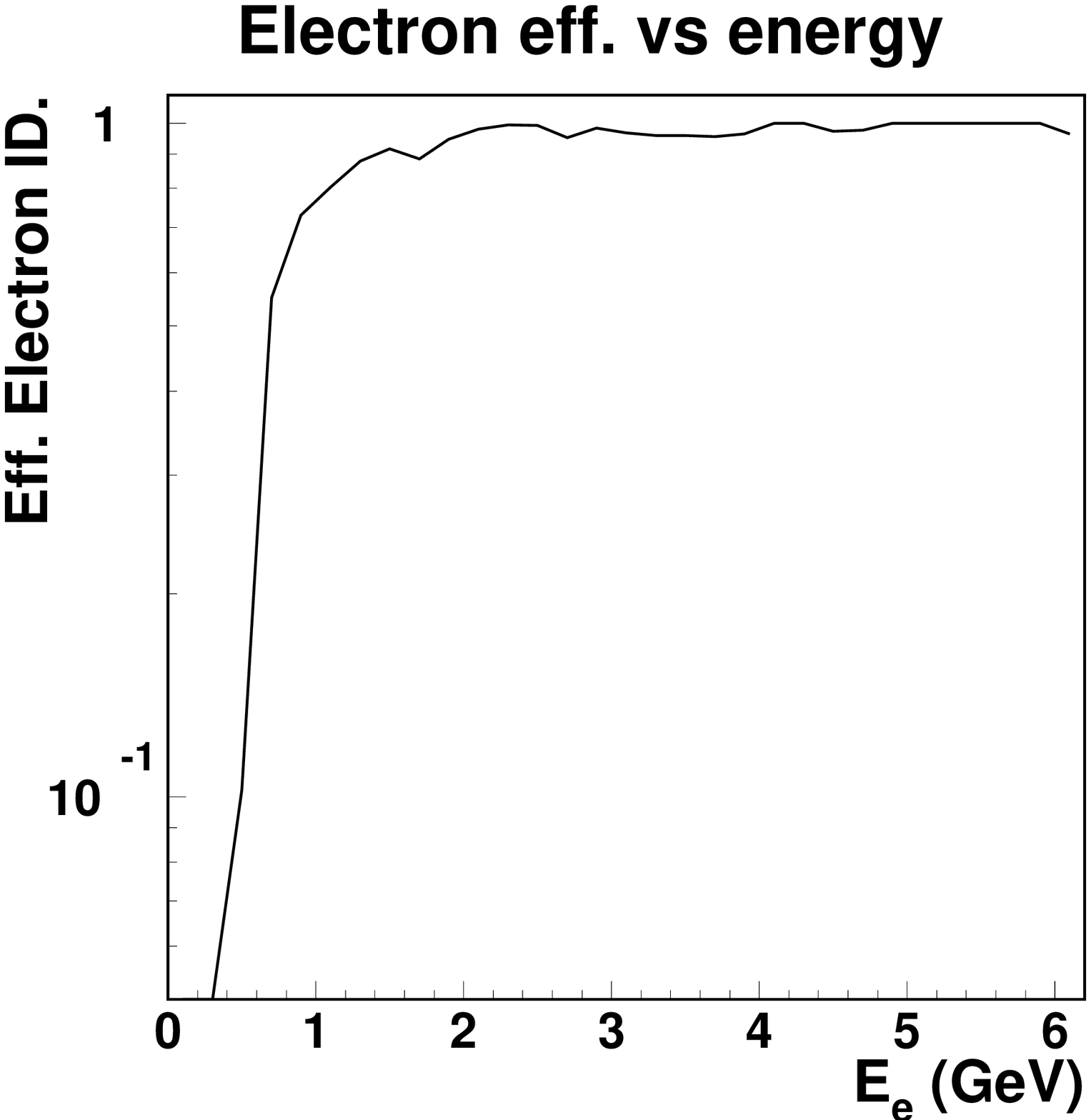} 
    \caption[$\pi^0$ misidentification probability and electron efficiency.]
    {On the left: the rate of misidentification of 
$\pi^0$ events as electrons versus total $\pi^0$ energy 
 for the calculations in this
paper. On the right: electron efficiency used in this calculation.  }
   \label{pirej}
  \end{center}
\end{figure}

This modest rejection can be obtained by first cutting all 
events with visible energy less than 500 MeV. Further 
rejection is obtained by 
getting rid of events with two showers each with energy greater than 
150 MeV separated by more than 9 degrees in angle and by cutting events 
with angle between the shower and the neutrino direction of greater than 
60 degrees; this was 
calculated using a fast Monte Carlo with appropriate angle and energy
resolution corresponding to a water \cerenkov{} detector. At high energies, 
above 3 GeV, a full simulation of a large water \cerenkov{} 
detector showed 
us that it is possible to obtain about a 50\% rejection based on the 
\cerenkov{} ring characteristics.  The overall rate of $\pi^0$ 
misidentification is shown in Figure \ref{pirej}. 

\begin{figure}
  \begin{center}
    \includegraphics*[width=\textwidth]{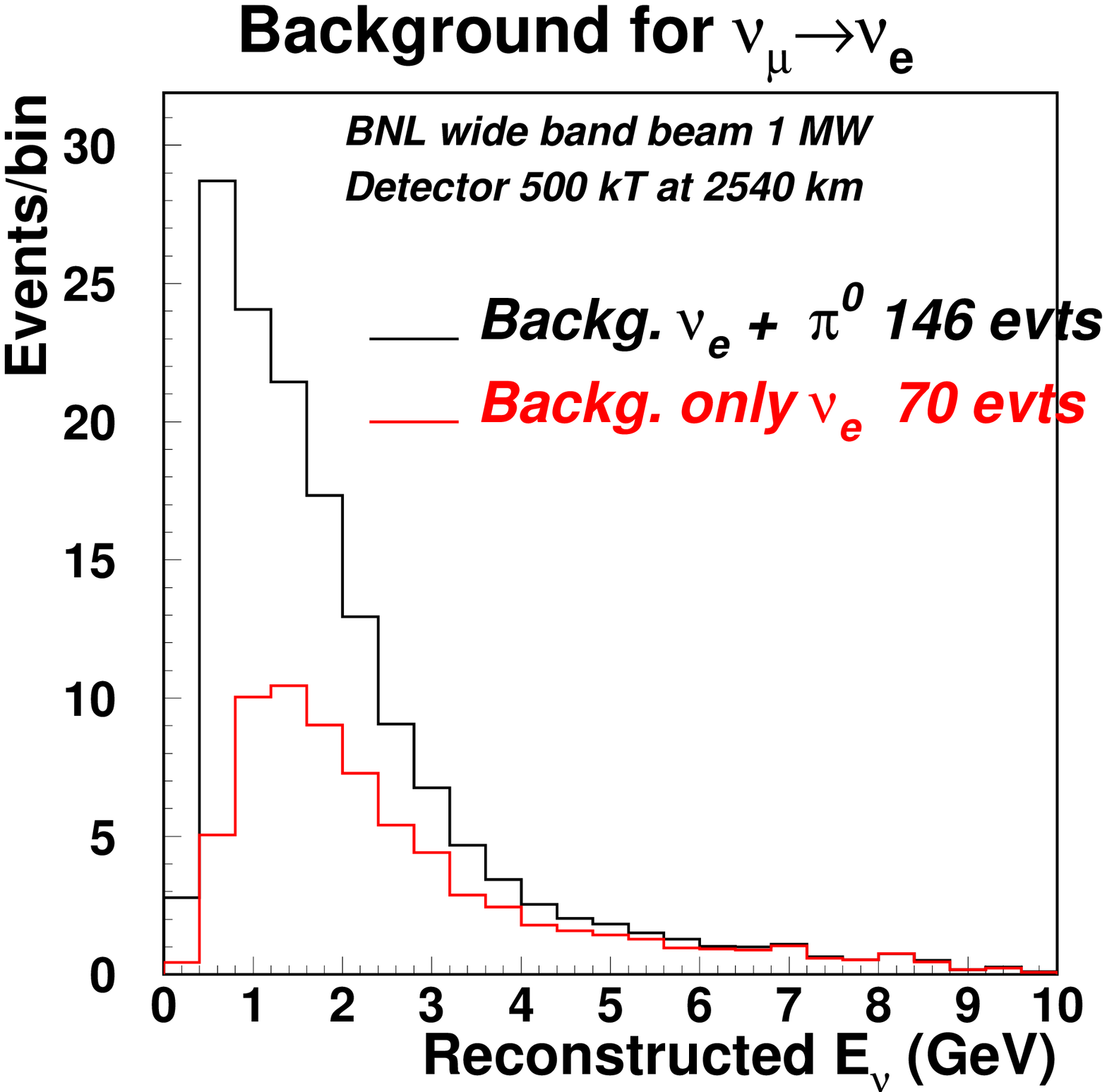}
    \caption[$\nu_\mu \to \nu_e$ background reconstructed energy spectrum.]
{ Spectrum of reconstructed electron neutrino energy (assuming 
quasielastic events) of the background for $\nu_\mu \to \nu_e$ search. 
This is for 1 MW beam power, 0.5 MT detectors mass and $5\times 10^7$ 
sec of running. 
The top histogram includes both the NC($\pi^0$) and electron 
contamination backgrounds. 
The electron neutrino contamination is also shown separately. }
    \label{allbck}
  \end{center}
\end{figure}

It should be noted 
that the advantage of the very long baseline is in applying 
a simple cut on the total visible energy to eliminate most of the 
background.  The rate of 
$\pi^0$ misidentification for neutral current events (Figure \ref{pi0e}) 
above 500 MeV is  6\%. 
The efficiency for electrons is shown on the right hand side of 
Figure \ref{pirej}. The efficiency for quasielastic 
electron neutrino events is  64\%  at 
energy less than 1.5 GeV. Above 1.5 GeV the efficiency is 90\%.  
Using appropriate resolution and  efficiency factors we obtain the 
predicted background spectrum of electron like showering events in 
Figure \ref{allbck}.
The reconstructed  electron energy and 
the angle of the electron with respect to 
the neutrino direction is used to reconstruct the neutrino energy assuming a 
quasielastic scattering event.   
 Figure \ref{allbck} includes backgrounds from  the 
neutral current single $\pi^0$ production off nucleon as well as coherent 
$\pi^0$ production off $O^{16}$, which has a much more energetic spectrum. 
The spectrum also includes the background from $\nu_e$
 contamination in the beam.

The predicted number of total background events
 is 146 with the beam-$\nu_e $ contamination 
accounting for 70 events. 
It should be remarked that above 2 GeV the background 
is dominated  by the beam-$\nu_e$ contamination: there are 
35 $\nu_e$ events versus 
17 $\pi^0$ events. This is despite the 
rather poor rejection of NC($\pi^0$) events 
at high energies. 
Below 2 GeV the background will be dominated by the  NC($\pi^0$) events: with 
35 $\nu_e $ events and 59  $\pi^0$ events. 
Therefore any error in the determination of the NC($\pi^0$) background including 
contamination from other neutral current background channels (which will have 
similar energy dependence) will not significantly 
affect the high energy region above 2 GeV  where we expect to see a distinct 
signal for electron neutrino appearance.

\subsection{Sensitivity to $\sin^2 2 \theta_{13}$ }

\begin{figure}
  \begin{center}
    \includegraphics*[width=\textwidth]{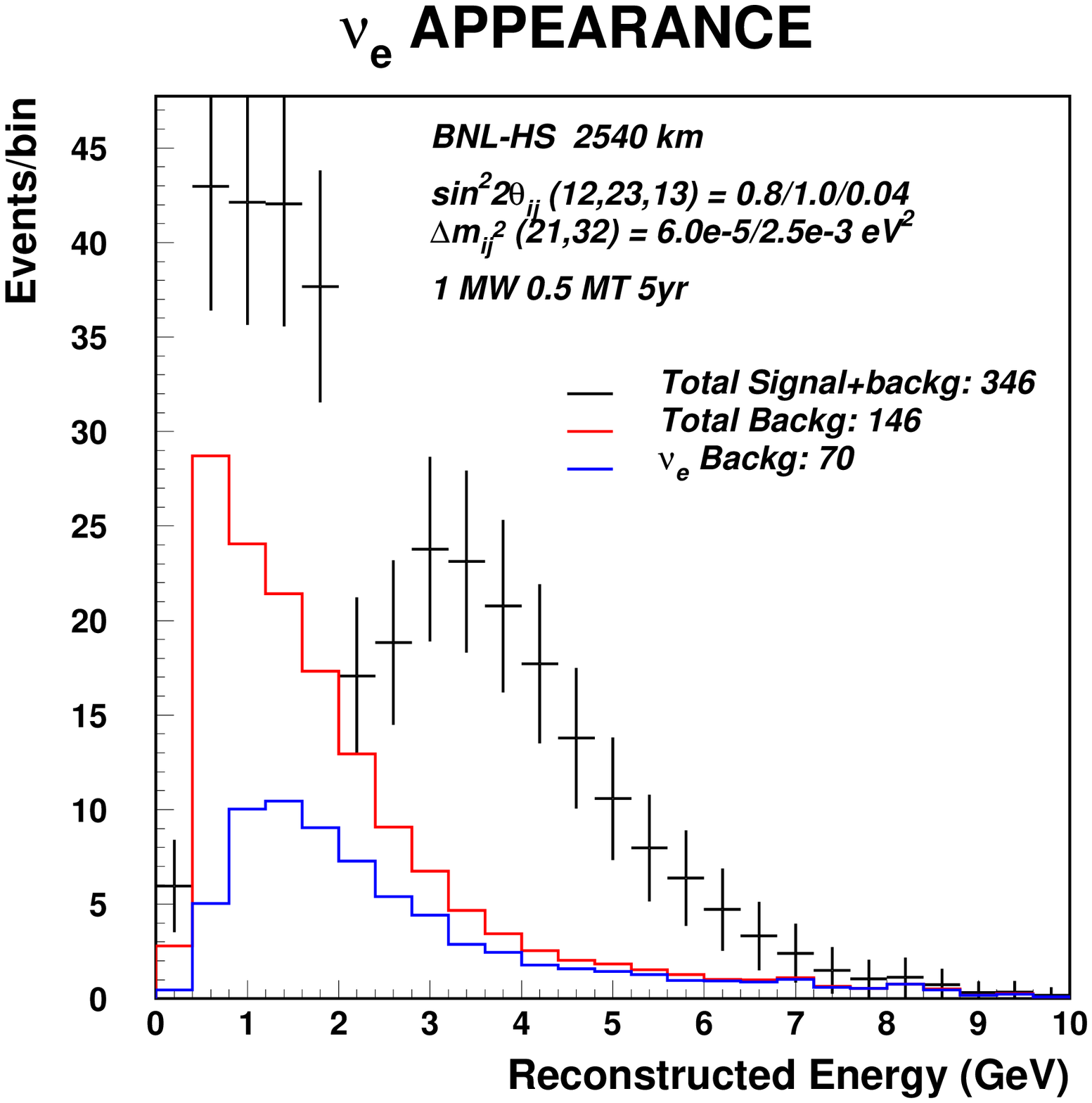}
    \caption[$\nu_\mu \to \nu_e$ appearance spectrum $\Delta m^2_{32} = 0.0025$]
    {Spectrum of detected quasi-elastic electron neutrino 
      charged current events in a 0.5 MT detector at 2540 km from BNL.
      We have assumed 1 MW of beam power and 5 nominal years of running.
      This plot is for $\mdmatm = 0.0025~\meV^2$.  We have assumed
      $\sin^2 2 \theta_{13} = 0.04$ and 
$\mdmsol = 6\times 10^{-5}~\meV^2$. 
 The error bars correspond to the
      statistical error expected in the bin. The spectrum includes 
effects of Fermi motion, energy resolution and efficiency. 
    }
    \label{nuenodesa}
  \end{center}
\end{figure}

\begin{figure}
  \begin{center}
    \includegraphics*[width=\textwidth]{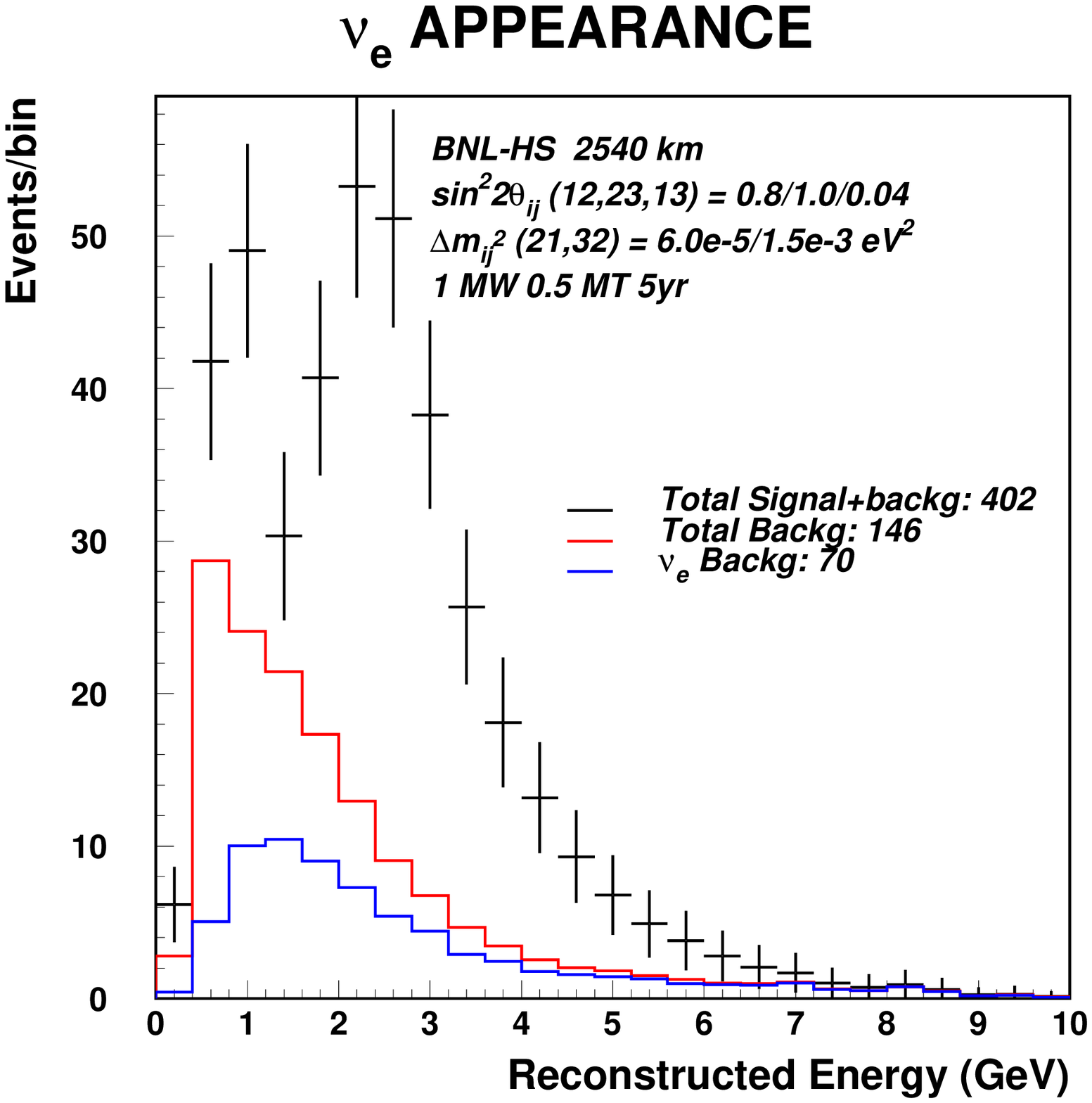}
    \caption[$\nu_\mu \to \nu_e$ appearance spectrum $\Delta m^2_{32} = 0.0015$]
    {Spectrum of detected quasielastic electron neutrino 
      charged current events in a 0.5 MT detector at 2540 km from BNL.
      We have assumed 1 MW of beam power and 5 nominal years of running.
      This plot is for $\mdmatm = 0.0015~\meV^2$.  We have assumed
      $\sin^2 2 \theta_{13} = 0.04$ and  
$\mdmsol = 6\times 10^{-5}~\meV^2$. }
    \label{nuenodesb}
  \end{center}
\end{figure}

Figures \ref{nuenodesa} and \ref{nuenodesb} show
 the spectrum of electron like 
events  that will be detected at 2540 km.  The signal for
$\Delta m^2_{32}=0.0025~eV^2$ and 
$\sin^2 2 \theta_{13} \sim 0.04$ will be about 200 events.
The advantages of the very long baseline are in obtaining a large
enhancement at higher energies and creating a nodal pattern in the
appearance spectrum. Both of these can be used to further improve the
sensitivity of the experiment. 
It should be noted that the value of $\Delta m^2_{32}$ will be 
known very precisely from the disappearance measurement; this value 
can then be used to precisely predict the shape of the spectrum 
of electron-like events. Unlike past experiments in 
which only a simple counting of signal over background was performed, 
 the node pattern in this experiment will  be a strong 
confirmation of $\nu_\mu \to \nu_e$. 
The broadband beam also allows for sensitivity over a broad range of 
  $\Delta m^2_{32}$. This can be seen in Figure \ref{nuenodesb}.

\begin{figure}
  \begin{center}
    \includegraphics*[width=\textwidth]{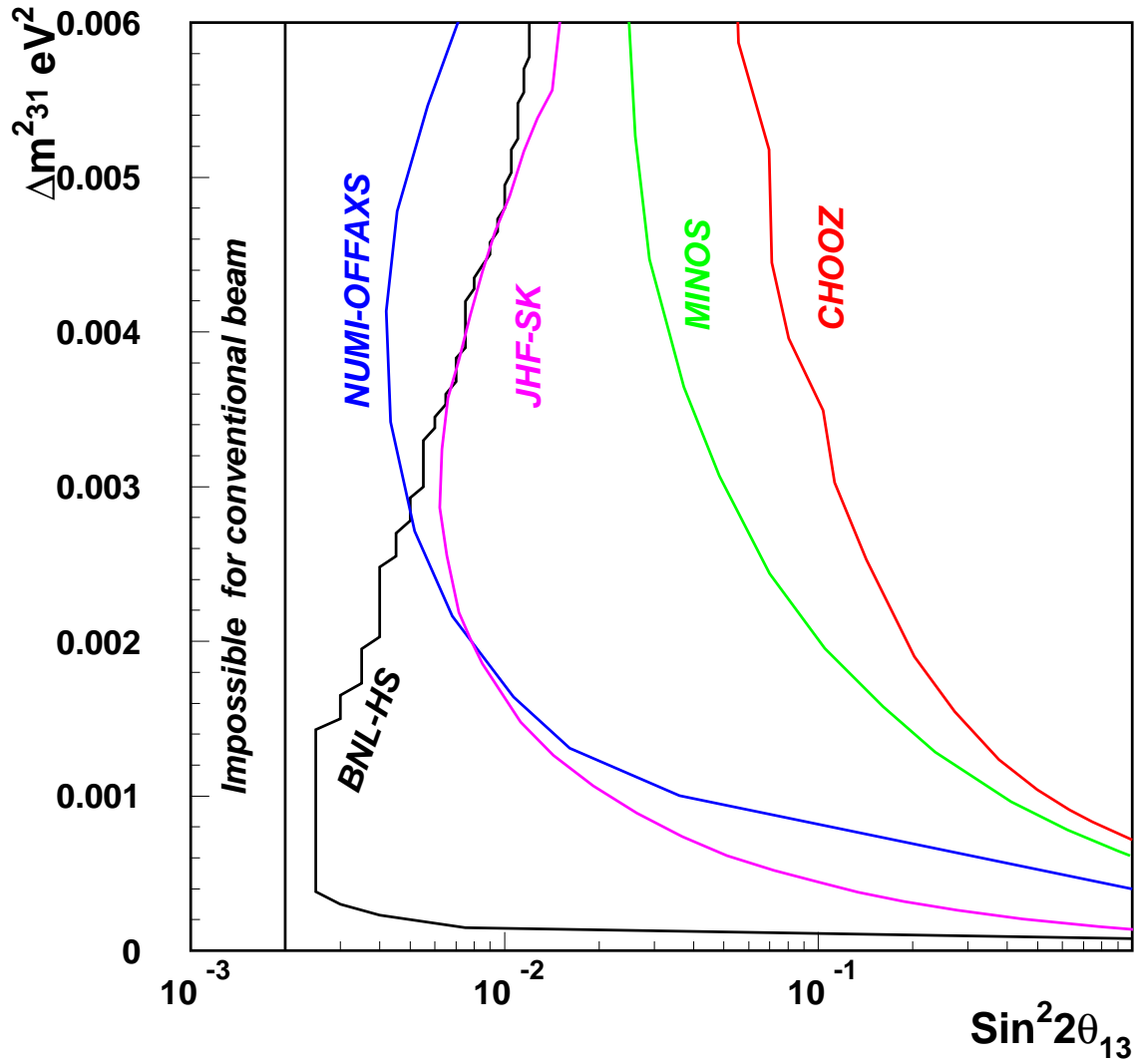}
    \caption[Sensitivity to $\sin^2 2 \theta_{13}$ and $\Delta m^2_{31}$]
{Expected 90\% confidence level upper limit
on $\sin^2 2 \theta_{13}$ versus $\Delta m^2_{31}$ for the 
BNL-to-Homestake experiment compared to other proposed 
experiments. The current limit from the CHOOZ reactor experiment 
is also shown on the same plot. 
    }
    \label{limit1}
  \end{center}
\end{figure}

We calculated the background electron spectrum assuming 
$\sin^2 2 \theta_{13}=0$; then we varied the 
parameters, $\Delta m^2_{31}$ and $\sin^2 2 \theta_{13}$, 
and calculated the $\chi^2$ with respect to the background 
spectrum. 
The other parameters in this calculation were set as follows: 
$\Delta m^2_{21}=6\times 10^{5}~eV^2$, 
$\sin^2 2 \theta_{12}=0.8$, 
$\sin^2 2 \theta_{23}=1.0$ and $\delta_{CP}=0$. 
We assumed that the remaining parameters will be well-known 
from other experiments. However, the small uncertainty
on $\Delta m^2_{21}$
will cause us to lose sensitivity to $\sin^2 2 \theta_{13}$ 
at values of $\Delta m^2_{32} < 0.001~eV^2$, outside the 
region favored by SuperK. 
For the calculation
 we assume a 10\% systematic error (in addition to the 
statistical error) on the background 
spectrum of events. This level of 
systematic uncertainty is attainable with a  modest sized near detector
and it compares well with proposals for other such experiments.  
 The 90\% 
confidence level upper limit obtained from this calculation is 
shown in Figure \ref{limit1}.  The same figure also shows 
the sensitivities of several other proposed  experiments as well as
the current best limit from the CHOOZ reactor 
experiment. The current upper limit at $\Delta m^2_{31} = 0.0025~eV^2$ 
is $\sin^2 2 \theta_{13} = 0.12$. 
It should be noted that if $\Delta m^2_{32}$ is lower 
the current limit becomes much poorer. 
(We will use 
the values $\sin^2 2 \theta_{13} = 0.04 $ and $\sin^2 2 \theta_{13} = 0.06$,
which are a factor 1/3 and 1/2 below the current limit as benchmark
points for some of the plots.)

The sensitivity shown in Figure \ref{limit1} can be divided in 
two regions: above $\Delta m^2_{32} = 0.0015~eV^2$
(in the parameter region preferred by the SuperK data)
 the electron 
spectrum shape will be very distinct and show at least 
two clear  nodes;
below $\Delta m^2_{32} = 0.0015~eV^2$ the statistics will be 
larger and we will get a better limit, however the signal will not 
have the distinct shape that will be a strong confirmation of 
an oscillation signal.  Moreover, the $\sin^2 2 \theta_{13}$
measurement in the lower region could be correlated with $\Delta m^2_{21}$.

\begin{figure}
  \begin{center}
    \includegraphics*[width=\textwidth]{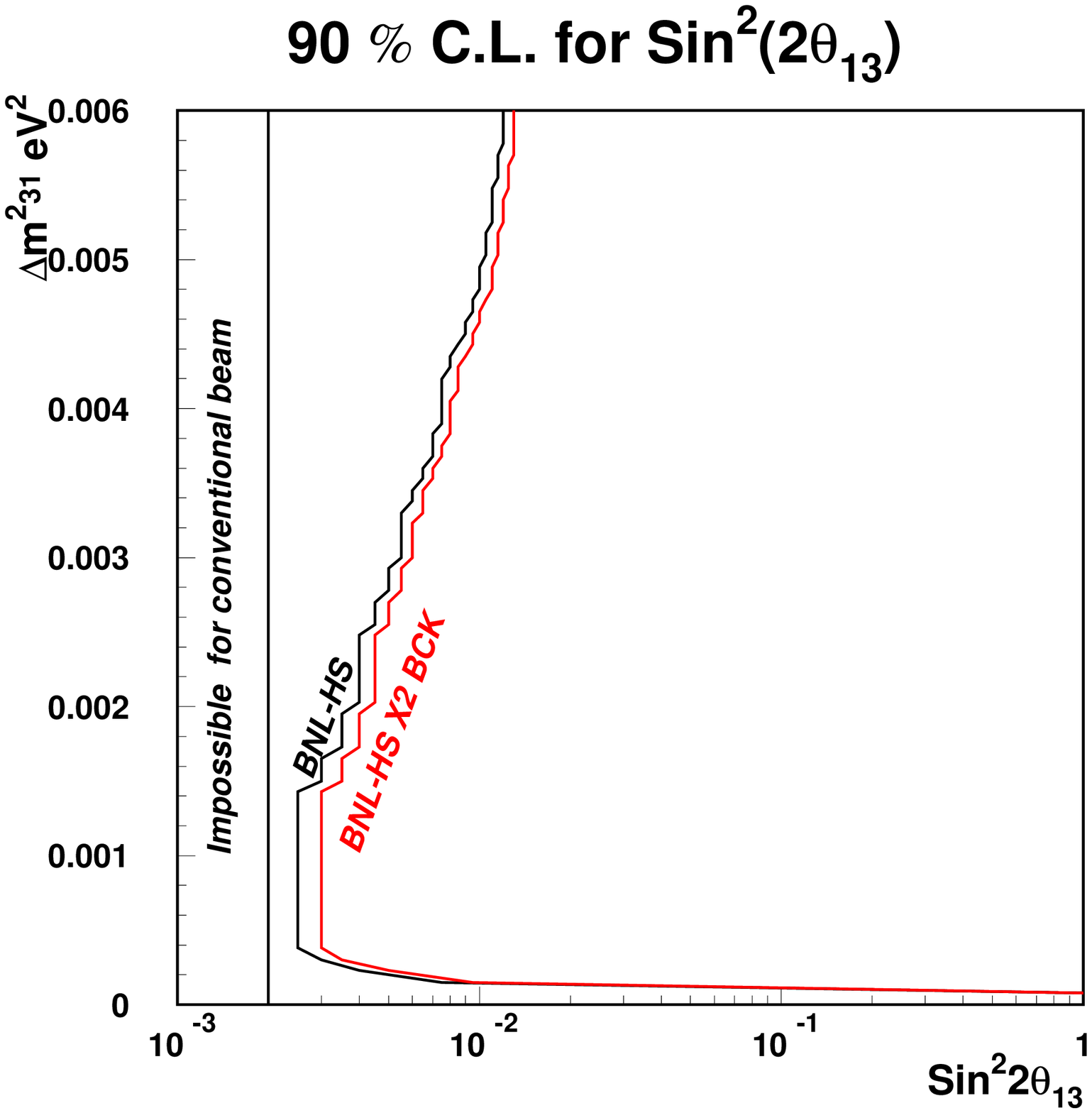}
    \caption[Sensitivity to $\sin^2 2 \theta_{13}$ and $\Delta m^2_{31}$ with $\times$2 NC background.]
    {Expected 90\% confidence level upper limit
on $\sin^2 2 \theta_{13}$ versus $\Delta m^2_{31}$ for the 
BNL-to-Homestake experiment. The two curves are 
with the background as predicted in Fig. \ref{allbck}
(the left hand curve) and 
assuming the neutral current background to be a factor of 
two larger (the curve to the right). 
    }
    \label{limit2}
  \end{center}
\end{figure}

The sensitivity for the BNL-to-Homestake experiment declines 
as $\Delta m^2_{32}$ becomes larger and the first oscillation 
node moves to higher energies where our spectrum has 
much lower flux. This can be improved by adding more focusing 
elements to the horn-produced beam to increase the high 
energy flux; however, this will increase the background for the 
lower energy events. We are in the process of   performing
 these  optimization studies to determine the best spectrum shape for 
this experiment. 
Lastly, we note that the sensitivity does not depend strongly on the 
amount of neutral current background. This is shown in 
Figure \ref{limit2} where we have calculated the 90\% confidence 
level upper limit assuming that the the neutral current background 
is twice as high as in Figure \ref{allbck}.  
This is because the spectrum is already dominated by the intrinsic 
$\nu_e$ background in the higher energy region above 2 GeV.
Therefore any additional NC background makes little difference to the
statistical sensitivity. 
Much higher NC background 
will  affect the spectrum below 2 GeV and this could lower 
the sensitivity to CP parameters as well as  $\Delta m^2_{21}$.

\subsection{Sensitivity to the CP violation parameter}

\begin{figure}
  \begin{center}
    \includegraphics*[width=\textwidth]{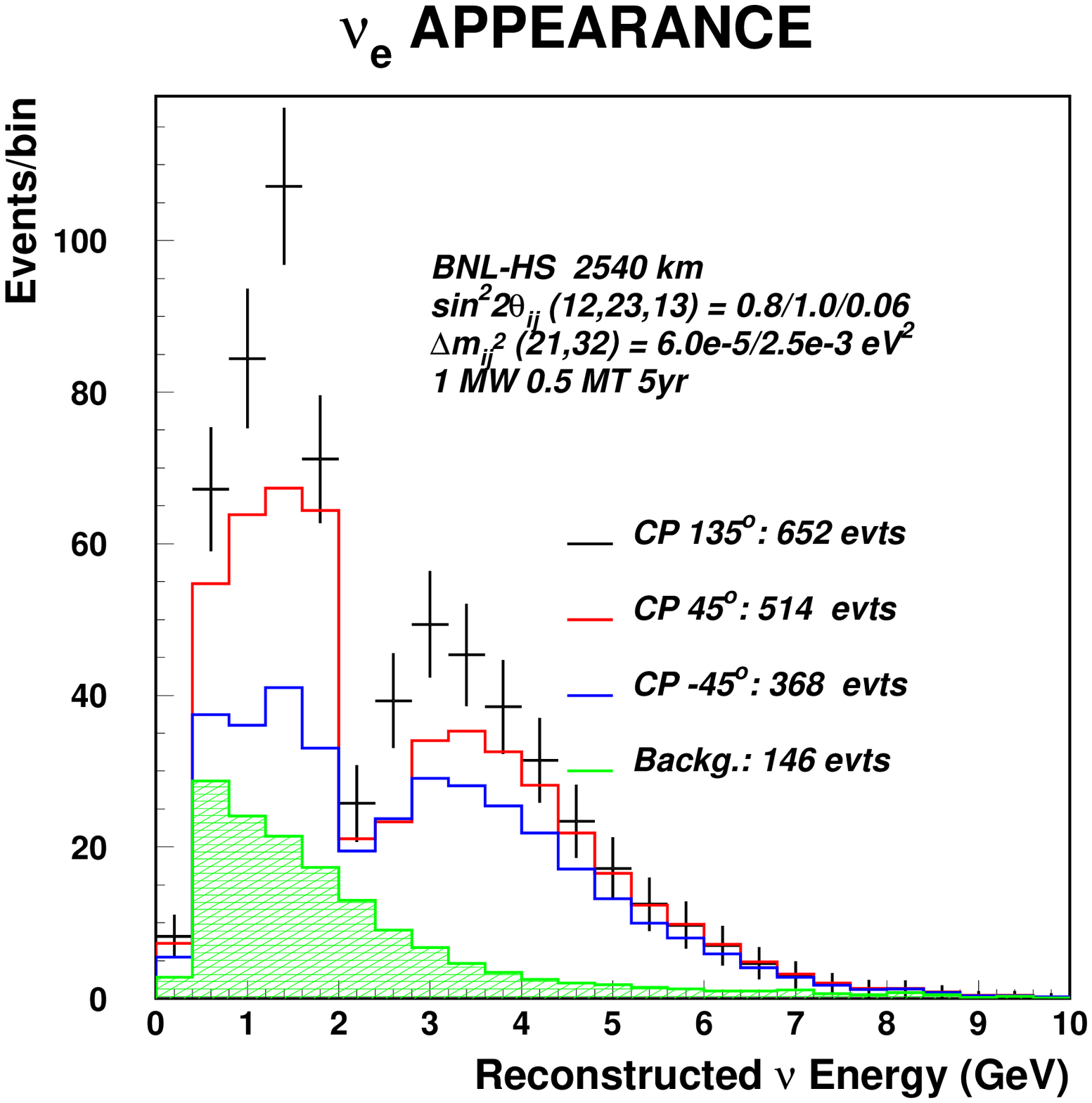}
    \caption[Expected $\nu_e$ appearance at various $\delta_{CP}$]
{The observed electron neutrino spectrum including 
background contamination for 
3 different values of the CP parameter $\delta_{CP}$. 
The error bars are for $\delta_{CP} = 135^o$; the errors 
bars indicate the statistical error on eah bin. 
 The red histogram 
below the error bars 
 is for $\delta_{CP} = 45^o$, and the blue histogram 
is for  $\delta_{CP} = -45^o$. 
The green hatched histogram shows just the background (Figure \ref{allbck}).
This plot is for $\mdmatm = 0.0025~\meV^2$.  
We have assumed
      $\sin^2 2 \theta_{13} = 0.06$ and  
$\mdmsol = 6\times 10^{-5}~\meV^2$. The values of 
$\sin^2 2 \theta_{12}$ and $\sin^2 2 \theta_{23}$ are set to 
0.8, 1.0, respectively.  }
    \label{3cp}
  \end{center}
\end{figure}

\begin{figure}
  \begin{center}
    \includegraphics*[width=\textwidth]{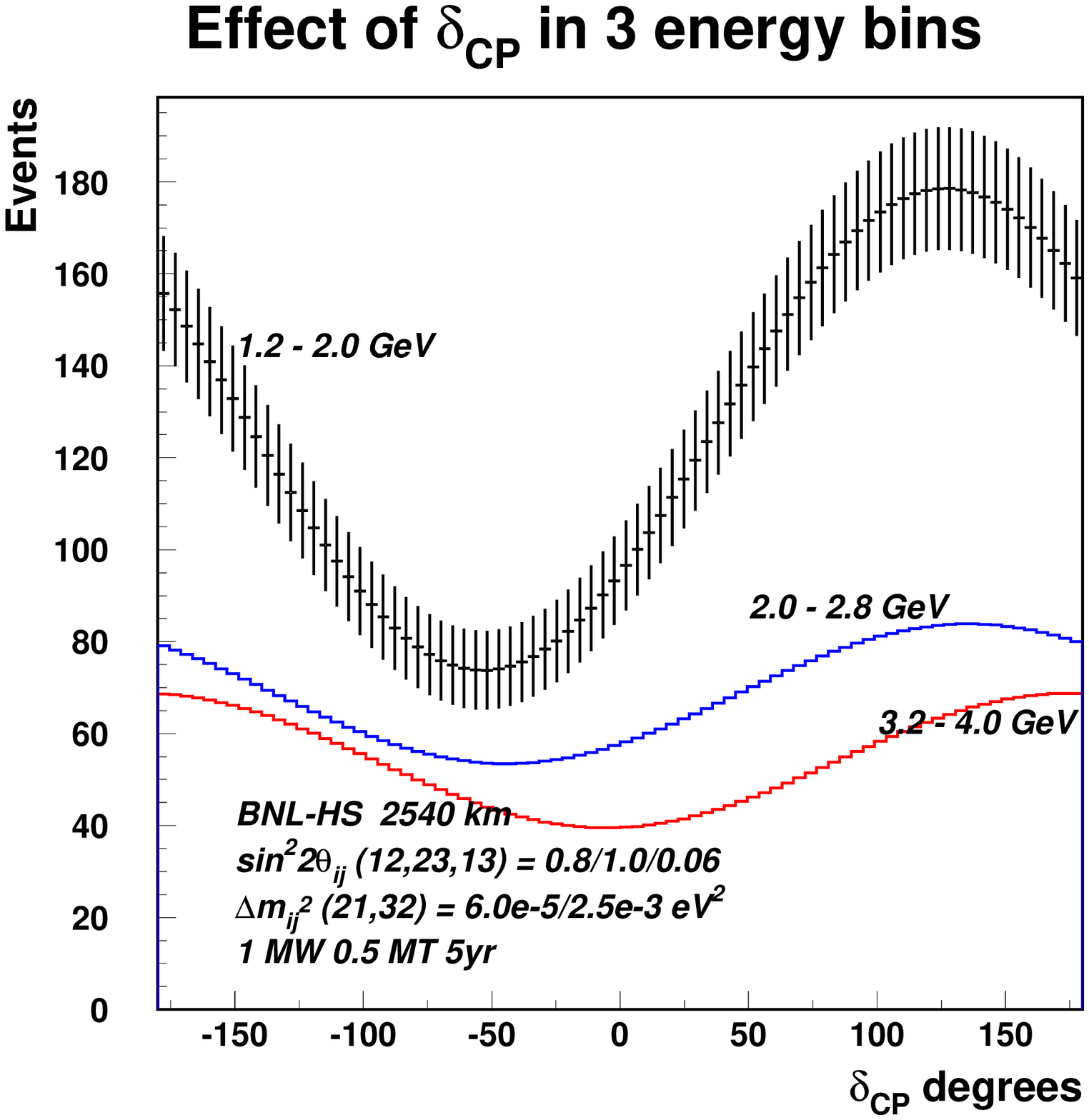}
    \caption[Event rate as function of $\delta_{CP}$ for various energies.]
{The event rate in 3 energy bins from 
Fig. \ref{3cp} as a function of $\delta_{CP}$. This plot also 
includes the background in each of the 3 energy bins. 
This plot shows that both the phase and the size of the modulation  
changes as we examine different energy bins. 
Thus a fit to the entire spectrum should give us good sensitivity 
to $\delta_{CP}$. 
    }
    \label{cp3bn}
  \end{center}
\end{figure}

As shown in Figure~\ref{cpasym}, the effect of CP violation grows linearly
as energy is decreased (or the baseline increased). For a very long
baseline experiment, it is possible to compare the signal strength in
the $\pi/2$ node versus the $3\pi/2$ or higher nodes. Such a
comparison will yield a measurement of the CP violation parameter 
$\delta_{CP}$. Such a measurement can be done with only  neutrino beam 
running over most of the parameter region (anti-neutrino running not 
necessary). 
  Any such
measurement of CP should eventually be augmented by data using a muon
 anti-neutrino beam in the same experiment. Nevertheless, 
we have calculated the sensitivity to CP parameter $\delta_{CP}$ with 
only neutrino running. 
In Figure \ref{3cp} we plot the reconstructed neutrino spectrum for 
electron-like events including background for 3 different values of 
$\delta_{CP}$. The effect of $\delta_{CP}$ is clearly large for 
the lower energy signal region as pointed out earlier. In Figure 
\ref{cp3bn} we further examine the effect of CP on the electron 
spectrum. This plot shows that both the size of the modulation 
and the phase shifts as we examine different energy bins. 
The phase shift is due to the presence of terms involving both 
$\sin \delta$ and $\cos \delta$ in the $\nu_\mu \to \nu_e$ 
probability over the entire spectrum. 
 The broadband beam, therefore,  
allows us to  fit  the entire spectrum and  gives
 us good sensitivity to $\delta_{CP}$ with much reduced correlation 
with $\sin^2 2 \theta_{13}$.

\begin{figure}
  \begin{center}
    \includegraphics*[width=\textwidth]{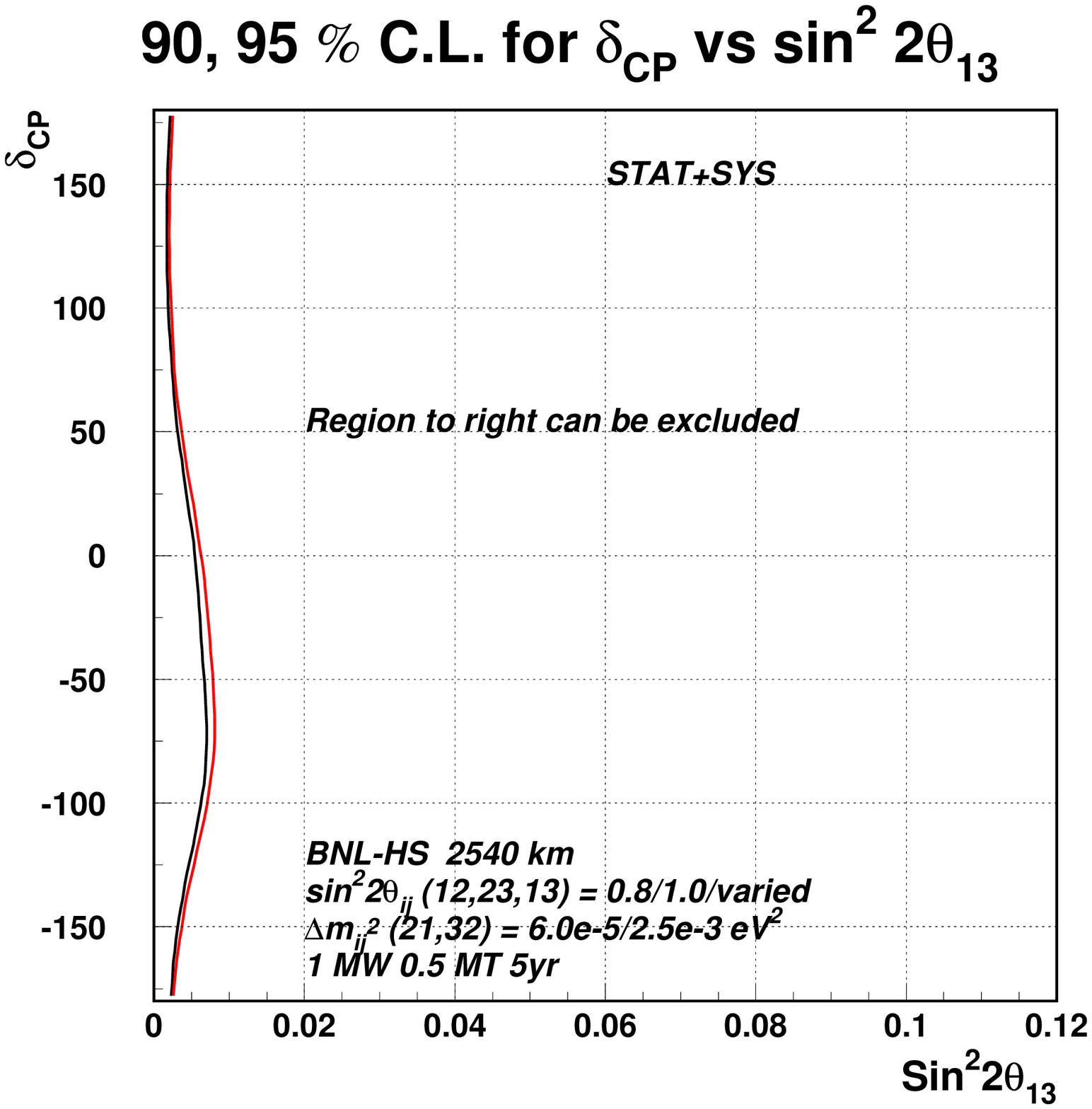}
    \caption[Sensitivity to $\delta_{CP}$ and $\sin^2 2 \theta_{13}$]
{90\% and 95\% confidence level upper limit in $\sin^2  2 \theta_{13}$ 
as a function of $\delta_{CP}$ 
if no excess of electron is found as expected 
for 
 $\mdmatm = 0.0025~\meV^2$, and   
$\mdmsol = 6\times 10^{-5}~\meV^2$. The values of 
$\sin^2 2 \theta_{12}$ and $\sin^2 2 \theta_{23}$ are set to 
0.8, 1.0, respectively.}
    \label{cpth1390}
  \end{center}
\end{figure}

It is clear from Figure \ref{3cp} that sensitivity to $\nu_\mu \to \nu_e$ 
 depends on both $\sin^2  2 \theta_{13}$ and $\delta_{CP}$. Therefore,
we have calculated the 90\% confidence level upper limit on 
$\sin^2 2 \theta_{13}$ as a function of $\delta_{CP}$ with all other 
parameters fixed in Figure \ref{cpth1390}. The region on the right
 hand side of the curves in Figure \ref{cpth1390} can be excluded
if no excess of electrons is found as expected
for the parameters shown in the figure.

\begin{figure} 
  \begin{center}
    \includegraphics*[width=\textwidth]{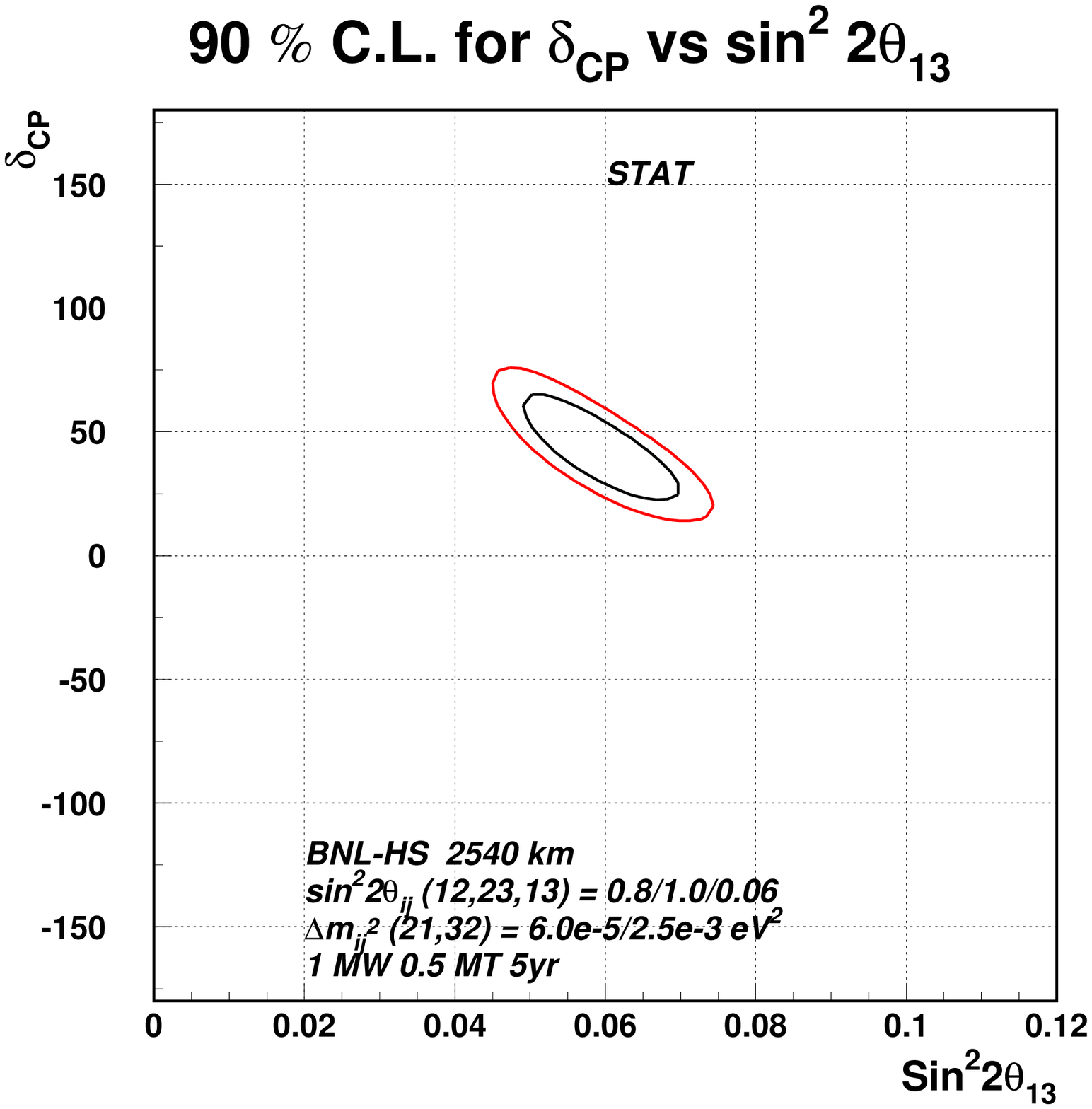}
    \caption[Expected  statistical uncertainties at test point of $\delta_{CP} = 45^\circ$ and $\sin^22\theta_{13} = 0.06$] 
{68\% and 90\% confidence level error contours in  $\sin^2  2 \theta_{13}$ 
versus $\delta_{CP}$ for statistical errors only. 
The test point used here is  
$\sin^2  2 \theta_{13}=0.06$ and $\delta_{CP}=45^o$.
 $\mdmatm = 0.0025~\meV^2$, and   $\mdmsol = 6\times 10^{-5} ~\meV^2$. The values of 
$\sin^2 2 \theta_{12}$ and $\sin^2 2 \theta_{23}$ are set to 
0.8, 1.0, respectively.  }
    
    \label{cpth1345st}
  \end{center}
\end{figure}

\begin{figure}
  \begin{center}
    \includegraphics*[width=\textwidth]{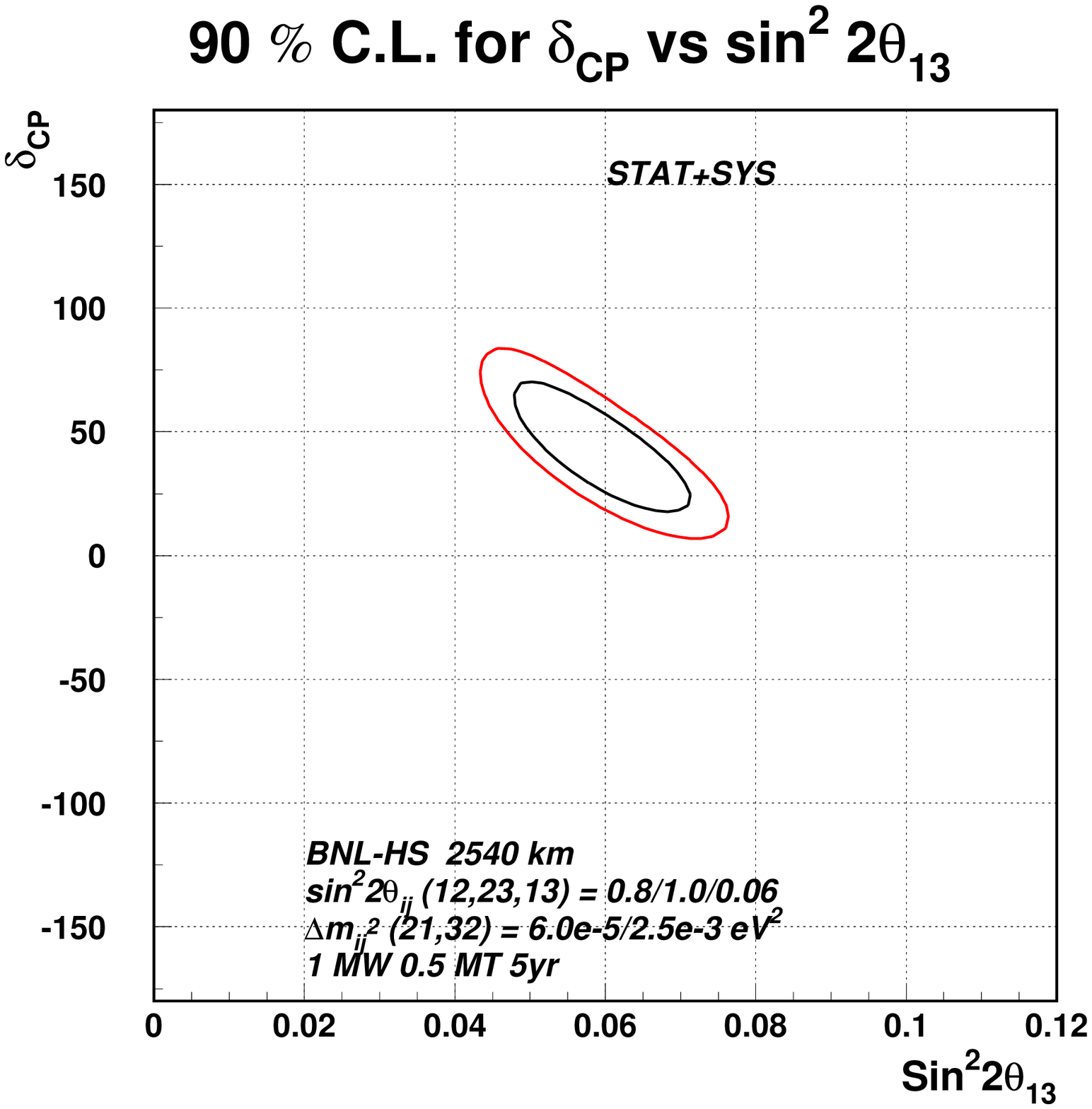}
    \caption[Expected  statistical + systematic uncertainties at test point of $\delta_{CP} = 45^\circ$ and $\sin^22\theta_{13} = 0.06$] 
{68\% and 90\% confidence level error contours in  $\sin^2  2 \theta_{13}$ 
versus $\delta_{CP}$ for statistical and systematic errors. 
The test point used here is  
$\sin^2  2 \theta_{13}=0.06$ and $\delta_{CP}=45^o$.
 $\mdmatm = 0.0025 ~\meV^2$, and   $\mdmsol = 6\times 10^{-5} ~\meV^2$. The values of 
$\sin^2 2 \theta_{12}$ and $\sin^2 2 \theta_{23}$ are set to 
0.8, 1.0, respectively.  }
    
    \label{cpth1345sy}
  \end{center}
\end{figure}

\begin{figure}
  \begin{center}
    \includegraphics*[width=\textwidth]{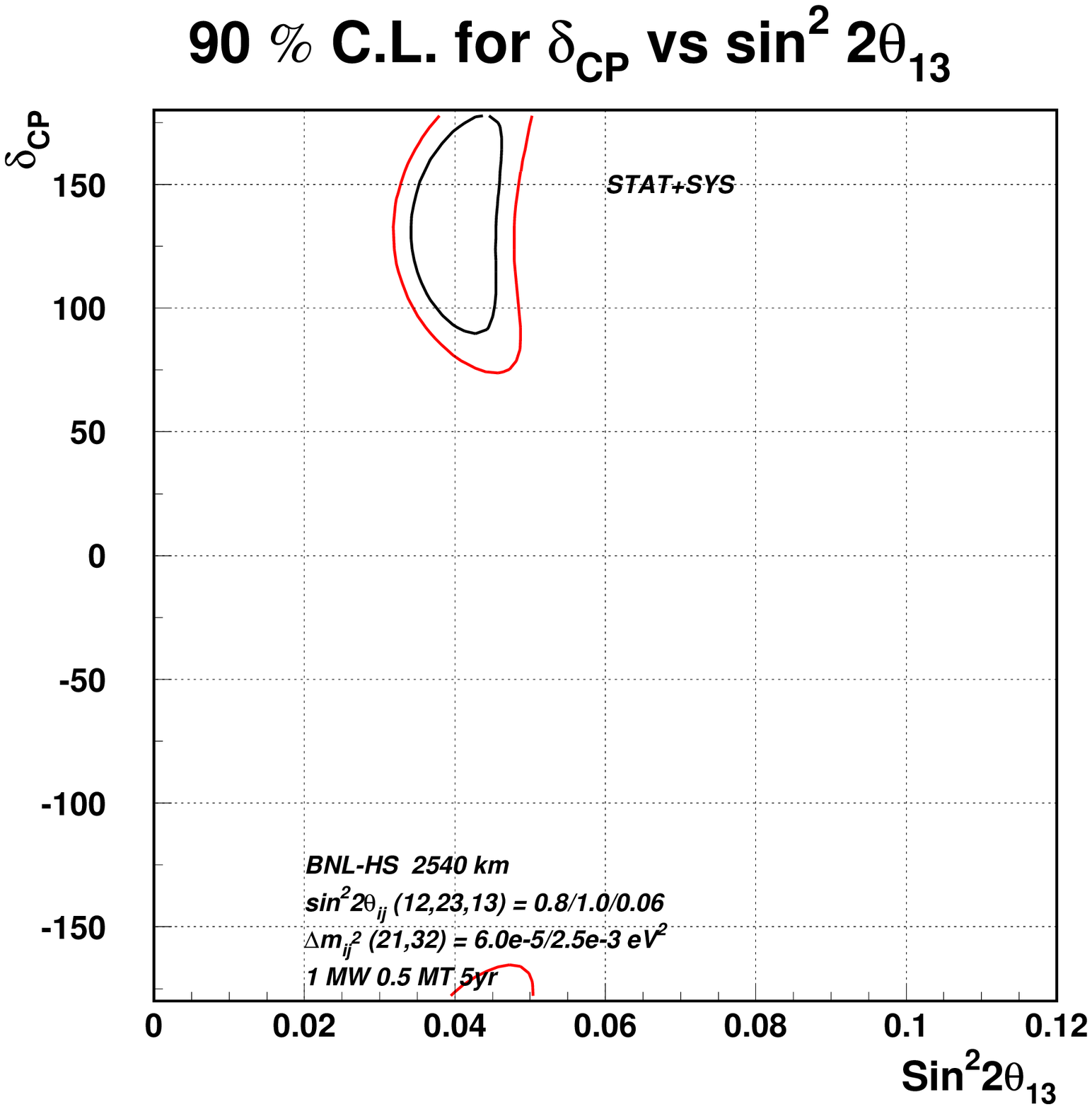}
    \caption[Expected  statistical + systematic uncertainties at test point of $\delta_{CP} = 135^\circ$ and $\sin^22\theta_{13} = 0.04$] 
{68\% and 90\% confidence level error contours in  $\sin^2  2 \theta_{13}$ 
versus $\delta_{CP}$ for statistical and systematic errors. 
The test point used here is  
$\sin^2  2 \theta_{13}=0.04$ and $\delta_{CP}=135^o$.
 $\mdmatm = 0.0025 ~\meV^2$, and   $\mdmsol = 6\times 10^{-5} ~\meV^2$. The values of 
$\sin^2 2 \theta_{12}$ and $\sin^2 2 \theta_{23}$ are set to 
0.8, 1.0, respectively.  }
    \label{cpvsth13-3}
  \end{center}
\end{figure}

\begin{figure}
  \begin{center}
    \includegraphics*[width=\textwidth]{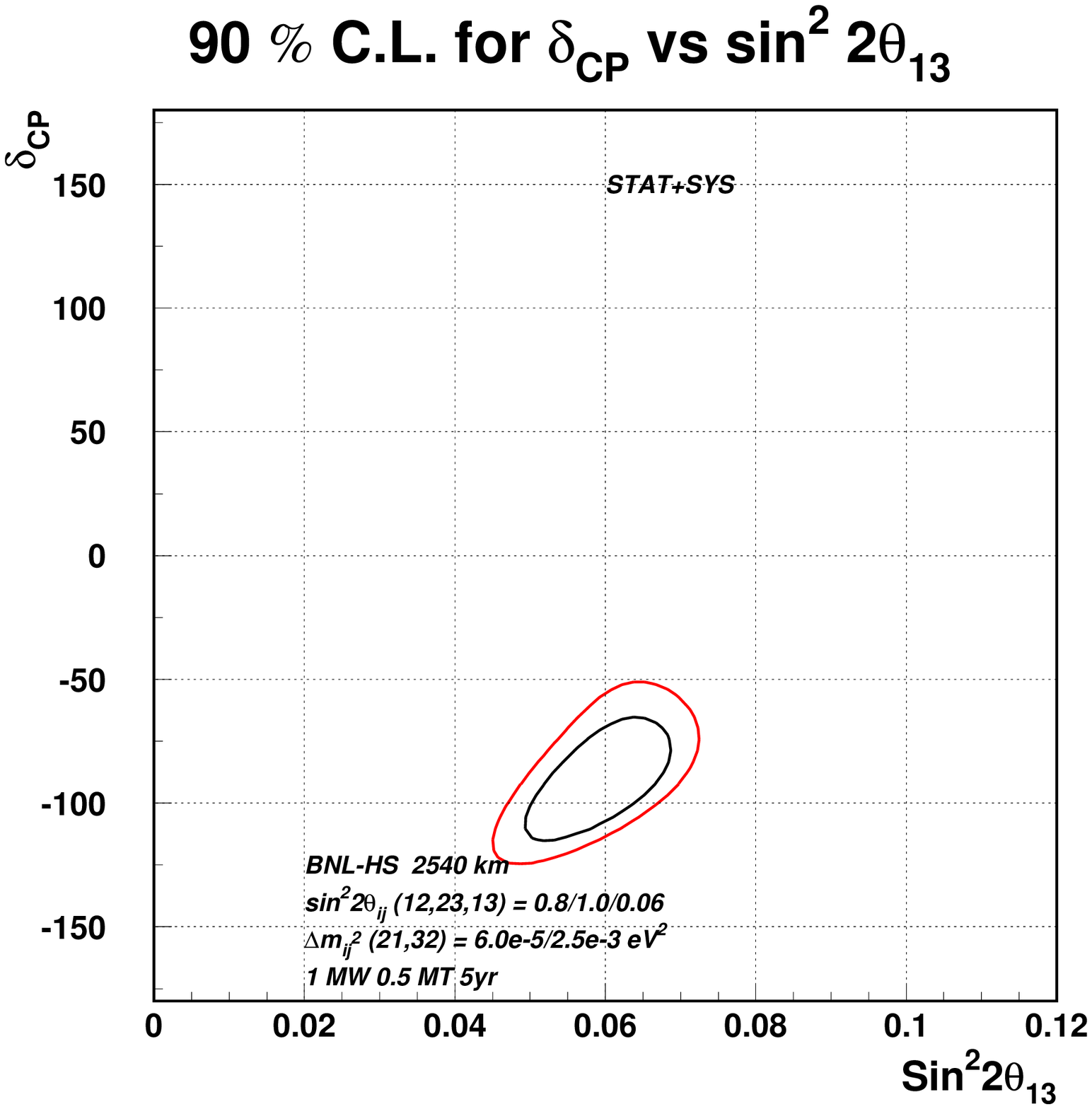}
    \caption[Expected  statistical + systematic uncertainties at test point of $\delta_{CP} = -90^\circ$ and $\sin^22\theta_{13} = 0.06$] 
{68\% and 90\% confidence level error contours in  $\sin^2  2 \theta_{13}$ 
versus $\delta_{CP}$ for statistical and systematic errors. 
The test point used here is  
$\sin^2  2 \theta_{13}=0.06$ and $\delta_{CP}=-90^o$.
 $\mdmatm = 0.0025 ~\meV^2$, and   $\mdmsol = 6\times 10^{-5} ~\meV^2$. The values of 
$\sin^2 2 \theta_{12}$ and $\sin^2 2 \theta_{23}$ are set to 
0.8, 1.0, respectively.  }
    \label{cpvsth13-4}
  \end{center}
\end{figure}

If $\sin^2 2 \theta_{13}$ is reasonably large then a good measurement of 
$\delta_{CP}$ is possible from the neutrino data alone. 
68\% and 90\% confidence level 
error contours  are shown in 
Figure \ref{cpth1345st} with statistical errors only for 
$\delta_{CP}=45^o$ and 
$\sin^2 2 \theta_{13} = 0.06$ (the other parameters are listed
 in the figure caption). 
Systematic errors on the background will mainly affect the 
low energy (0.5 to 2 GeV) 
 region, which has large sensitivity to the CP parameter. 
We  have calculated the 
error contours assuming 10\% systematic uncertainty on the background in 
Figure 
\ref{cpth1345sy}. We believe that with the use of a near detector
as well as clearly tagged  background events we can achieve 
 10\% determination of the expected background.    
Figures \ref{cpvsth13-3} and \ref{cpvsth13-4} show the expected error
contours at $\sin^2 2 \theta_{13} = 0.04, \delta_{CP}=135^o$ and 
$\sin^2 2 \theta_{13} = 0.06, \delta_{CP}=-90^o$, respectively.  
Two important observations considering these results are:
if we perform the measurement without using a wide band beam in a 
narrow region of $L/E$ the result will have a severe correlation 
between $\sin^2 2 \theta_{13}$ and $\delta_{CP}$; this correlation is 
broken by the use of a wide band beam.  Secondly,  
the expected error on $\delta_{CP}$ is $\pm 20^o$ over a wide 
range of $\sin^2 2 \theta_{13}$; it can be improved considerably 
with modest amount of anti-neutrino data running.  
We will examine the consequences of the anti-neutrino running 
in an update to this paper.  

For the result in this section on the CP measurement we have assumed that the
values of $\Delta m^2_{21}$ and $\sin^2 2 \theta_{12}$ will be 
well known. The measurement of $\delta_{CP}$ is, of course, 
correlated to these quantities. On the other hand, we could 
fit the observed electron distribution for the quantity $J_{CP}\times 
\Delta m^2_{21}$ to simply detect  the presence of CP-violating terms 
in the spectrum without attempting to measure $ \delta_{CP}$. 
We will examine these and other subtleties in the next update to 
this paper.

\subsection{Sensitivity to mass hierarchy }

\begin{figure}
  \begin{center}
    \includegraphics*[width=\textwidth]{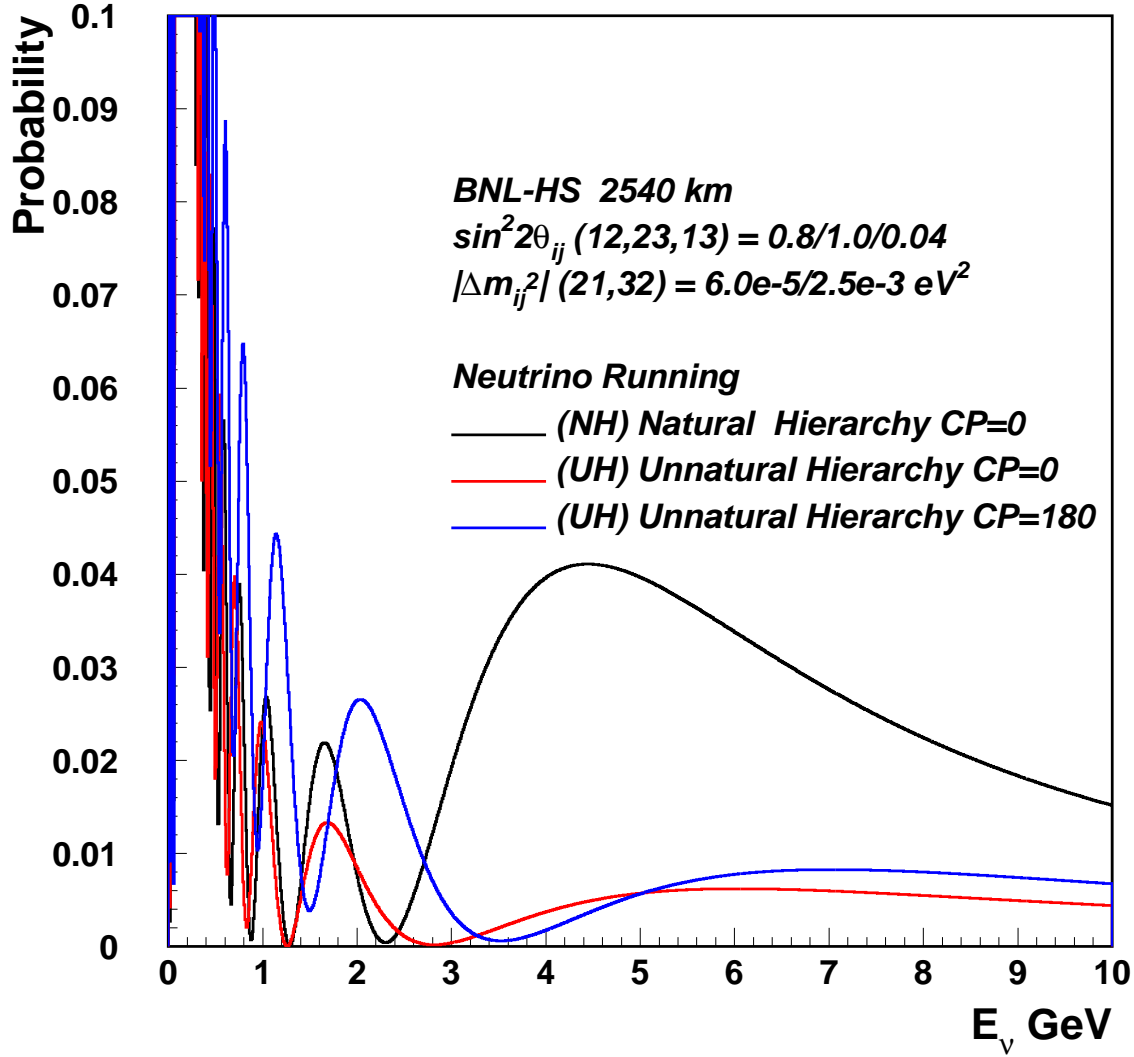}
    \caption[P($\nu_\mu \to \nu_e$) with natural and unnatural mass hierarchy at $\delta_{CP} = 0,180^\circ$]
{ Probability for $\nu_\mu \to \nu_e$ oscillations 
as a function of neutrino energy for a baseline of  2540 km. 
The three curves correspond to regular mass hierarchy (RH) with 
$\delta_{CP} = 0^o$ (black), irrational mass hierarchy (IRH) with 
$\delta_{CP} = 0^o$ (red), and irrational mass hierarchy (IRH) with 
$\delta_{CP} = 180^o$ (blue). The other parameters are indicated in 
the figure.
}
    \label{nurevm}
  \end{center}
\end{figure}

\begin{figure}
  \begin{center}
    \includegraphics*[width=\textwidth]{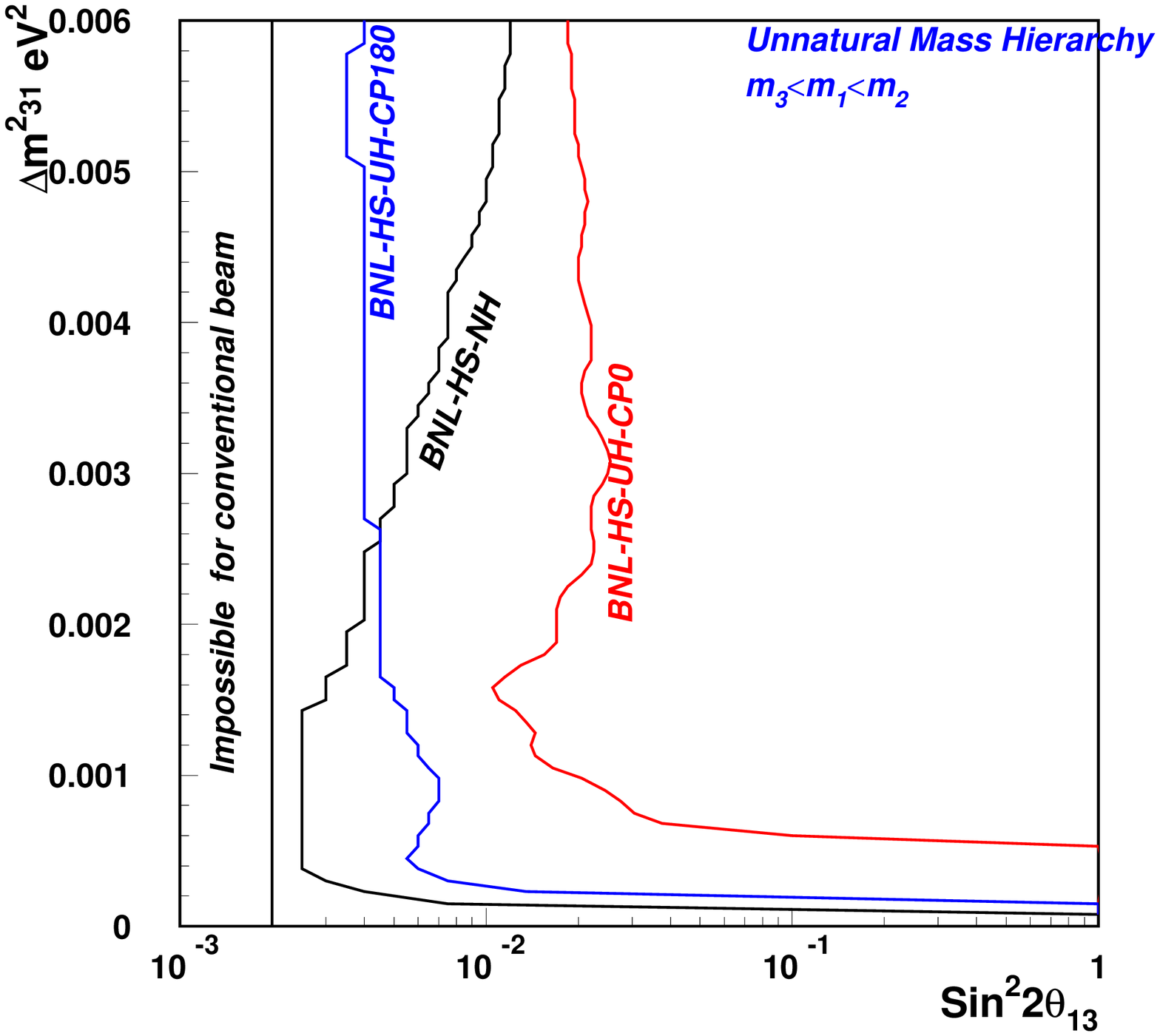}
    \caption[Sensitivity to $\Delta m^2_31$ and $\sin^22\theta_{13}$ for natural and unnatural mass hierarchy at $\delta_{CP} = 0,180^\circ$]
{Expected 90\% confidence level upper limit
on $\sin^2 2 \theta_{13}$ versus $\Delta m^2_{31}$ for the 
BNL-to-Homestake experiment for the UH hypothesis 
for running with neutrinos for 5 years. 
We have used $\delta_{CP} = 0^o$ and $\delta_{CP} = 180^o$
for the two curves labeled BNL-HS-UH-CP0 and BNL-HS-UH-CP180, respectively. 
The limit that can be obtained for the NH possibility with $\delta_{CP} = 0^o$ 
is also shown labeled BNL-HS-NH.
}
    \label{sin13revm}
  \end{center}
\end{figure}

There are three possible neutrino mass hierarchies possible 
with the existing data on atmospheric and solar neutrinos. 
For most of this paper we have assumed 
the normal mass hierarchy (NH): $m_3 > m_2 > m_1$. 
The reversed mass hierarchy (RH), $m_1 > m_2 > m_3$, 
will be  ruled out if the preferred Solar-LMA solution is confirmed
in the near future. The LMA solution depends on  $m_2 > m_1$
through the MSW mechanism. 
The third possibility, $m_2 > m_1 > m_3$, called the unnatural 
hierarchy (UH), will result in a very different appearance spectrum 
in the case of the BNL to Homestake experiment. This is illustrated in
Figure \ref{nurevm}; the UH possibility causes a suppression $\nu_\mu \to \nu_e$
oscillation in the high energy region.  However, the second oscillation 
maximum is still present and it is quite sensitive to the CP phase.
In the case of UH, therefore, we will still obtain reasonable
sensitivity to $\sin^2 2 \theta_{13}$ with neutrino running, but it will 
depend strongly on $\delta_{CP}$ 
 as shown in Figure \ref{sin13revm}.

\begin{figure}
  \begin{center}
    \includegraphics*[width=\textwidth]{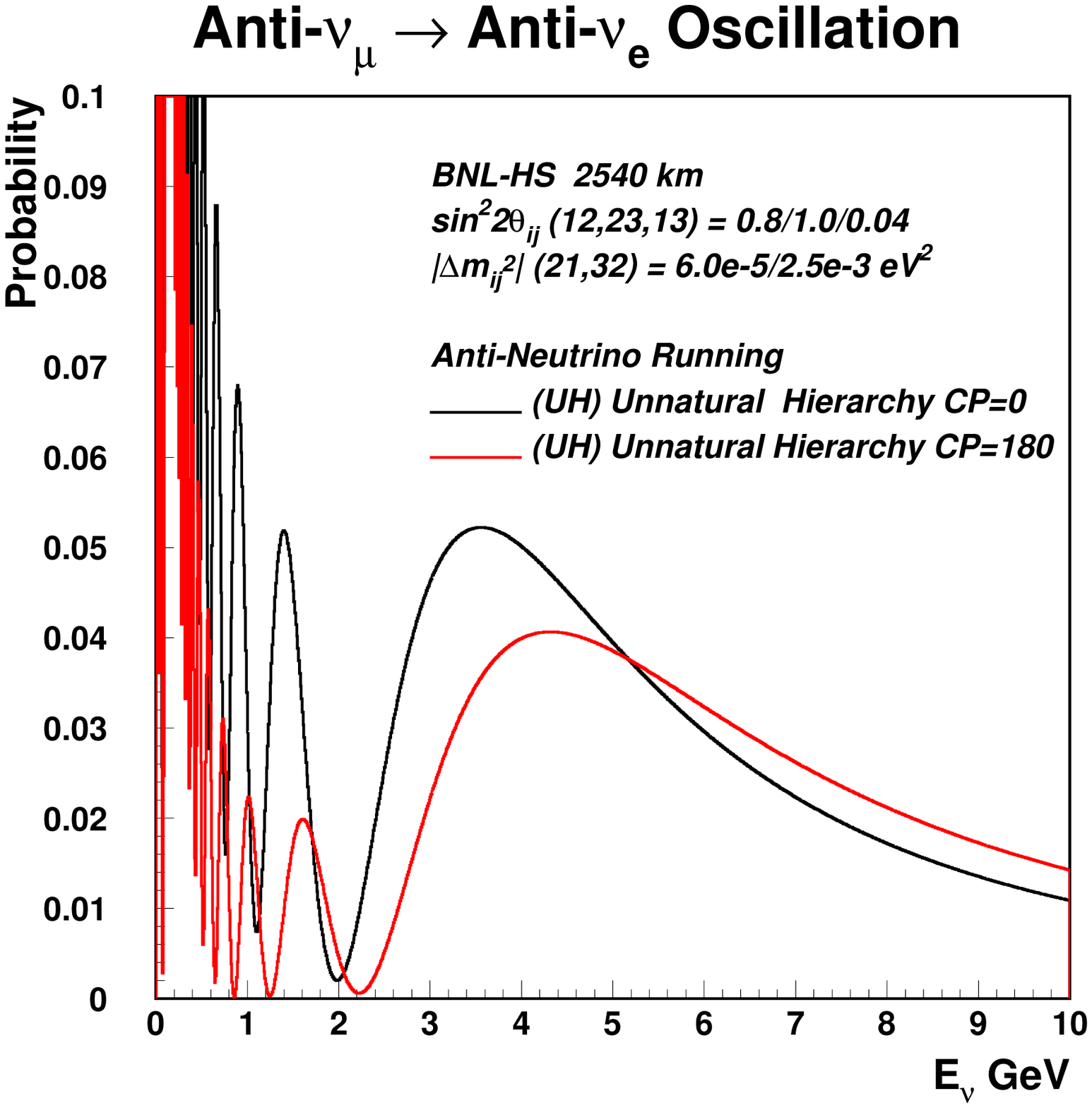}
    \caption[P($\bar\nu_\mu \to \bar\nu_e$), unnatural hierarchy, $\delta_{CP} = 0,180^\circ$]
{ Probability for $\bar\nu_\mu \to \bar\nu_e$ oscillations 
as a function of anti-neutrino energy for a baseline of  2540 km. 
The two curves correspond to  unnatural  mass hierarchy (UH) with 
$\delta_{CP} = 0^o$ (black), and unnatural mass hierarchy (UH) with 
$\delta_{CP} = 180^o$ (red). 
The other parameters are indicated in 
the figure.
}
    \label{anurevm}
  \end{center}
\end{figure}

For a large region of parameter space, the UH and NH possibilities can be 
separated with good significance using the spectrum obtained from the
neutrino running only.  Nevertheless, anti-neutrino running may be essential
if $\sin^2 2 \theta_{13}$ is small. The probability of $\bar\nu_\mu \to \bar\nu_e$
for the UH case in the of anti-neutrinos is shown in Figure \ref{anurevm}.  In
the UH case the oscillation probability is enhanced in the high energy
($> 3$~GeV) region.  This could be detected easily by changing the polarity 
of the horn focussed beam to make an anti-neutrino beam.

\begin{figure}
  \begin{center}
    \includegraphics*[width=\textwidth]{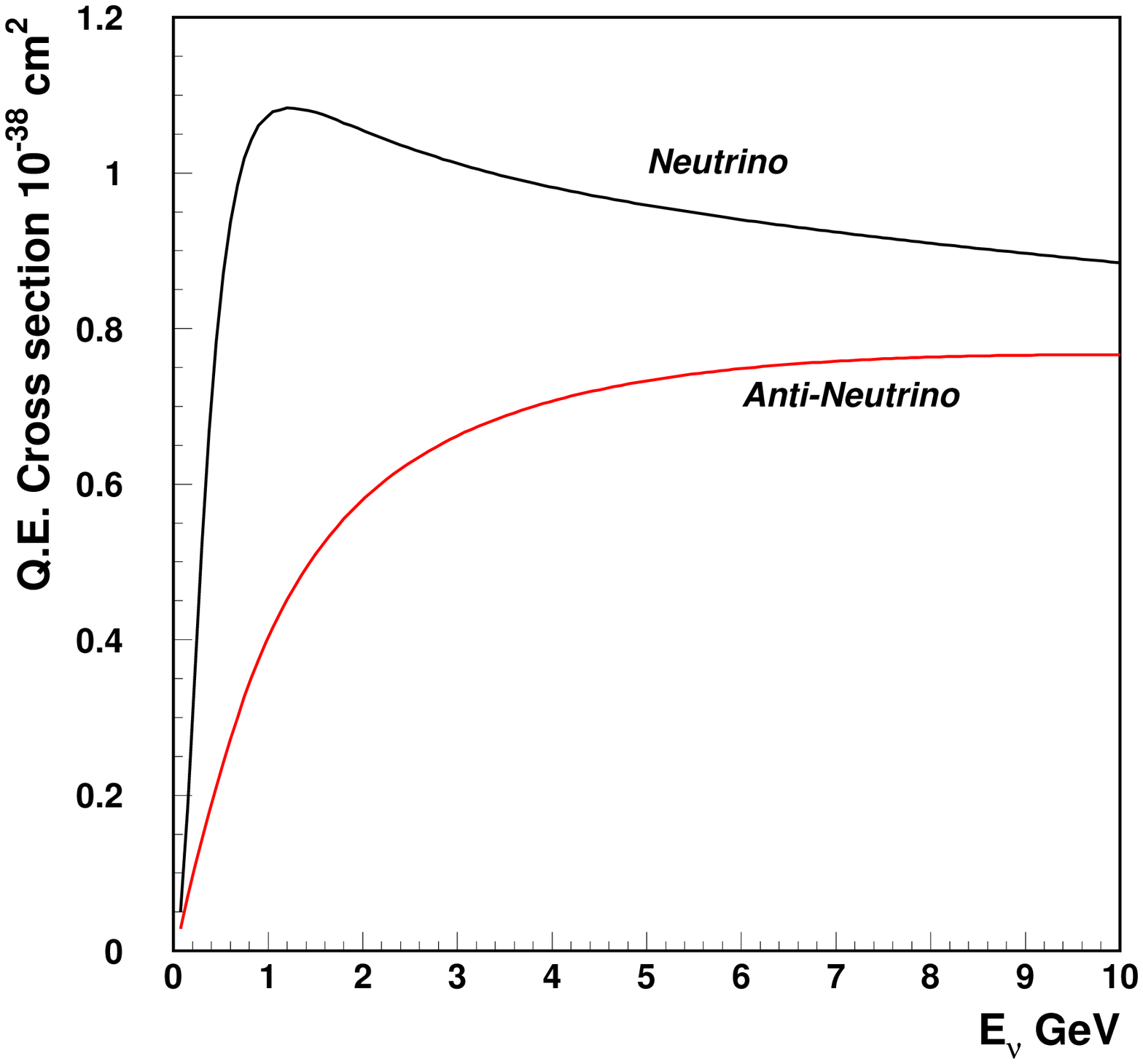}
    \caption{
Cross section for quasielastic events. $\nu_e + n \to e^- + p$ 
for neutrinos and $\bar\nu_e + p \to e^+ + n$ for anti-neutrinos.  
}
    \label{qecrs}
  \end{center}
\end{figure}

For this report we have concentrated on first running the beam 
with the neutrino
polarity.  In an updated to this report we will examine the event rates,
and sensitivities for anti-neutrino running.  Nevertheless, we can make
a few remarks based on experience from \cite{e734d}. 
 The horn focussed anti-neutrino 
flux will be about 80\% of the neutrino flux.  However, the event rate from 
anti-neutrino will be suppressed because of the lower cross section.  The 
event rate will also have about 10\% contamination from neutrinos.  An 
important feature, however, for the very long baseline experiment can be 
seen in Figure \ref{qecrs}, which shows the cross section for quasielastic 
events for neutrinos and anti-neutrinos.  In the interesting energy region
about 3 GeV where we expect the matter enhanced signal for anti-neutrino
running, the quasielastic cross section for anti-neutrino running is about 
70\% of the neutrino cross section.  This implies that the sensitivity
to $\sin^2 2 \theta_{13}$ in the UH case using anti-neutrinos could be
quite good with similar amount of running as in the neutrino case for NH.

\subsection{Sensitivity to $\Delta m^2_{21}$ }

\begin{figure}
  \begin{center}
   \includegraphics*[width=\textwidth]{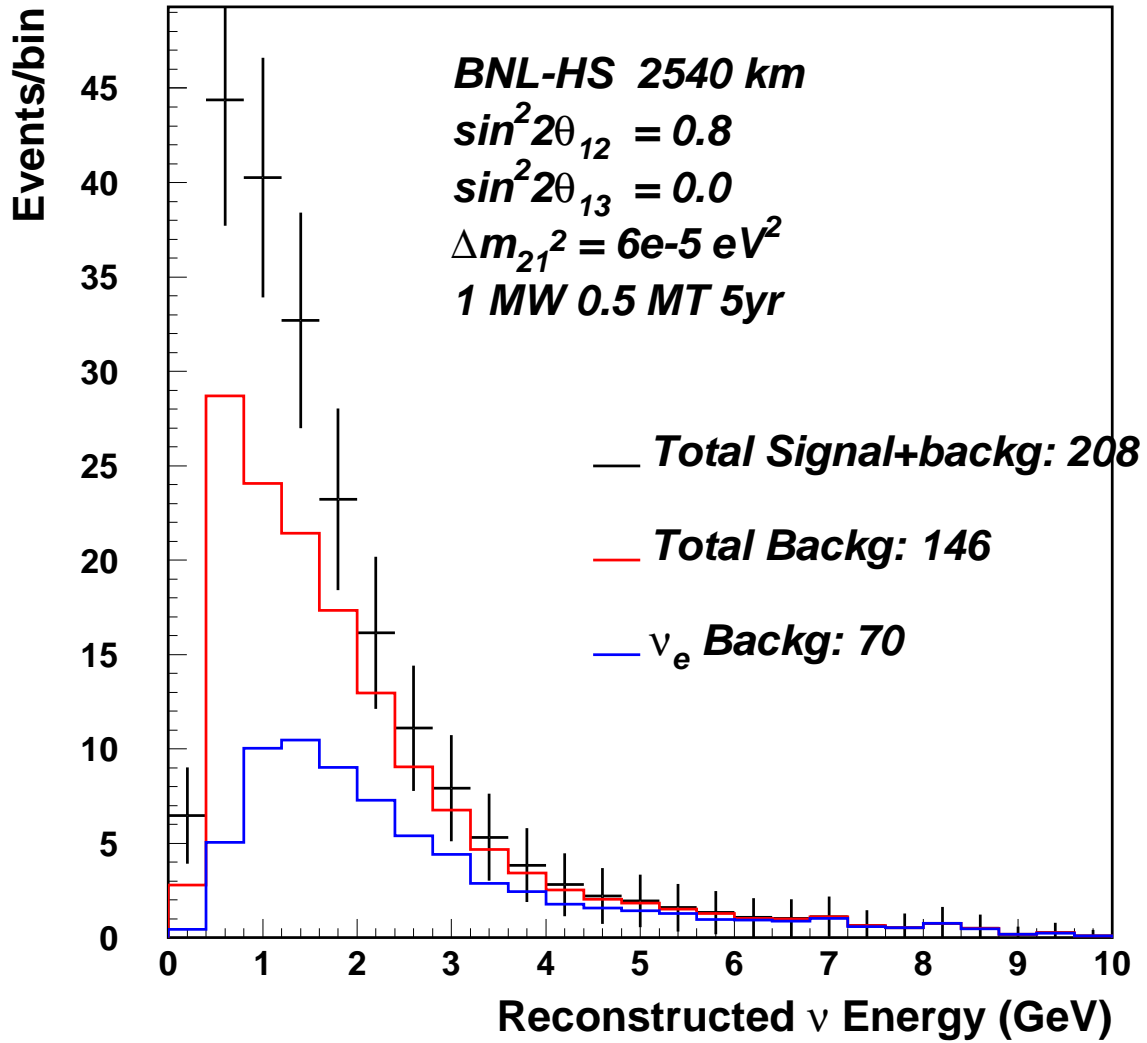}
\caption{
 Spectrum of electron-like events for 
$\sin^2 2 \theta_{13}=0$. The other important parameters are 
$\Delta m^2_{21} = 6\times 10^{-5}~eV^2$  and 
$\sin^2 2 \theta_{12} =0.8$.
}
\label{sol1}
\end{center}
\end{figure}

\begin{figure}
\begin{center}
    \includegraphics*[width=\textwidth]{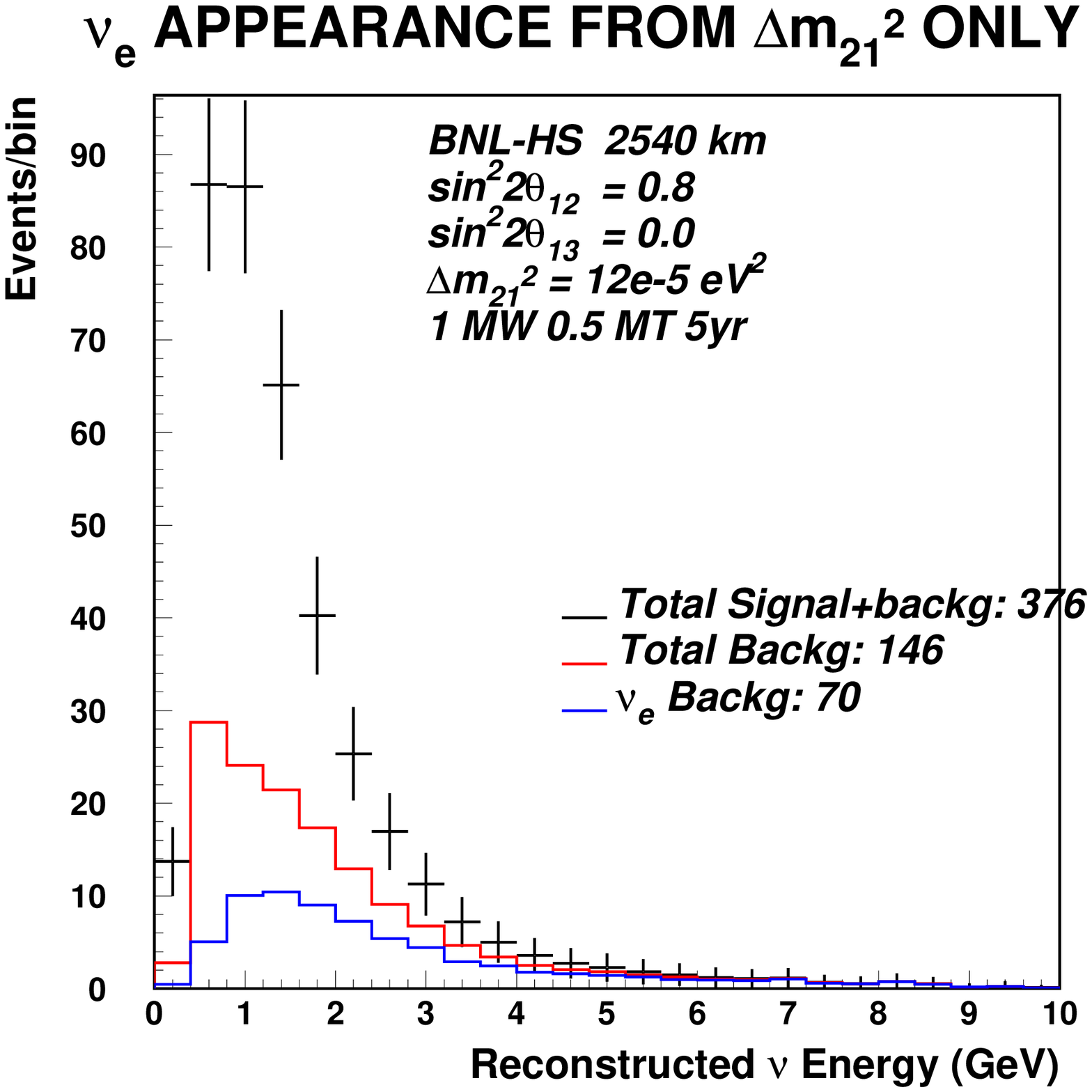} 
    \caption{
 Spectrum of electron-like events for 
$\sin^2 2 \theta_{13}=0$. The other important parameters are 
$\Delta m^2_{21} = 6\times 10^{-5}~eV^2$  and 
$\sin^2 2 \theta_{12} =0.8$.     }
    \label{sol2}
  \end{center}
\end{figure}

\begin{figure}
  \begin{center}
    \includegraphics*[width=\textwidth]{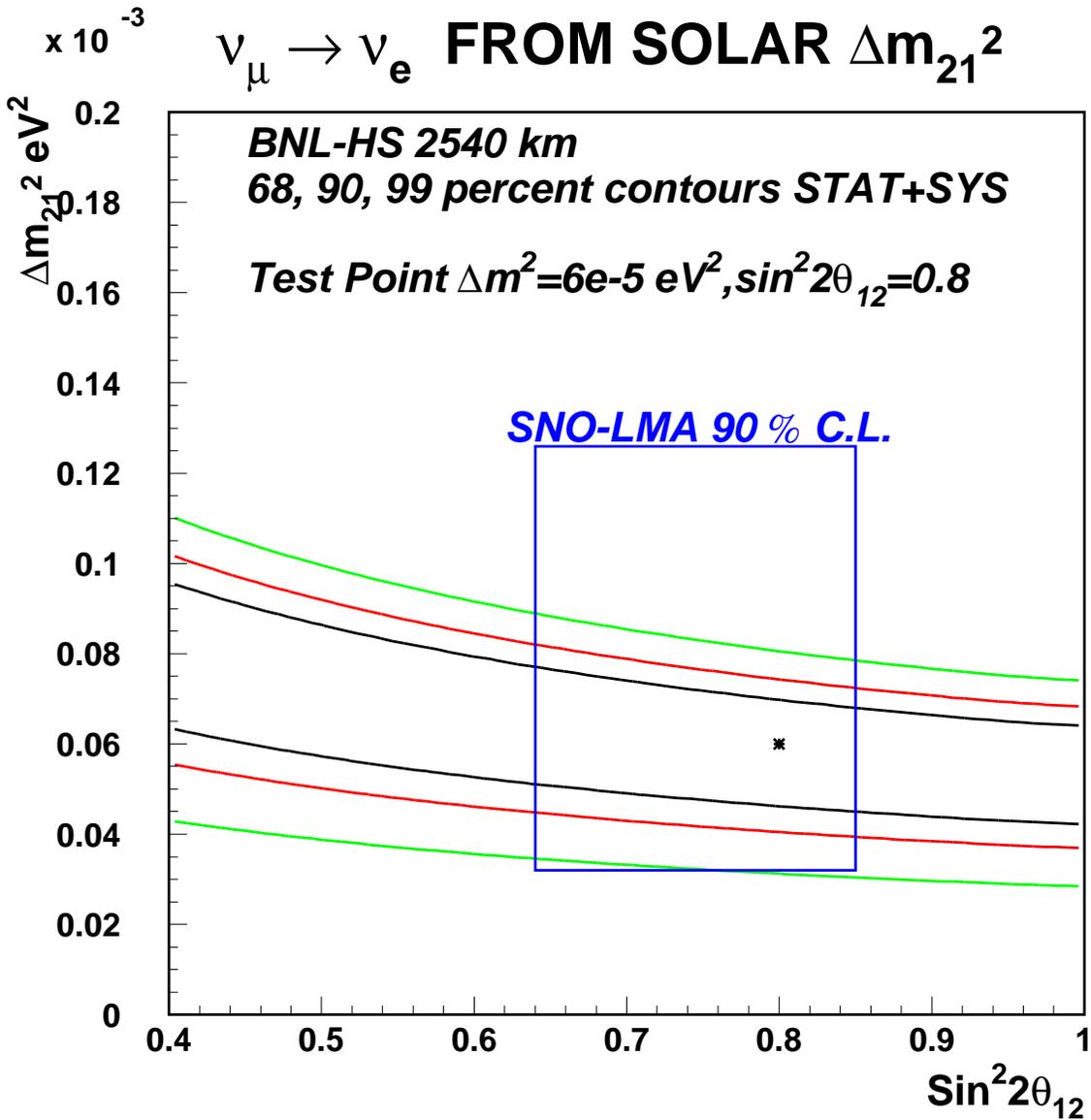} 
    \caption[68, 90, and 99 percent confidence level contours for a measurement 
at the LMA best fit point.]
{68, 90, and 99 percent confidence level contours for a measurement 
at the LMA best fit point. Both statistical and systematic errors are
included. We assume a 10\% systematic error on the background.  
    }
    \label{sol3}
  \end{center}
\end{figure}

The distance of 2540 km  is sufficient
 to obtain an appreciable signal for $\nu_\mu \to \nu_e$ 
because of the dominant mixing due to $\Delta m^2_{21}$ and
$\sin^2 2 \theta_{12}$ if the LMA (Large Mixing Angle) 
solution holds  for 
the solar neutrino anomaly. 
This is shown in Figures \ref{sol1} and \ref{sol2}.
The parameters for the best fit point in the LMA solution
contour were used for Figure \ref{sol1}. An excess of 62 events 
is expected in the lower part of the energy spectrum.
If the true value of $\Delta m^2_{21}$ is at the upper end of the 
LMA solution ($12.0\times 10^{-5}~eV^2$) then a rather large excess
of  230 events is expected. This signal can result in 
a reasonably good measurement of $\Delta m^2_{21}$; at the LMA 
best fit point the expected accuracy is $\pm 20\%$. The confidence 
level contours are shown in Figure \ref{sol3} where the LMA allowed 
contour is approximated as a rectangle. Statistical and 10\% systematic 
error on the background  are included in this determination.   
The accelerator experiment by itself will yield a result with 
a  correlation between 
$\Delta m^2_{21}$ and $\sin^2 2 \theta_{12}$; therefore another 
experiment must provide a measurement of $\sin^2 2 \theta_{12}$ to give the 
best result on $\Delta m^2_{21}$.  

\begin{figure}
  \begin{center}
    \includegraphics*[width=\textwidth]{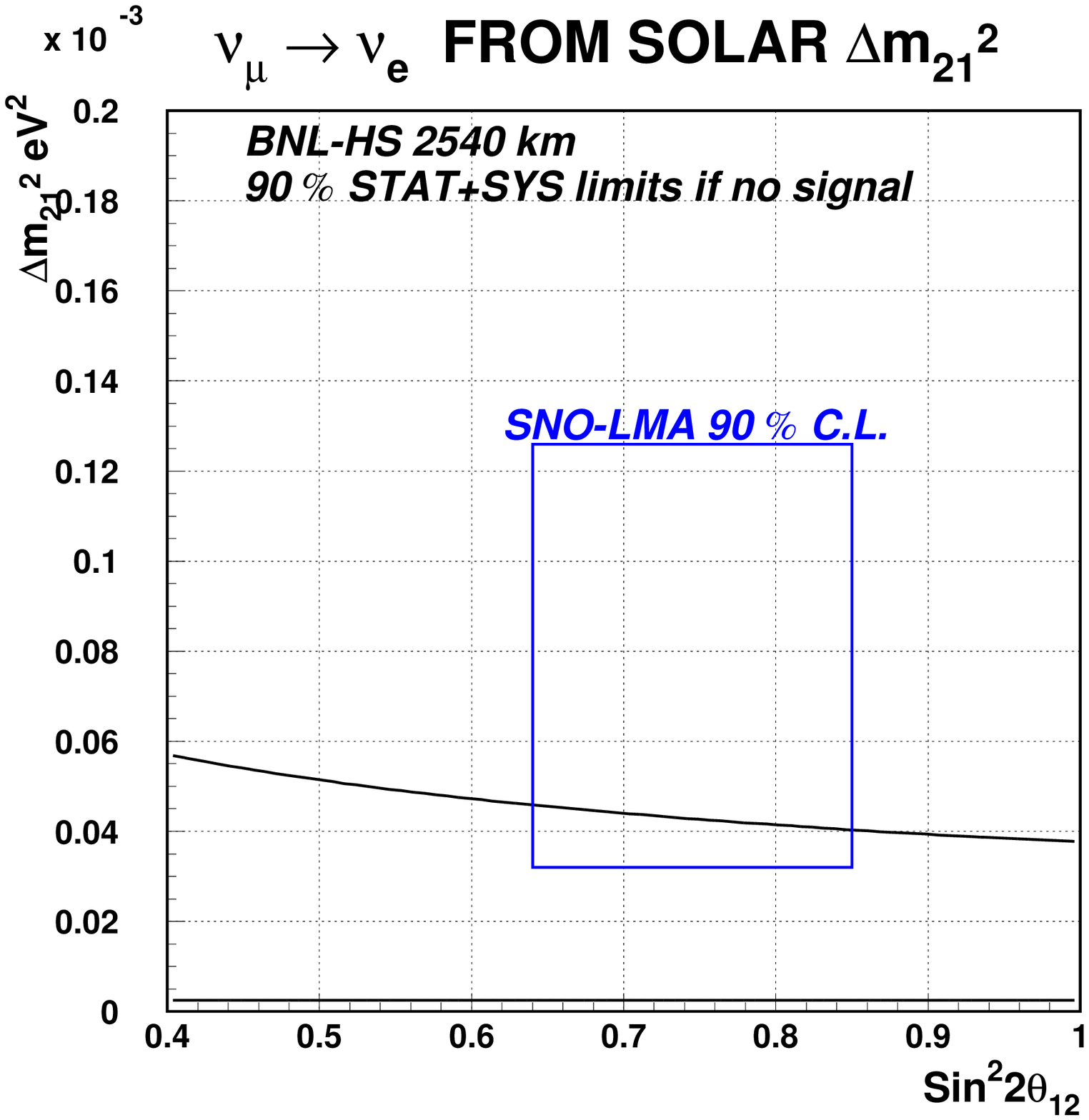} 
    \caption[Expected 90\% confidence level limit on $\Delta m^2_{21}$ 
versus $\sin^2 2 \theta_{12}$ if there is no excess of electron-like
events.]
{ Expected 90\% confidence level limit on $\Delta m^2_{21}$ 
versus $\sin^2 2 \theta_{12}$ if there is no excess of electron-like
events. Both statistical and systematic errors are included. 
    }
    \label{sol4}
  \end{center}
\end{figure}

If there is no excess of electron-like events in the spectrum 
such as Figure \ref{sol1} then an upper limit can be obtained 
on the parameters $\Delta m^2_{21}$ versus $\sin^2 2 \theta_{12}$.
Such a 90\% confidence level 
limit is shown in Figure \ref{sol4}. This limit was obtained using
statistical errors and a 10\% systematic error on the background.
This experiment can cover most of the LMA solution; if the background
can be measured better or 
suppressed further then all of the LMA region could be 
covered. 

Such a measurement of the parameters governing the solar neutrino 
anomaly in the $\nu_e$ appearance mode is qualitatively very different
from measurements in the SNO experiment or long baseline  reactor 
experiments such as KAMLAND \cite{kamland} and confirms the neutrino
oscillation picture in a useful new mode.

\subsection{The Experimental Strategy and Program}

For most of this document we have shown the possible results of 
running with a neutrino beam for a total of about 5 years
of practical operation.  
The actual running conditions will, of course, 
depend on the physics results that will be produced as 
the experiment progresses. The  great advantage of this 
proposed program is that it is so rich in its physics reach  that 
 important new physics results will come forth after every year of running.  
These results, in turn, will determine the character of running for the 
next period. 

For example, after only about 2 years of running
we will have a very accurate determination of $\Delta m^2_{32}$ 
and $\sin^2 2 \theta_{23}$. At this stage, if we see a substantial
peak in the electron spectrum at the expected energy, we will 
have strong evidence for the natural mass hierarchy (NH) and then
continue running with neutrinos to detect the presence of 
CP violating terms. But if we see no electron signal, we 
would then switch to antineutrino running. Since at higher energies 
($> 3 $ GeV), the antineutrino quasielastic cross section is 
70\% of the  neutrino quasielastic cross section, it will not take
a very large amount of antineutrino running to see a peak in the 
positron spectrum at the expected energy 
if $\sin^2 2 \theta_{13}$ is $> 0.01$ with the unnatural 
mass hierarchy.  Determination of the 
mass hierarchy is by itself a major goal of this experimental 
program.  

If the value of $\sin^2 2 \theta_{13}$ is too small, one will still 
complete the experimental program to determine $\Delta m^2_{21}$ 
in the appearance mode, a qualitatively different mode
from the SNO or the KAMLAND experiments, and a confirmation of 
the oscillation picture we now have.

It should also be pointed out that although the absolute determination
of $\Delta m^2_{32}$ may be limited by systematic errors on the 
energy scale, this systematic error is eliminated when we compare the 
values of $\Delta m^2_{32}$ obtained from neutrino versus 
anti-neutrino running. Such a comparison will yield a truly unique 
test of CPT conservation and more new physics. 

Other unexpected results cannot be ruled out because of the 
spectacular physics reach of such an experiment. These results 
will influence the running conditions as well as 
future accelerator, beam, and detector  upgrade paths. 

We also emphasis that this all-inclusive neutrino oscillations
program can be completed in a single facility. Other proposed 
methods that feature a sequential experiments approach 
will take much longer to perform and will ultimately 
cost more.  This is an important strategic point for 
the US particle physics program.

\subsection{Detectors for the very long baseline experiment} 

The conversion of Homestake Gold Mine in Lead, South Dakota, into the
National Underground Science and Engineering 
Laboratory (NUSEL), tentatively to take
place in the next few years, will provide a unique opportunity 
for a program of
very-long baseline neutrino oscillation experiments.  As explained
above, these experiments are possible only 
due to the length of the baseline,
2540 km, from the Brookhaven National Laboratory (BNL) to Lead.
It is proposed that the NUSEL facility will accommodate either an
array of detectors or a single monolithic one both with total masses
approaching 1 Megaton. Most of these will be water \cerenkov{} detectors
that can observe neutrino interactions in the desired energy range
with sufficient energy and time resolution \cite{3m}.  Details of
underground construction of these detectors is provided in Appendix
II.

An alternative to Homestake also exists at the Waste Isolation Pilot
Plant (WIPP) located in an ancient salt bed at a depth of $\sim 700$ 
meters
near Carlsbad, New Mexico. 
 One advantage of the WIPP site is that it is owned
by the DOE and now has a program of underground science.
We note that the recent Neutrino Factory Study \cite{study2}
 at BNL identified the
WIPP site as one possible location for a far detector,
and the current BNL neutrino beam could use the same concept.  
The distance from BNL to WIPP is about 2880
km.  The cosmic ray background will be higher at WIPP because the
facility is not as deep as Homestake, which has levels as deep as $\sim
2500$ meters.  The increased background, although undesirable, is not an
insurmountable problem. However, the mechanical design of a large
cavity in a salt bed has to be very different because of the slow
movement of salt that causes a cavity to slowly collapse. 

\begin{figure}
\begin{center}
    \includegraphics*[width=\textwidth]{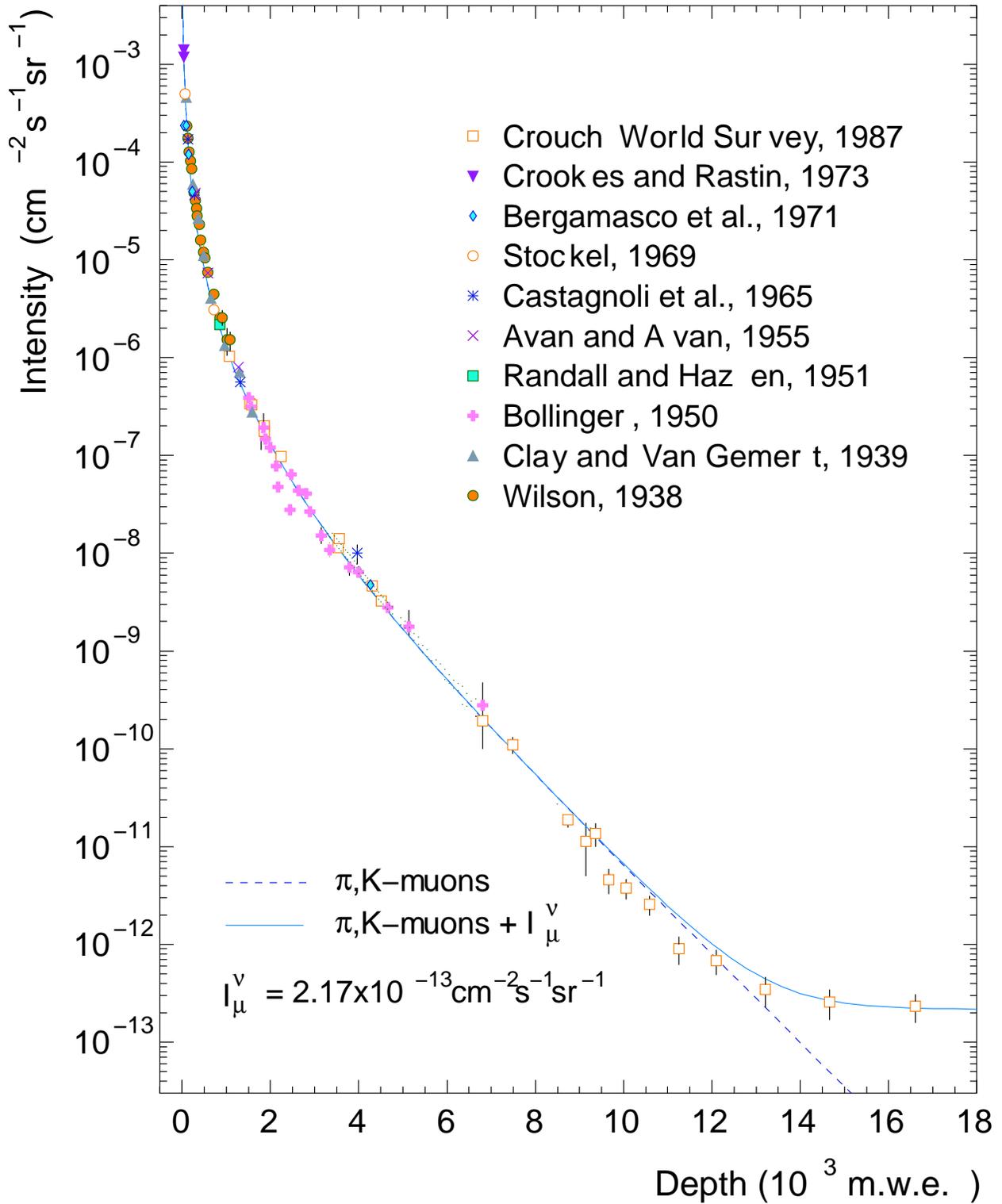}
\end{center}
\caption{Cosmic ray muon intensity as a function of depth in meters water equivalent (m.w.e) (from ref \cite{muonflux}).}

\label{muon_depth}
\end{figure}

The issue of depth is not one that greatly impacts this experiment.  A
modest overburden is needed so that the detector is not swamped by
cosmic ray muon events and thus overwhelmingly dead.  Because the beam
spill times are well known, simple timing gates suffice to remove
virtually all direct cosmic ray backgrounds.  This method is
successfully used in the K2K experiment.  The background from cosmic
ray muon spallation events as well as electrons from muon decay are
both well below the analysis energy threshold.  However, there is a
very rich array of physics that a large underground detector can do in
the time between beam spills many of which either benefit from or
require depths of greater than 3000 meters water equivalent\cite{uno}.
Figure \ref{muon_depth} shows the cosmic ray muon intensity as a
function of depth.

\begin{figure}
\begin{center}
    \includegraphics*[width=\textwidth]{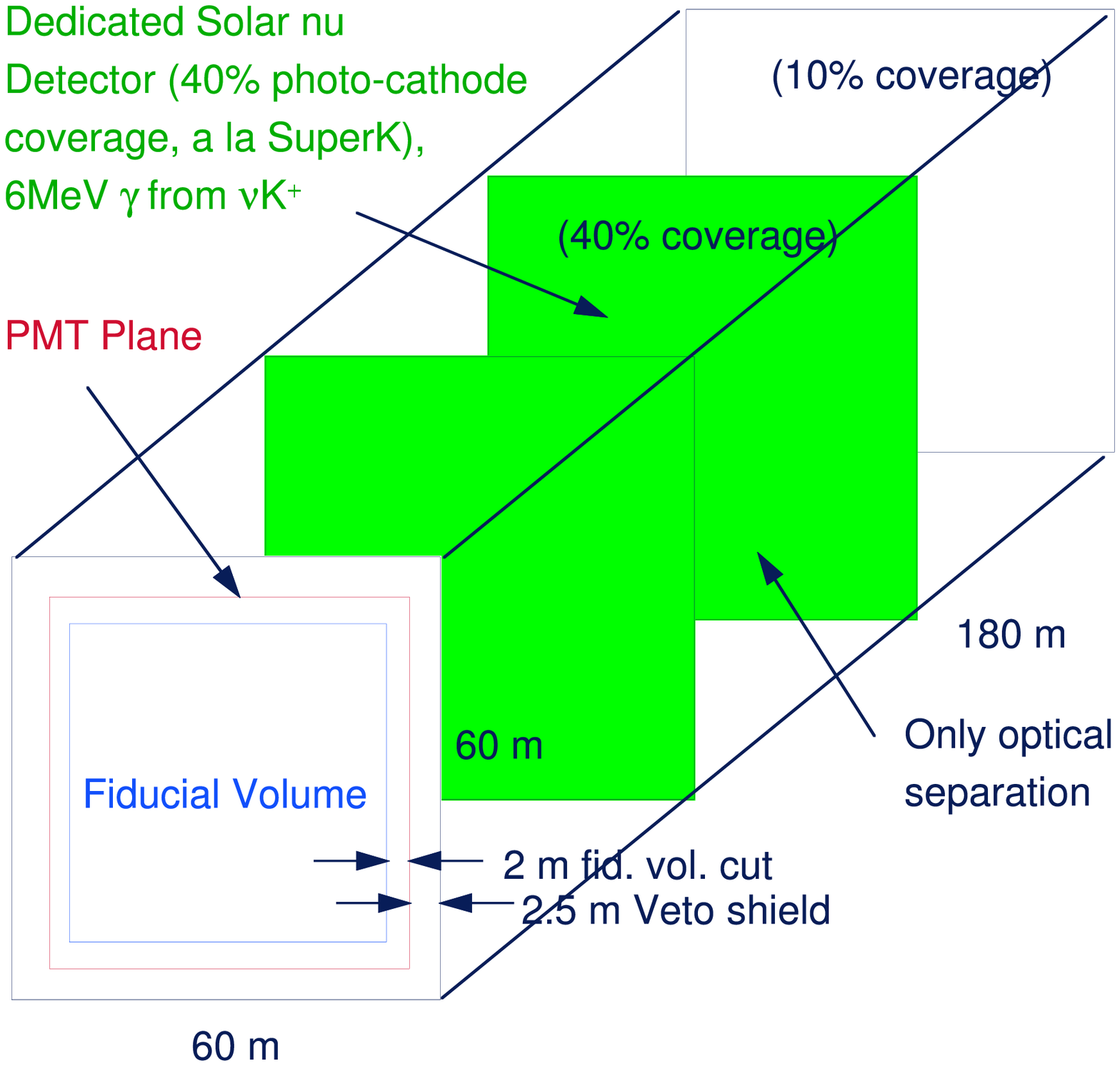}
\end{center}
\caption{Conceptual design of baseline UNO detector (from ref \cite{uno}).}

\label{uno_detector}
\end{figure}

In this report we will not address the detailed issues of detector
design and cost.  A more detailed study of a very large water \cerenkov{}
detector has been done by the UNO collaboration \cite{uno}.  Figure
\ref{uno_detector} shows a conceptual design drawing of their detector
layout.

Another option for detector technology is a liquid Argon (LAR)
time projection chamber.  
Although a  massive LAR detector (500 kT)  
cannot be ruled out at this stage, a near LAR detector
 to precisely 
measure the beam spectrum appears to be a very attractive 
possibility. 

The viability of a large liquid argon detector 
is presently being demonstrated
by the ICARUS collaboration \cite{ICARUS} in cosmic-ray tests of a 
300-ton module located on the Earth's surface. 
Currently, a study is in progress to site the LANNDD, 70 KT liquid
Argon detector at WIPP \cite{lannddp, franco1}.
 The key issue at this stage is of safety and
a proposal to the DOE to study this is in preparation. 
The LANNDD detector can be used for  neutrino physics, as well as 
the search for proton decay and other astro-particle
physics goals.  Currently, the ICARUS detector at the Gran Sasso is
being constructed with a 3kT detector as a goal.  The operation of
this detector will provide key information for the eventual
construction of LANNDD and for the neutrino physics identified in this
paper.
A magnetized liquid argon detector would give the maximal discrimination
against backgrounds in a neutrino beam, would enhance the ability to
perform CP violation experiments, and would permit use of a beam
produced by a solenoid focusing scheme \cite{skahn}
that contains both neutrinos and antineutrinos.  An R\&D experiment is
proposed to use a prototype liquid argon detector in
a magnetic field to determine the sign of electrons via analysis of
their electromagnetic showers up to several \GeV{} \cite{argonprop}.

%
%

\vspace{1ex}

  \section{AGS Upgrade} 


The Alternating Gradient Synchrotron (AGS) at BNL is 
presently the world's highest
intensity, multi-GeV proton accelerator and is a natural 
candidate for the proton
driver needed to provide multi-megawatt proton beams (superbeams) for 
the next generation of neutrino oscillations research program 
in the U.S.  Taking this qualitative
fact to the next level, accelerator scientists at BNL have 
created a credible and 
effective plan for upgrade of the AGS to the 1 MW proton source 
needed by the neutrino program advocated in this paper. 
 The increase is a factor of 6 from the present 
0.17 MW beam power level. 
 Furthermore, this plan could be time
phased to evolve in stages from a 0.4 MW source available in a few years to an
ultimate capability of 
up to 4 MW if such driver power is  needed to complete the neutrino
research program.  At present, we believe a 1 MW source will be adequate for the
foreseen program.

Our planned upgrade path would begin with the addition of a 1 GeV
superconducting extension to the existing 200 MeV Cu \linac{} that currently feeds
the Booster ring.  The resulting 1.2 GeV hybrid \linac{} would bypass the Booster
and inject directly into the AGS.
The purpose here is to eliminate the need for six complete Booster cycles to
fill
the AGS and to inject all the needed 1.2 GeV protons in about 0.7 milliseconds. 
This
step increases the average AGS power from 0.17 MW to 0.4 MW, enough to credibly
begin 
the proposed neutrino oscillations program.  By next adding new power supplies
for the AGS ring, plus added RF power to rapidly accelerate the beam to 28 GeV,
the AGS will be operational at the 1 MW power level.  Further upgrades could
increase the power level 
to as high as 4 MW if this becomes necessary.

\begin{figure}[ht]
\begin{center}
\includegraphics*[width=0.9\textwidth]{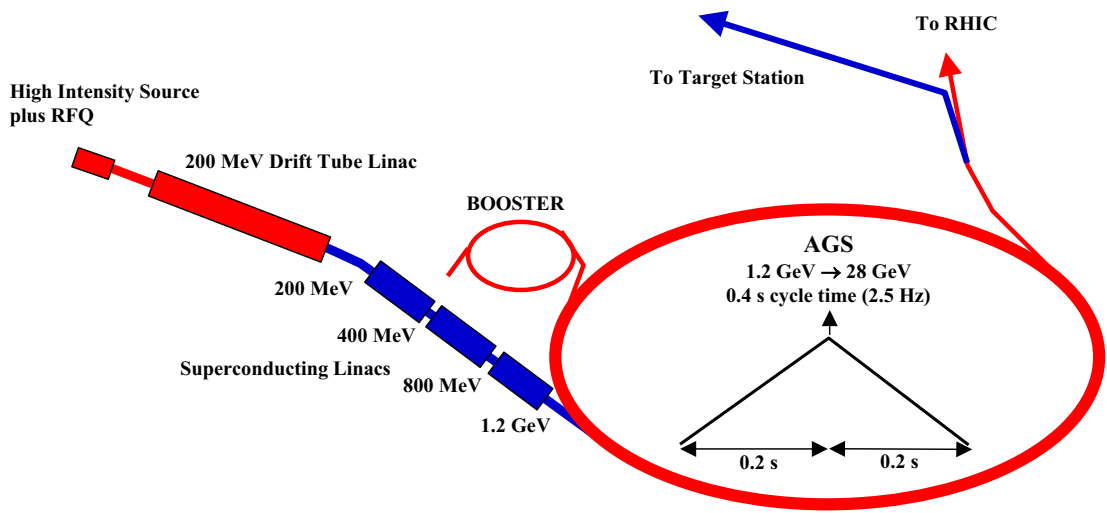}
\end{center}
\caption{\label{fig1}AGS Proton Driver Layout.}
\end{figure}

We also note that the technical basis for the proposed upgrade has been
documented in
a recent study for a muon storage ring, "Feasibility Study-II of a Muon-Based
Neutrino 
Source", June 14, 2001 \cite{study2}.  Here we present a brief summary of the parameter
lists for the required AGS upgrade, along with a summary of the direct costs that were
derived in the muon storage ring study. 
  The 1 MW requirements are summarized in 
Table \ref{tab:tab1} and a layout of the upgraded AGS is shown in Figure \ref{fig1}.



\begin{table}[ht]
\caption{\label{tab:tab1}AGS Proton Driver Parameters.}\smallskip
\begin{tabular}{@{}llll}
\hline 
Total beam power & 1 MW & Protons per bunch & $0.4\times 10^{13}$ \\
Beam energy & 28 GeV & Injection turns & 230 \\
Average beam current & 42 $\mu$A & Repetition rate & 2.5 Hz \\
Cycle time & 400 ms & Pulse length & 0.72 ms \\
Number of protons per fill & $9\times 10^{13}$ & Chopping rate & 0.75 \\
Number of bunches per fill & 24 & \linac{} average/peak current & 20/30 mA \\
\hline 
\end{tabular}
\end{table}

\subsection{Superconducting \linac{}}
The superconducting \linac{} (SCL) accelerates the proton beam 
from 200 MeV to 1.2 GeV.  The presented configuration 
follows a similar design described in detail in 
\cite{Proton:ref1} and \cite{acc:ref2}.  All 
three \linac{}s are built up from a sequence of 
identical periods.  The major parameters of the 
three sections of the SCL are given in Table \ref{tab:tab2}. 
The low energy section operates at 805 MHz and accelerates 
proton from 200 to 400 MeV.  The following two sections, 
accelerating to 800 MeV and 1.2 GeV respectively, operate
 at 1.61 GHz.  A higher frequency is desirable for 
obtaining a larger accelerating gradient with a more 
compact structure and reduced cost.  The SCL will be 
operated at $2$K for the assurance of reaching the 
desired gradient.

\begin{table}

\begin{center}

\caption{\label{tab:tab2}General Parameters of the SCL.}\smallskip
\begin{tabular}{@{}llll}
\hline
\linac{} Section & LE & ME & HE \\
\hline
Average Beam Power, kW & 7.14 & 14.0 & 14.0 \\
Average Beam Current, $\mu$A & 35.7 & 35.7 & 35.7 \\
Initial Kinetic Energy, MeV & 200 & 400 & 800 \\
Final Kinetic Energy, MeV & 400 & 800 & 1200 \\
Cell Reference $\beta_0$ & 0.615 & 0.755 & 0.887 \\
Frequency, MHz & 805 & 1610 & 1610 \\
Cells/Cavity & 8 & 8 & 8 \\
Cavities/Cryo-Module & 4 & 4 & 4 \\
Cavity Internal Diameter, cm & 10 & 5 & 5 \\
Total Length, m & 37.82 & 41.40 & 38.32 \\
Accelerating Gradient, MeV/m & 10.8 & 23.5 & 23.4 \\
Cavities/Klystron & 1 & 1 & 1 \\
Norm. rms Emittance, $\pi$mm-mrad & 2.0 & 2.0 & 2.0 \\
Rms Bunch Area, $\pi^o$MeV (805 MHz) & 0.5 & 0.5 & 0.5 \\
\hline
\end{tabular}
\end{center}

\end{table}

\subsection{Upgrade to 4 MW}
The AGS-based neutrino superbeam can be further 
upgraded to 4 MW by: 1) increasing the \linac{} 
energy to 1.5 GeV, 2) increasing the AGS intensity to 
$1.8\times 10^{14}$ ppp, and 3) increasing the 
AGS rep rate to 5.0 Hz.  The associated problems 
in beam dynamics, power supply, RF system, beam 
losses and radiation protection are under study 
and appear to be feasible if such a 
capability is required by the physics experiments.

\subsection{Cost of the AGS upgrade}

A preliminary cost of upgrading the accelerator complex to 1 MW is 
shown in Table \ref{agscost}. 
This upgrade could be done in phases if required by the 
funding plan.  We are still in the 
process of creating a detailed staging plan.

\begin{table}
\caption{Preliminary direct  costs of upgrading the AGS  to 1 MW. 
These costs do not include EDIA, contingency, and overheads.}\smallskip 
  \begin{center}
\begin{tabular}{|l|r|}
\hline 
1.2 GeV Superconducting \linac{}  &   \\
LE SC \linac{}   &   \$36.1 M      \\
ME SC \linac{}    &  \$25.9 M   \\
HE SC \linac{}    &  \$28.2 M    \\
\hline 
AGS upgrades  &     \\
AGS Power Supply & \$32.0 M \\
AGS RF upgrade &   \$8.6 M  \\
AGS injection channel & \$ 3.7 M \\
Full turn extraction & \$ 5.5 M  \\
\hline 
Total  & \$140 M \\
\hline 
\end{tabular}

\label{agscost}
  \end{center}
\end{table}

\section{Neutrino Beam Design} 

\begin{figure}
  \begin{center}
    \includegraphics*[width=0.9\textwidth]{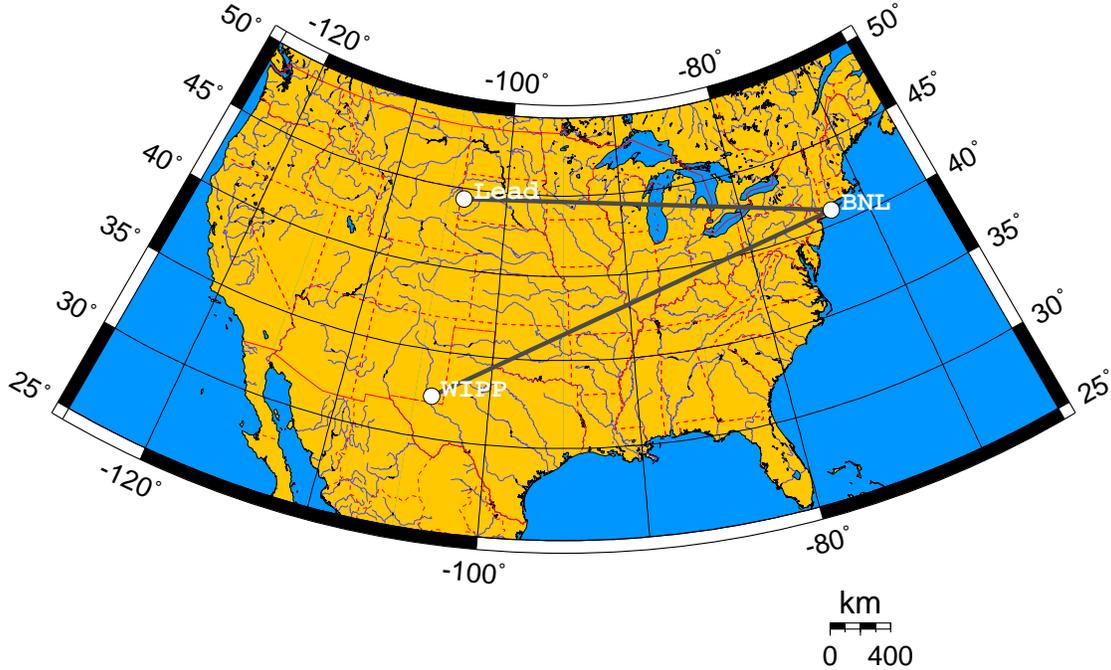}
    \caption[BNL-Homestake and BNL-WIPP baselines]
    {Possibilities for very long baselines from BNL.
      The distances from BNL to  Lead (Homestake), and
      WIPP are  2540, and 2880 km, respectively. }
    \label{blines}
  \end{center}
\end{figure}

The geographic location of BNL on one side of the continent allows us
to send beams to a variety of distances including very long baselines
of 2000 km or more.  This is shown in Figure \ref{blines}.  The
distances from BNL to Lead, SD (Homestake), 
and WIPP in NM
 are 2540 and 2880 km, respectively.  The respective dip angles
are 11.5, and 13.0 degrees.  The difficulty of building the
beam and the cost increases with the dip angle but all these angles and 
distances are feasible.

\begin{figure}
  \begin{center}
    \includegraphics*[width=0.8\textwidth]{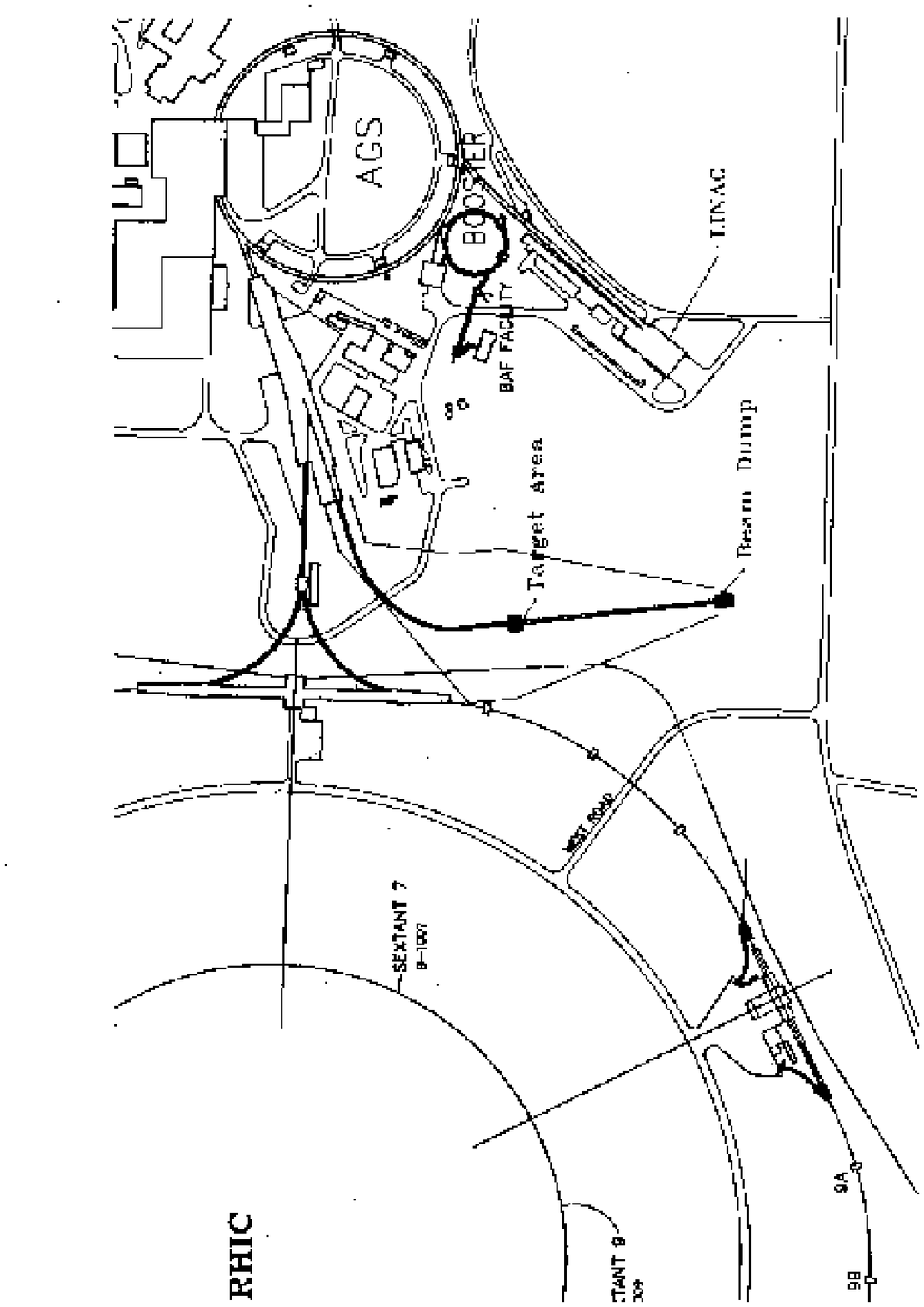}
    \caption[Beam line layout on BNL site]{ The beam line for sending a neutrino beam to Homestake mine,
      South Dakota.
      This same beam line can be adapted for any far location in the Western 
      direction.}
    \label{planview}
  \end{center}
\end{figure}

\begin{figure}
  \begin{center}
    \includegraphics*[width=\textwidth]{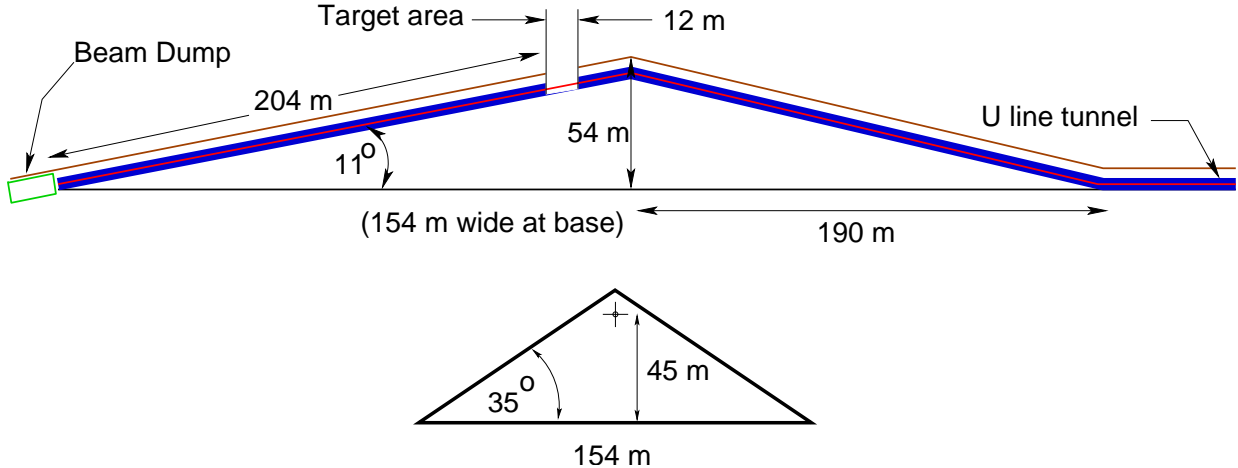}
    \caption[Schematic diagram of neutrino beamline hill]{Elevation view of the neutrino beam line to 
      Homestake, South Dakota.  For a nearer location a much smaller
      hill can be constructed.  In this beam we assume a decay tunnel
      length of 200 m. For a shorter tunnel the cost of the hill will
      reduce as shown in Table \ref{bcost}. }
    \label{eleview}
  \end{center}
\end{figure}

Our conceptual design for a beam to Homestake is 
shown in Figures \ref{planview} and \ref{eleview}.
It can  be adapted to any far location in the Western direction. 
Our design addresses a number of issues.
At BNL we are constrained to keep the beam line above the water 
table which is at a shallow depth ($\sim$ 10 m) 
  on Long Island. Therefore the beam has to be constructed on 
a hill that is built with the appropriate 11.5 degree slope. 
Fortunately, it is relatively easy, and inexpensive to 
build such hills on Long Island because of the flat, sandy 
geology. It is important to keep the height of the hill 
low so that the costs are not dominated by its
construction. 
The proton beam must be elevated to
a target station on 
  top of the hill. The cost of the hill can be lowered  
by bending the proton beam upwards as quickly as possible.
We have, however, chosen the design and the bend angle used 
for the RHIC injection lines in our proposal because 
the RHIC injection lines have well known costs. 

The proposed  fast-extracted proton beam line in the 
U-line tunnel will be a
spur off the line feeding RHIC. It will   turn almost due west, a few
hundred meters before the horn-target building. In addition to its 90
degree bend, the extracted proton beam will be bent upward through
13.76 degrees and then down by 15 degrees 
to strike the proton target.  The downward 11.30 degree
angle of the 200 meter 
 meson decay region will then be aimed at the
4850 feet level of the Homestake laboratory. This will require
the construction of a 54 meter hill to support the target-horn building,
so as to  avoid any penetration of the water table.  At its midpoint
(about Lake Michigan) the center of the neutrino beam will be roughly
120 km below the Earth's surface.


\subsection{Optimization of the wide band spectrum}

\begin{figure}
  \begin{center}
    \includegraphics*[width=0.8\textwidth]{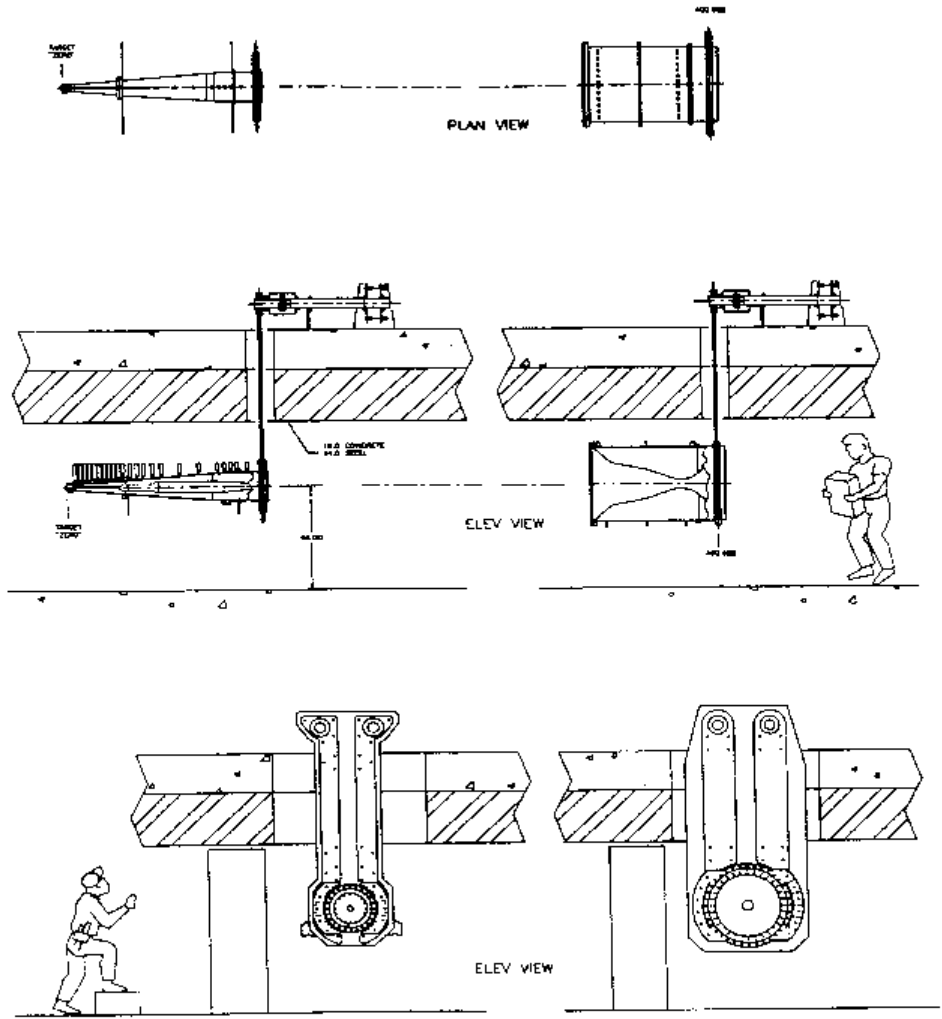}
    \caption{The design of the horn focusing system used for 
      the E734 experiment adapted from the E889 proposal.}
    \label{horns}
  \end{center}
\end{figure}

\begin{figure}
  \begin{center}
  \includegraphics*[width=\textwidth]{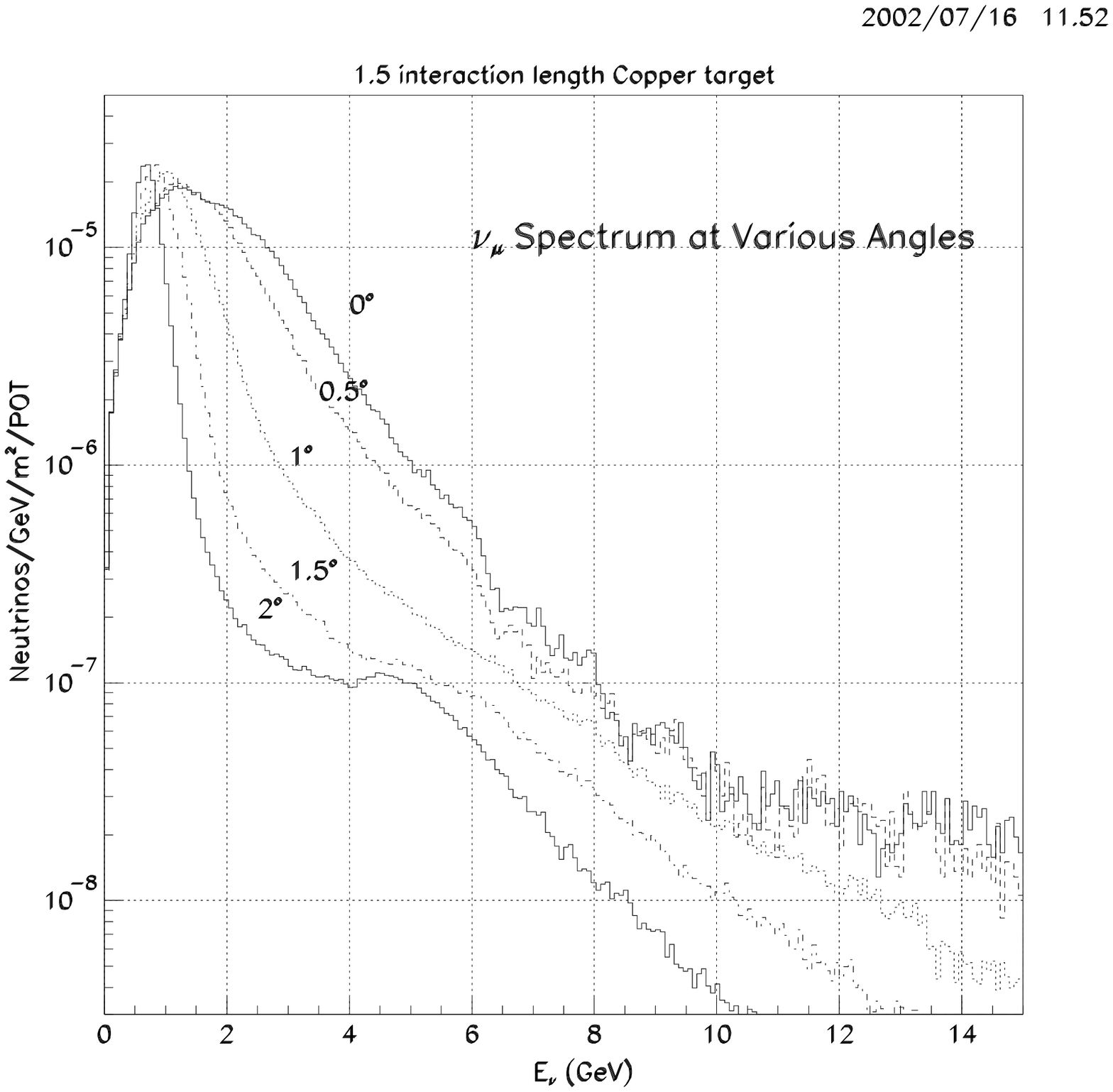}
  \caption[Neutrino energy spectrum with copper or Super-Invar target]
{Wide band horn focussed neutrino spectrum for 28 \GeV{} protons
on a copper target. The spectrum is approximately the same if 
Super-Invar is used as target material.   
    Spectra of neutrinos are calculated at various angles with respect to the
    200 m decay tunnel axis at the AGS and at a distance of 1 km from the target.} 
  \label{bnlspec}
  \end{center}
\end{figure}

For this report we have attempted to optimize the beam for 
the Homestake distance (2540 km). However, our optimization 
process could be applied to any  distance. 
As already explained, the ideal beam for 
Homestake will be a broadband beam that covers $\sim$0.5 GeV to 
$\sim$7.0 GeV range. The $\nu_\mu \to \nu_e$ process through 
$\Delta m^2_{21}$ (solar oscillations) will generate a sizable effects
at the lowest energies. The energy range $1-3$ GeV will be important 
for the detection of CP violation. The energy region $3-5$ GeV 
contains the first  matter enhanced (for neutrinos with regular
 mass hierarchy) $\nu_\mu \to \nu_e$ oscillation maximum. Recall
 that the highest energies are important for establishing 
the existence of $\nu_\mu \to \nu_e$ signature because this region is
free from the neutral current $\pi^0$ background and should have 
very good efficiency for the signal.  Lastly, the energy region $6-7$ GeV 
is important for the $\nu_\mu$ disappearance measurement. 

\begin{figure}
  \begin{center}
    \includegraphics*[angle=180,width=0.5\textwidth]{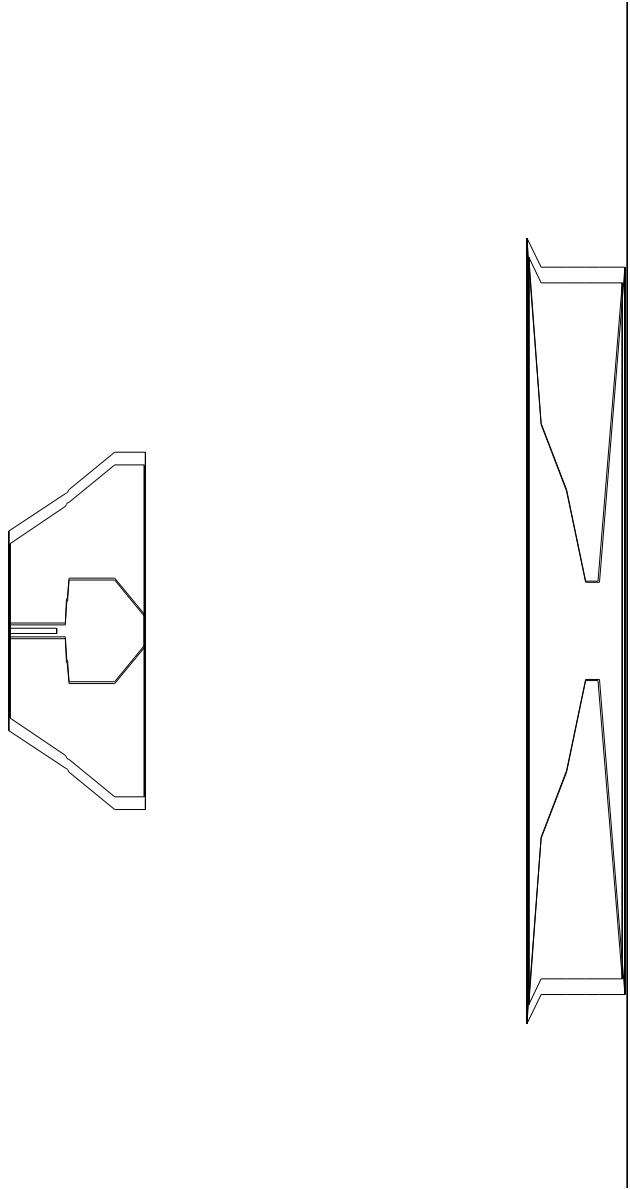}
    \caption[Geometry of the focusing horns]
    { The horn geometry in the GEANT simulation.
The vertical and horizontal scales are in the ratio of 1 to 13.  The beam
  is incident from the left.}
    \label{horns2}
  \end{center}
\end{figure}

To obtain a broad band neutrino spectrum we have adapted the standard 
scheme of multiple parabolic horns, each one focuses a different pion 
momentum region. The difficulty with this approach is that the lowest energy pions
we need to capture and focus are approximately 1-2 GeV and come 
from a long target. 
 Figures \ref{horns} and \ref{horns2}
 shows the design of the target and horn geometry for 
a conventional wide band  neutrino beam, similar
to that   used in previous experiments at 
BNL, such as  E734.   
The E734 design used a water cooled 1.5 interaction length copper target. 
The calculated energy distributions of a $\nu_{\mu}$ beam produced by
28 \GeV{} protons is shown in Figure \ref{bnlspec}~\cite{e889}.  The
$0^\circ$ calculation has been shown consistent with neutrino beam
data~\cite{e734}. A copper target will not survive the 
1 MW intensity proton beam that we propose. Therefore, 
both new materials and new focusing geometries must be considered.

\begin{figure}
  \begin{center}
  \includegraphics*[width=\textwidth]{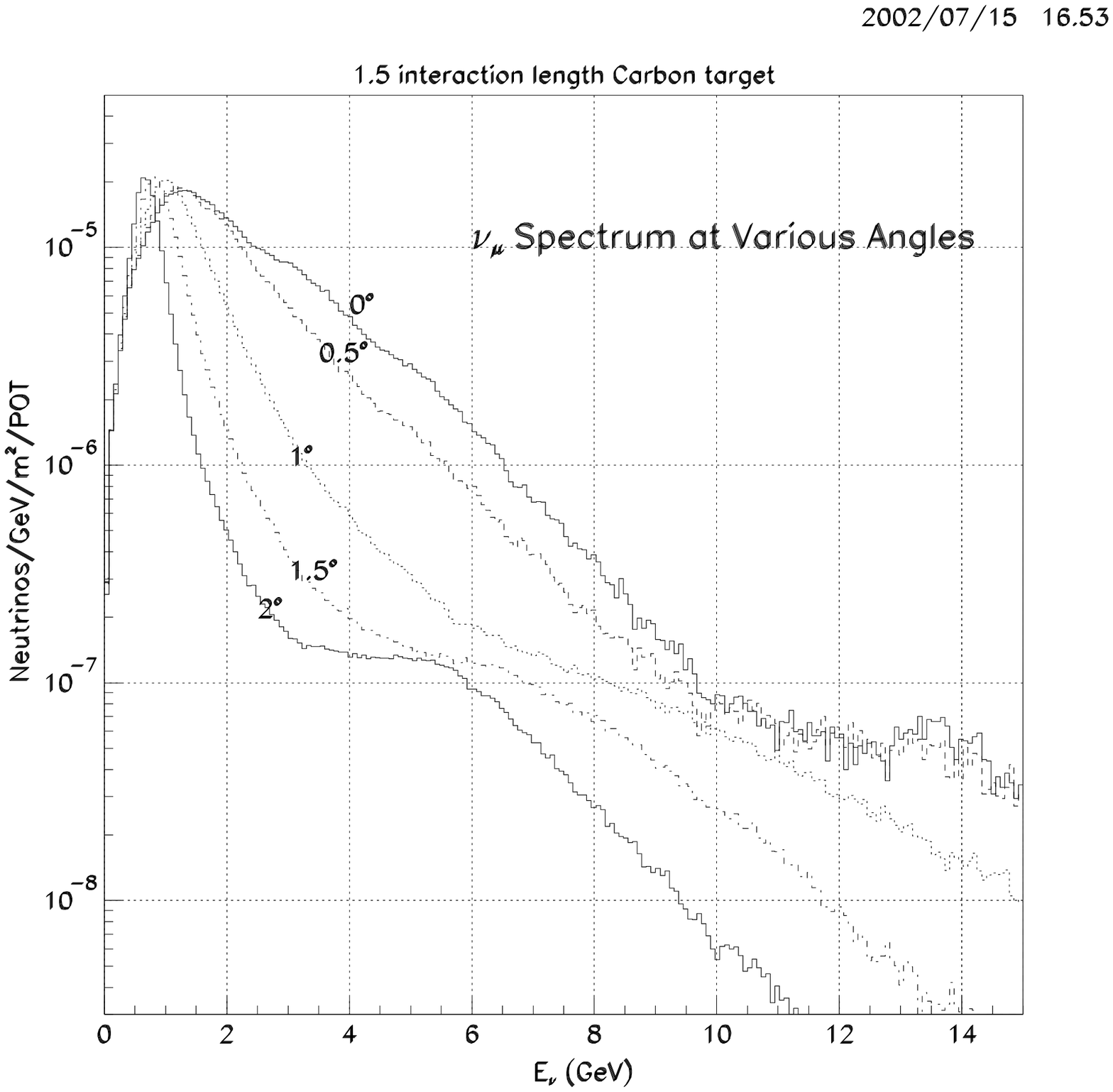}
  \caption[Optimized $\nu_\mu$ neutrino energy spectrum with graphite target.]
{Wide band horn focussed muon neutrino spectrum for 28 \GeV{} protons
on a graphite target. The spectra of neutrinos are calculated at various 
angles with respect to the 200 m decay tunnel axis and at a distance of 1 km 
from the target.} 
  \label{ctarg1}
  \end{center}
\end{figure}

\begin{figure}
  \begin{center}
  \includegraphics*[width=\textwidth]{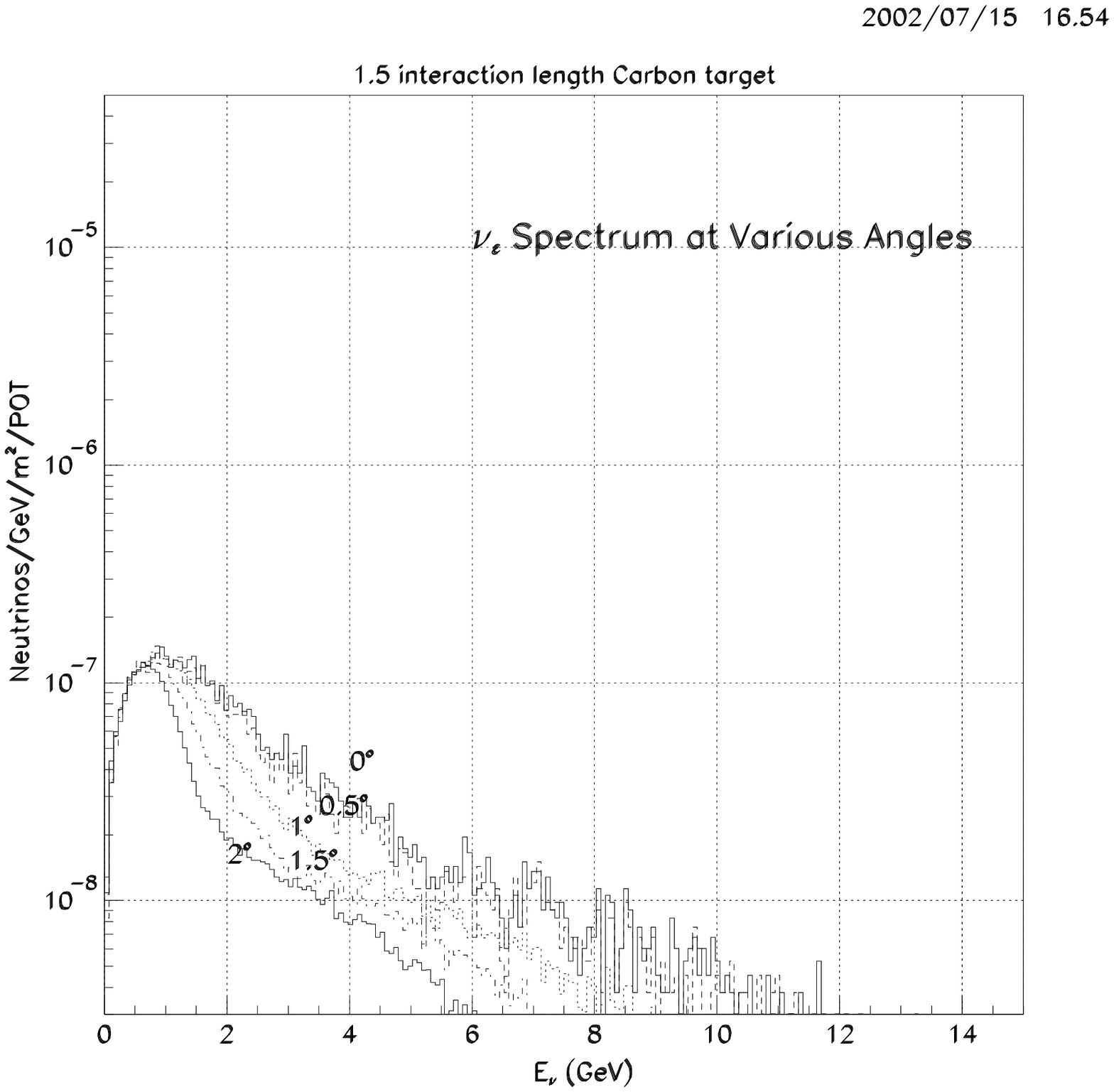}
  \caption[Optimized $\nu_e$ neutrino energy spectrum with graphite target.]
  {Wide band horn focussed electron 
neutrino spectrum for 28 \GeV{} protons
on a graphite target. 
    Spectra of $\nu_e$ are calculated at various angles with respect to the
    200 m decay tunnel axis and at a distance of 1 km from the target.} 
  \label{ctarg2}
  \end{center}
\end{figure}

We discuss the target in much more detail in a later section. 
The two main issues in the target design are the target material and the  
space available 
for cooling. If a dense material, such as Super-Invar, is 
used then the spectrum will be approximately the same as shown in 
Figure \ref{bnlspec}.  A better approach is to use graphite 
as the target material and modify the horn geometry to 
allow for a longer target (Figure \ref{horns2}). 
The result of these modifications is shown 
in Figure \ref{ctarg1}.   The electron neutrino contamination is 
shown on the same scale in Figure \ref{ctarg2}. We have used a 
1.5 interaction length graphite target. 
As shown in the figures, the flux resulting from a graphite target is 
considerably higher in the 3.5 to 8 GeV region. 
There is no significant change in the ratio of electron type neutrinos
to muon type neutrinos between a graphite and a copper target.
We have used the neutrino flux from Figures \ref{ctarg1}
and \ref{ctarg2} for the calculation of event rates and backgrounds in 
the rest of this report.

There is a large 
($\sim 50\%$)  model dependent 
uncertainty on the neutrino flux at high energies ($>4$ GeV). 
In particular the hadron production model in  MARS gives lower 
flux than in GEANT~\cite{mars}.   This 
uncertainty will most likely be resolved by new experiments~\cite{harp, e910}
 in the near future.

Further work on the
optimization of this spectrum for the very long baseline experiment is
ongoing.  Further optimization focuses on enlarging the horns to accept more
lower energy pions so that the flux near 0.5 GeV can be enhanced, using
an evacuated or helium filled decay tunnel, and
as using the hadron hose \cite{numihose} to capture more higher energy 
particles. 

\subsection{Target Station}

To use the 1 MW proton driver proposed for BNL, serious 
consideration must be given to the target selection.
It is desirable to choose a solid target for generating a high intensity
neutrino beam.  
For pion production with high power proton beams, target
integrity becomes an important issue.  Up to now, the production
of secondary particles has been limited to proton beams with
average beam power on the order of 100 to 200 kW.  We now
have to consider a target which can survive a 1 MW (or greater)
average proton beam power.  For a 28 GeV proton beam, 1 MW
beam power implies $2.23\times 10^{14}$ proton/sec.  For a rep-rate
of 2.5 Hz we then must consider nearly 100 TP per spill.  
A number of options have been considered and investigated both in terms of the 
material selection as well as the feasibility of target configuration.
In evaluating the target choices the following concerns are being addressed:

\begin{itemize}
\item Heat removal from the target.
\item Survivability of the target intercepting energetic, 
 high intensity proton bunches.
\item Irradiation issues
\item Engineering integration issues
\item Heat generation and removal from the horn
\item Horn mechanical response
\end{itemize}  

Findings of a number of recent studies \cite{study2}, including
experimental results from AGS Experiment E951 \cite{e951}, 
on target issues for the muon collider/neutrino factory 
project are taken into consideration in this effort. 

Figures \ref{hkirk2} and \ref{hkirk3} show the spectra of $\pi^+$ and $\pi^-$ 
that are produced from a 2-interaction length target for various materials. 
For a conventional neutrino beam, the useful part of the pion spectrum is
in the energy region above 2 GeV.  For this reason, high-Z
targets are no longer advantageous and low-Z targets are preferred.

\begin{figure}
\begin{center}
\includegraphics*[width=\textwidth]{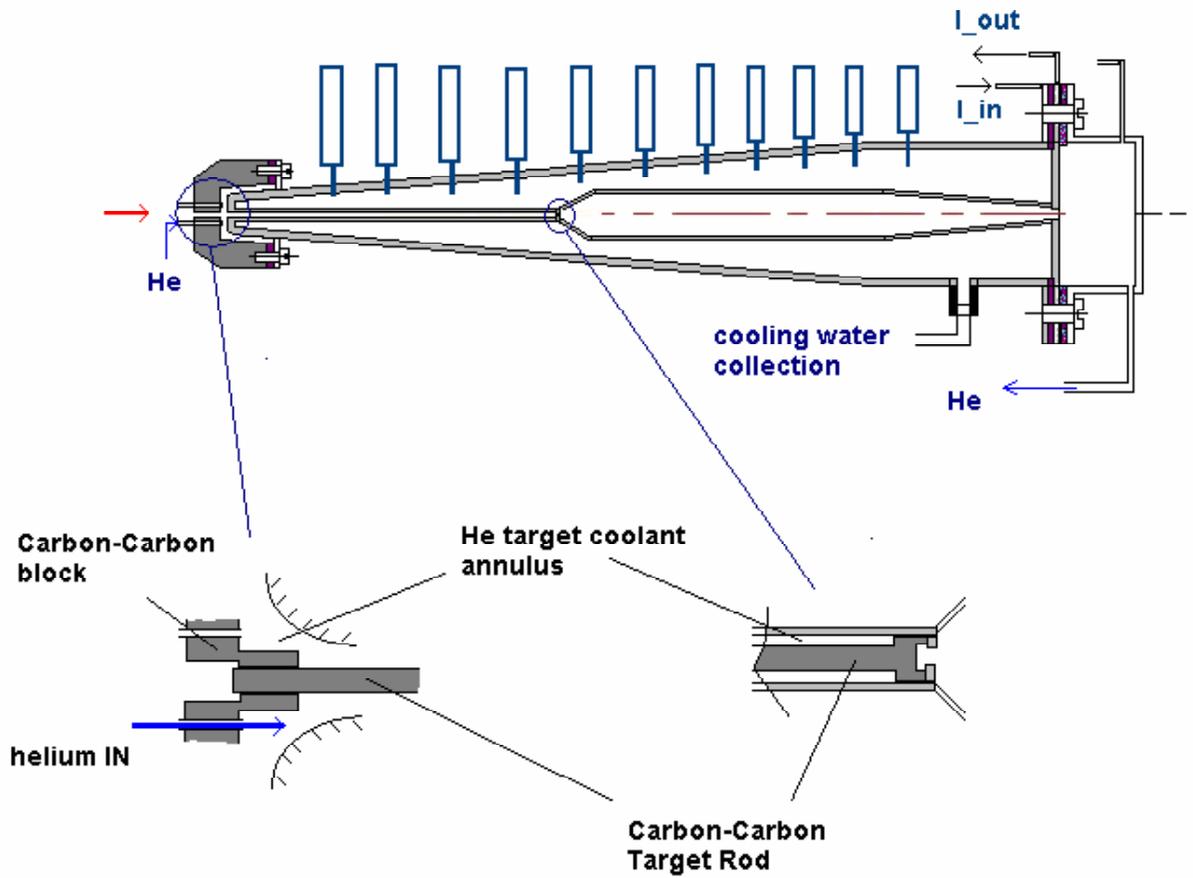}
\caption[Details of first horn and target positioning.]
{Sketch of the first horn with the graphite target mounted.  The
target is cooled by helium.  The horn is cooled by spraying water on
the conducting surface.}
\label{Horn}
\end{center}
\end{figure}

\begin{figure}
  \begin{center}
  \includegraphics*[angle=270,width=\textwidth]{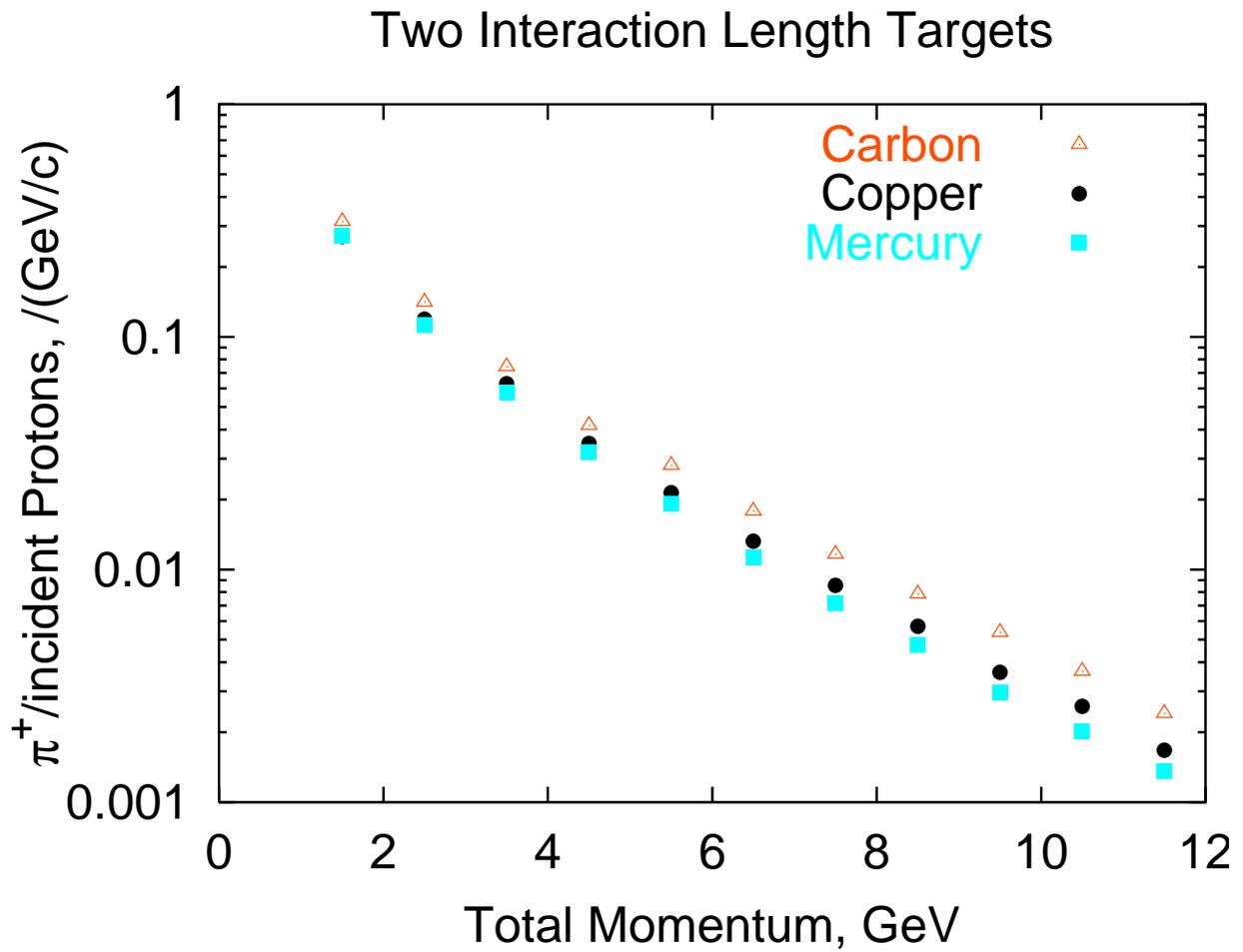}
  \caption[$\pi^+$ production in various targets]
{The number of $\pi^+$ per incident proton is shown as a function
of its momentum for carbon, copper and mercury targets.  The target is two
interactions lengths long for each material.}
  \label{hkirk2}
  \end{center}
\end{figure}

\begin{figure}
  \begin{center}
  \includegraphics*[angle=270,width=\textwidth]{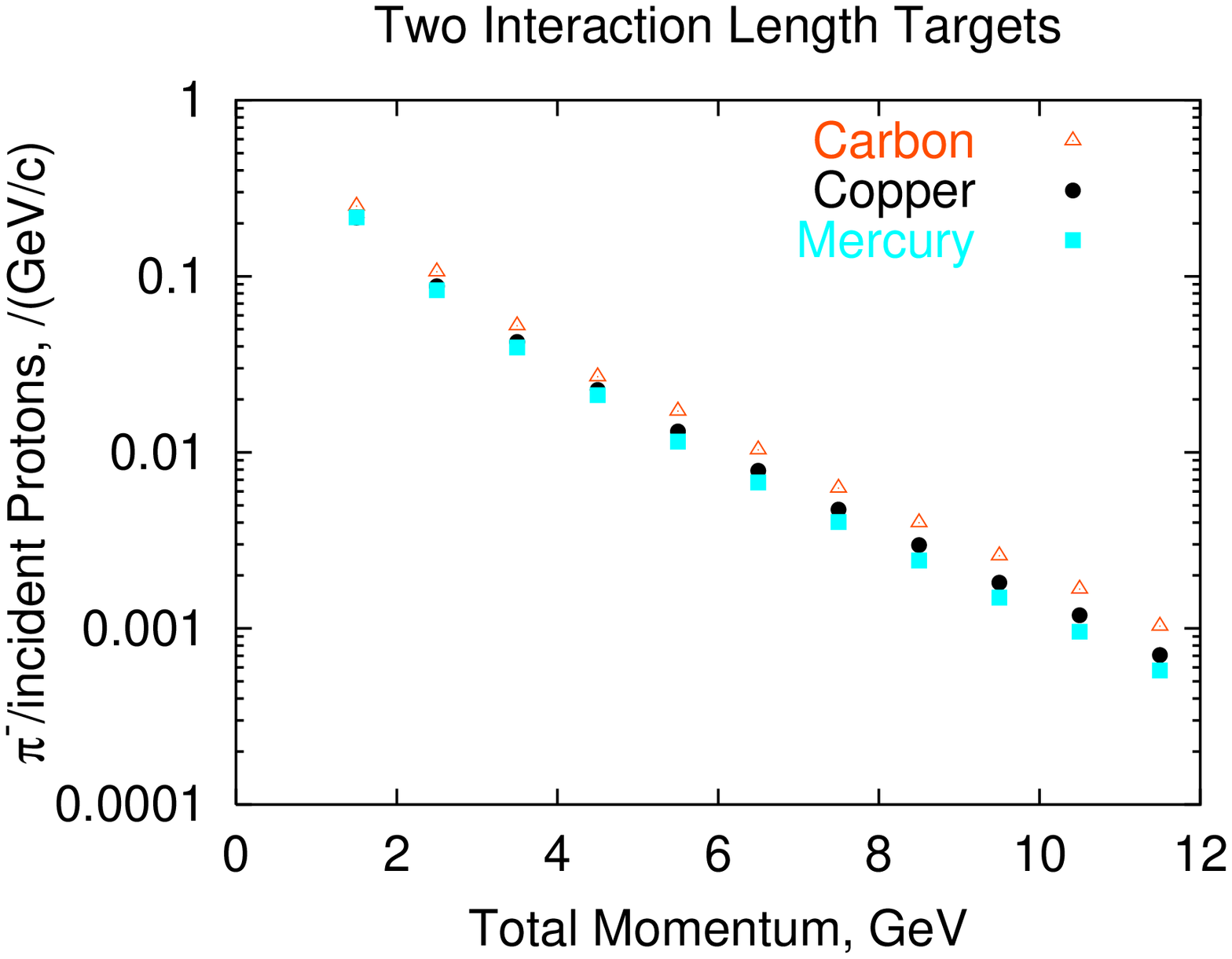}
  \caption[$\pi^-$ production in various targets]
{The number of $\pi^-$ per incident proton is shown as a function
of its momentum for carbon, copper and mercury targets.  The target is two
interactions lengths long for each material.}
  \label{hkirk3}
  \end{center}
\end{figure}

In addition to maximizing the flux, the target/horn configuration
  must  survive the thermal shock induced by
 the beam and the high current.
Specifically, the target scheme must (a)
 ensure the removal of the deposited beam energy within
 the 400 ms period and (b) survive the thermally induced
 elastodynamic stresses that are expected to be comparable
 to the mechanical strength of most common materials.
 Similar concerns are valid for the horn,
 itself, which will be subjected to rapid heating and, as a result,
 high levels of thermal stress that will propagate in its volume.
 In order to satisfy the first requirement, several
 cooling scenarios are being investigated such as 
edge-cooling, forced helium cooling in the space between the target
and the horn, and radiation cooling. All of these 
schemes present challenges stemming from integration
 with the horn in a limited space. To
 satisfy the second requirement, materials must
 be selected such that they can withstand and
 attenuate the thermal shock and be radiation resistant.
 To address this, low-Z carbon based materials such as graphite and 
carbon-carbon composites are being considered. These materials, while 
they have a lot of promise,  present some challenges.
Figure \ref{Horn} shows the target mounted in the first horn.  Also 
the helium cooling system for the target and the water cooling manifold 
for the horn are shown.

Two different forms of carbon, 
ATJ graphite and a carbon-carbon composite are considered as candidate
target materials. These two types have been exposed to the AGS beam in 
the E951 experiment\cite{e951}.  The carbon-carbon composite
is a 3-D woven material that exhibits extremely low thermal expansion
below 1000$^o$C and responds like graphite above that. 
Preliminary studies on the feasibility of using 
carbon-based targets for this neutrino beam have been conducted. 
Specifically, utilizing the energy deposition 
estimates from MARS for 1 mm and 2 mm RMS beam spots (corresponding to 
3 mm and 6 mm radii of target), the thermal shock response and the 
survivability potential of the target were studied. The total 
energy deposited on the target (and which needs to be removed between 
pulses) is 5.1 kJ for the 1mm spot and 7.3 kJ for the 2mm spot.

Since the 1 mm RMS beam spot is the most serious case, it is examined in 
detail. For the 100 TP beam the peak energy density is of the order of 
720 J/gram. 
This is expected to lead to instantaneous temperature increases of 
$\sim 1000^\circ$C.  A detailed finite-element analysis that involves both 
the horn and the target needs to be performed so the heat removal of the 
system can be optimized and, most importantly, so the thermal shock 
stresses can be computed.  A material with a small thermal expansion
should experience smaller thermal stresses.  However, carbon-carbon composite
materials exhibit an 
increasing thermal expansion at higher temperatures. This behavior of the 
material needs to be examined further. If the high temperature performance of
this material is not satisfactory, a larger beam spot size could be used. 
From energy density considerations, a 2 mm rms beam spot would have a peak 
temperature rise per pulse that is less than a third of the 1 mm rms case. 
This would ensure that the material will be well within the safe zone.
Cooling of the front-end is achieved by maintaining the temperature at the
surface of the first 4 cm to 27$^o$C.


\begin{figure}
  \begin{center}
  \includegraphics*[angle=270,width=\textwidth]{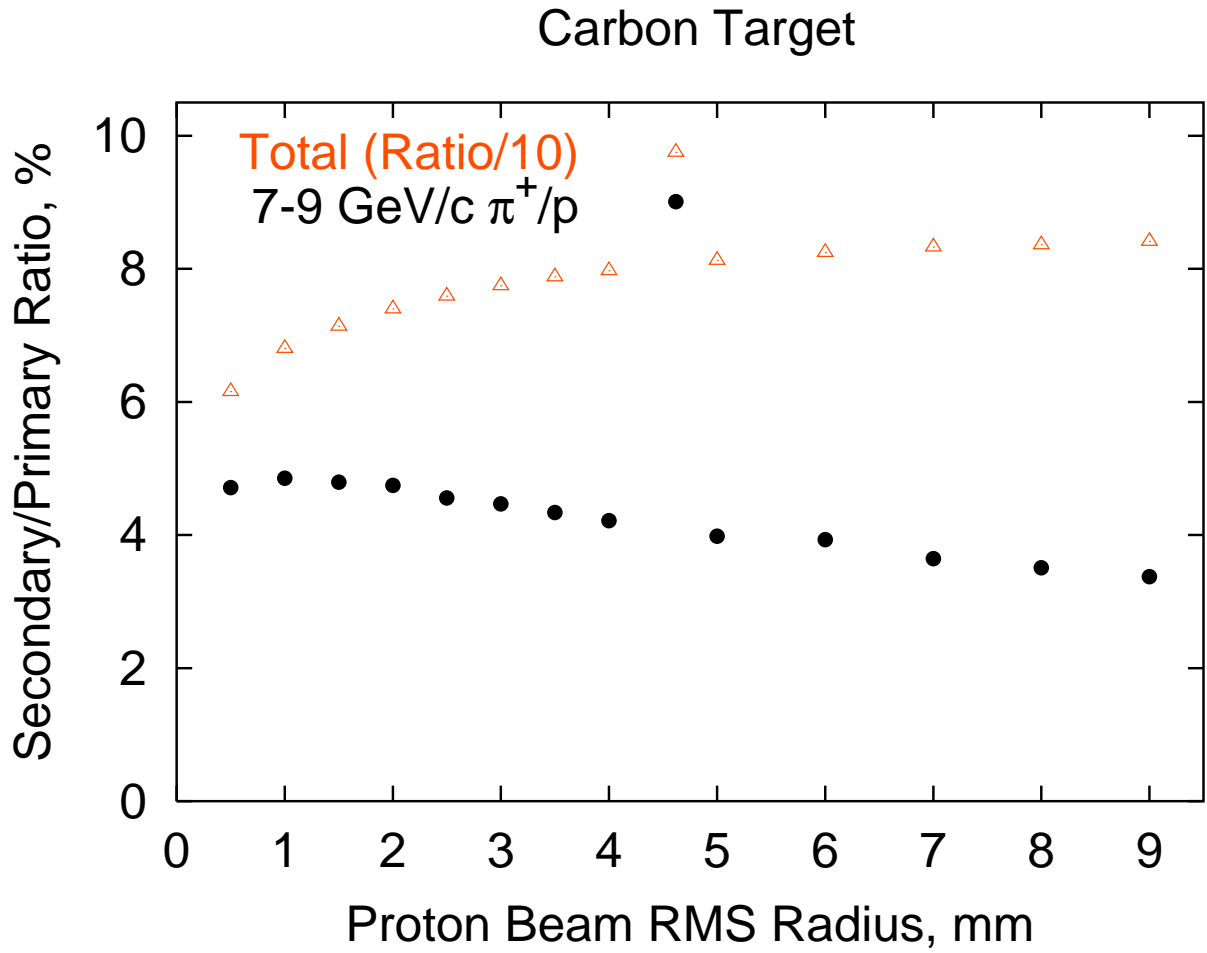}
  \caption[Secondary/Primary ratio {\it vs.} production radius.]
{The ratio of the numbers secondaries to the number of primaries
is shown as a function of RMS beam radius.  The target radius is assumed to 
be three times the RMS beam radius and the target length is 1.5 
interaction lengths.}
  \label{hkirk4}
  \end{center}
\end{figure}

\begin{figure}
  \begin{center}
  \includegraphics*[angle=270,width=\textwidth]{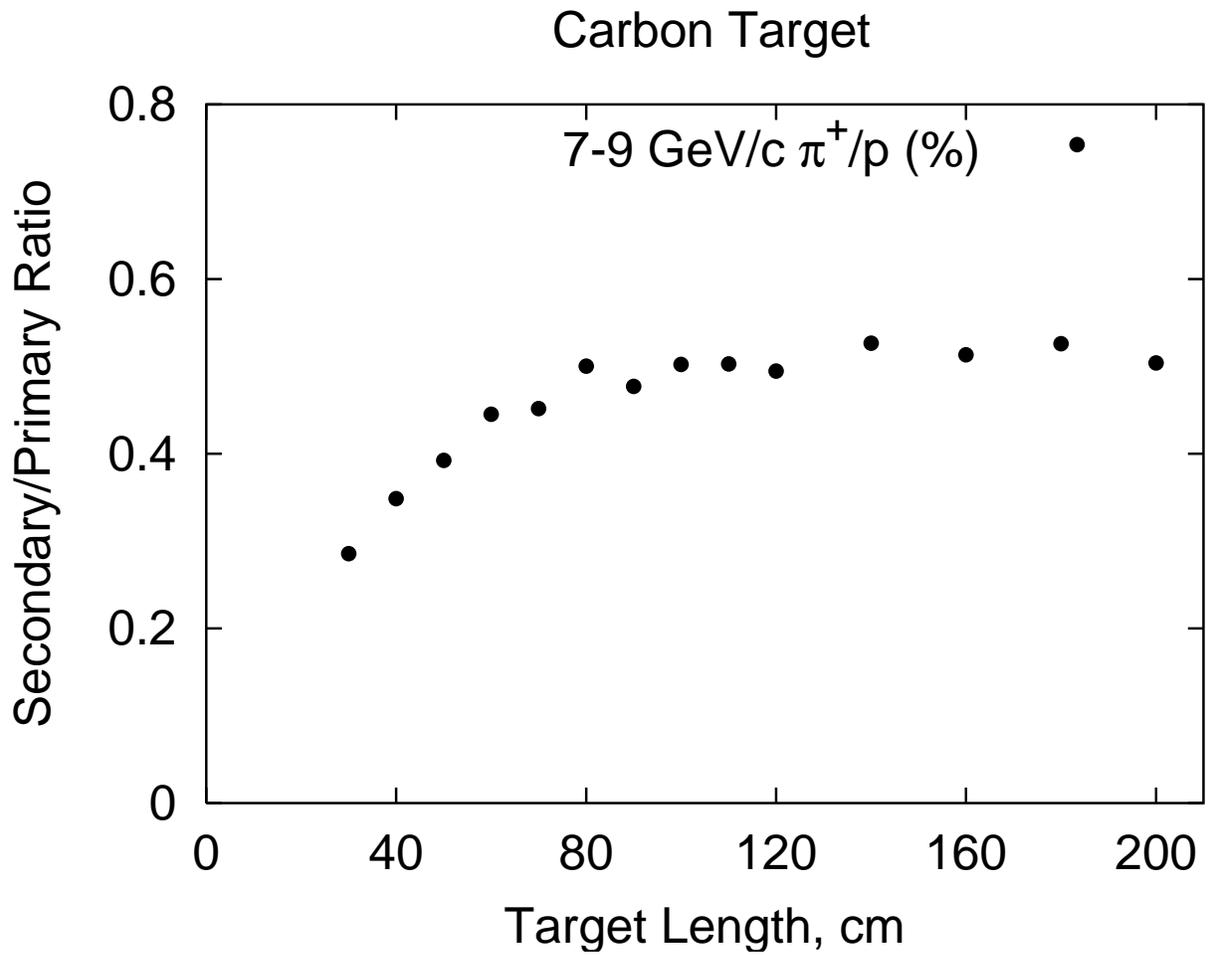}
  \caption[Secondary/Primary ratio {\it vs.} target length.]
{The ratio of the number of secondaries to the number of primaries
is shown as a function of the target length for a target radius of 6 mm and 
a RMS beam size of 2 mm.}
  \label{hkirk5}
  \end{center}
\end{figure}

We examine the optimal geometry for high-energy pion production
utilizing a carbon target.  In
Figure \ref{hkirk4} we see the result of varying the radius of a
1.5 interaction length (60 cm) long carbon target as we
varied the proton beam radius.  For this
analysis the target radius was constrained to 3 times the proton 
beam rms radius.  We note that although the total secondary
pion production increases with radius,  the desired high-energy
portion of the production spectra is enhanced with smaller beam 
spot sizes.  In Figure \ref{hkirk5} we fix the beam/target radius
at (2mm/6mm) and find that the production of 7-9 GeV pions increases
with target length up to about 80 cm (2 interaction lengths) and
then remains essentially constant up to 2 m.

\begin{figure}
  \begin{center}
  \includegraphics*[angle=270,width=\textwidth]{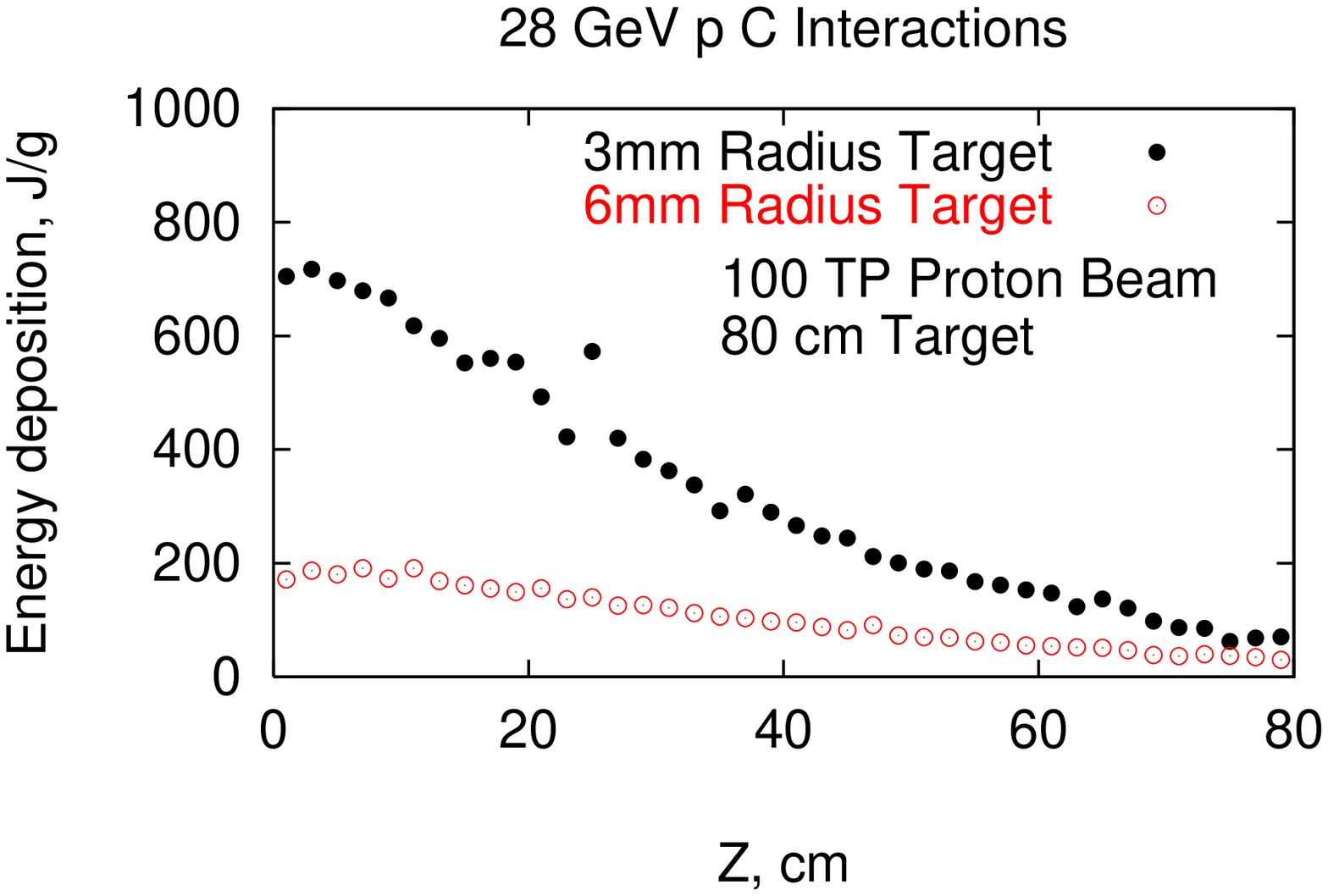}
  \caption{The energy deposition is shown as a function of target axial 
position for a 28 GeV 100 TP beam.}
  \label{hkirk6}
  \end{center}
\end{figure}

We now explore the impact of bringing 100 TP protons/spill
onto a carbon target.  For this analysis we utilize MARS to calculate
the energy deposition due to the hadronic showering within the target.
We examine the two cases of 3 mm and 6 mm radius targets in Figure 
\ref{hkirk6}.
We note the peak energy deposition density occurs near the entrance
of the target and has the respective values of 700 and 200 J/g.  As
a figure of merit, 300 J/g is considered the danger regime where
metal targets suffer damage due to the propagation of thermal generated
pressure waves through the material.  There is, however, evidence that
carbon can withstand energy depositions in this
regime.  The best evidence to date comes from experience in the NUMI target
development program.  The NUMI carbon target is designed to expect
390 J/g peak energy deposition.  A NUMI target test, performed in
1999, utilized a specially focussed beam to produce energy depositions
in the range of 400 to 1100 J/g without any external evidence of
target breakup.

The secondary particle shower resulting from the interaction of 
primary protons with the low-Z target will add to the transient heat
load of the horn. This shower will be less significant for low-Z 
targets than for high-Z targets. However, its effect will be examined,
and added to the electric resistance heat load estimated above.

There will be activation of the target and horn structure due to 
secondary and primary particles. This activation 
will be primarily due to spallation products and neutrons generated 
in the secondary shower. The survival of the 
primary target in the radiation field needs to be examined. This 
can only be carried out experimentally using a prototypic proton 
beam on samples of the appropriate target material. The change in 
physical properties including, thermal expansion coefficient, elastic 
modulus, and yield strength, need to be examined as a function of 
proton fluence.

In the current option the target is an 80-cm long cylindrical rod with 
12 mm diameter sizes.  The 12 mm diameter target is chosen to intercept 100
TP, 2 mm rms proton beam.  With this beam size, the total energy deposited as 
heat in the target is 7.3 kJ with peak temperature rise of about 280$^\circ$C.
Heat will be removed from the target through forced convection of 
helium through the outside surface.  This is a good solution for the 
1 MW beam power.

\subsection{Cost of the neutrino beam}

\begin{table}
\caption{Preliminary direct cost (FY02\$M) of building the neutrino beam with 200 meter decay
tunnel.  These costs do not include EDIA contingency, and overhead.}\smallskip
  \begin{center}
\begin{tabular}{|l|l|r|}
\hline
Item &  basis & cost \\\hline 
Proton transport & RHIC injector & \$11.8 M  \\ 
Target/horn & E889 &  \$3 M  \\
Installation/Beam Dump & New & \$2.6 M  \\
Decay Tunnel & E889 & \$0.4 M  \\
Conventional const. (hill) & New & \$8 M  \\  
Conventional const. (other) & E889 & \$9 M \\
\hline 
Total &      & \$35 M  \\
\hline 
\end{tabular}
\label{bcost}
  \end{center}
\end{table}

A preliminary estimate of the direct costs without  
burdens is shown in Table \ref{bcost}. The costs are based 
on the the RHIC injector work, as well as the E889 proposal and 
the neutrino factory study.  
The conventional construction costs are 
dominated by the size of the hill which is 
approximately proportional to 
the third power of the decay tunnel length. In our cost
estimate we assume that we will bury the beam dump underground
to reduce the height of the hill.  
It is assumed that the target station shielding can be 
retrieved from existing resources. 
We have also estimated the cost assuming a 200 m long decay tunnel.
The spectra shown in Figure \ref{bnlspec} are based on 
this 200 m tunnel filled with air. 

\vspace{1ex}





\section{ Conclusion}
\vspace{1ex}

    The possibility of making a low cost, very intense   high
    energy proton source  at the Brookhaven Alternating Gradient
  Synchrotron  (AGS) 
    along with  the forthcoming
    new large underground detectors at either  the National Underground
    Science and Engineering Laboratory (NUSEL) in Homestake, South Dakota or at the Waste
    Isolation Pilot Plant (WIPP) in Carlsbad, New Mexico, allows us to
    propose a program of experiments that will address fundamental
    aspects of neutrino oscillations and CP-invariance violation.  This
    program of experiments is unique because of the very long
     baseline of more than 2500 km from Brookhaven National
    Laboratory to the underground laboratories in the West, the high
    intensity of the proposed conventional neutrino beam, and the
    possibility of constructing a very large array of water \cerenkov{}
    detectors with total mass approaching 1 megaton. 
  This report examined   the design and
    construction of the necessary AGS upgrades and the new neutrino beam
    which will  have a proton beam of power $\sim$1.0~MW.
  We have examined the potential physics reach of 
  such an experiment. We used the running scenario of 
   a 1 MW AGS, 500 kT of fiducial
  detector mass, and $5\times 10^7$ secs of running time. With these 
  conditions, we conclude that such an experiment is capable of 
  precisely measuring $\Delta m^2_{32}$ and $\sin^2 2 \theta_{23}$;
  it has excellent sensitivity to $\sin^2 2 \theta_{13}$ with a
  signal spectrum that is very distinctive. Moreover, if $\sin^2 2 \theta_{13}$
  is sufficiently large ($> 0.01$) the experiment is sensitive to 
  the CP-violation parameter $\delta_{CP}$ with only neutrino running. 
  With the additional option of running in anti-neutrino mode, the
  experiment will be able to 
distinguish between the cases $\Delta m^2_{31} > 0$
  versus $\Delta m^2_{31} < 0$ using distinctive distortions to the observed 
electron or positron spectrum.  Lastly, the very long baseline will allow the 
  measurement of  $\Delta m^2_{21}$ with approximately the same  
  resolution as KAMLAND but in the $\nu_\mu \to \nu_e$ appearance channel 
  if the LMA solution 
  is correct for the solar neutrino deficit.


The AGS complex is unique because it can be upgraded simply by
increasing the repetition rate 
 of the machine.  This ability allows us the
flexibility to continuously upgrade the facility to as much as 4.0
MW~\cite{roser01}. In this proposal we have examined upgrades up to 1.0 MW.
The direct costs of such an upgrade are estimated to be approximately \$140M. 
This compares well with the estimated costs for the detectors and the 
neutrino beam-line. 
 Neither the
duration of the construction period nor the anticipated cost of the
improvements to the BNL AGS complex is large in relation to plans and
expenditures now usual for major apparatus in high energy and
elementary particle physics.
  Moreover, the
improvements to the AGS and the new beam line will be available for
carefully chosen other physics (for example, rare muon and kaon 
decays as well as muon EDM measurements)\cite{edm, rsvp}, 
 while advancing our understanding of the neutrino section.

\newpage 
\section{Appendix I Working Group Charge and Assignments}

\begin{verbatim}
Director's Office

                                                            Building 510F
                                                            P.O. Box 5000
                                                     Upton, NY 11973-5000
                                                       Phone 631 344-5414
                                                        Fax 631  344-5820
                                                            tkirk@bnl.gov

date: December 1, 2001 

to: S. Aronson, M. Harrison, D. Lowenstein,
R. Palmer, V. Radeka W. Marciano, M. Diwan and W.T. Weng 

from: T. Kirk
Associate Laboratory Director, 

HENP subject: Neutrino R&D Working Group Charge and Assignments

Attached, please find the Charge to the Neutrino R&D Working Group
that we have discussed.  As agreed, Bill Marciano will be the Neutrino
Team Leader, Milind Diwan will be the Physics Goals and Detector Team
Leader and Bill Weng will be the Accelerator and Beam Systems Team
Leader.  The recruitment of working participants on the teams will be
the responsibility of the team leaders, aided by the department heads
and myself.  The composition of the R&D teams will not be limited to
BNL employees.  In fact, the participation of outside physicists in
the study will have obvious benefits for the next stage of the work
which is expected to be the establishment of a formal collaboration
and the creation of a formal proposal to the funding agency or
agencies to build and operate a neutrino beam and detector system and
carry out an experimental neutrino physics program.  If the work gets
off to a promising start and the physics prospects appear to be
sufficiently compelling, it is possible that the initiation of the
collaboration and the start of a related proposal may overlap the R&D
study in time.  Such an outcome could also have benefits for the
timely advance of neutrino physics.

We are initiating neutrino R&D work without explicit funding for this
purpose.  Accordingly, the R&D work should be regarded as part of the
participants research activity, work that is generally supported by
the Laboratory research mission in high energy and nuclear physics.  I
expect that the department heads will help and support the teams to
carry out the work within their capabilities.  This has already been
discussed and agreed to.  If conflicts arise about the allocation of
internal resources and priorities between the needs of the R&D study
and other activities of the departments that cannot be settled between
the team leaders and the department heads, I will establish a forum
for reconciliation of the conflict.  I believe we are all aware of the
importance to the Laboratory of a successful outcome for this work and
we will expend our efforts accordingly.

Attachment (1)

Cc: P. Paul

\end{verbatim}

\newpage 

\begin{verbatim} 

Charge to the BNL Neutrino R&D Working Group

December 1, 2001

BNL intends to initiate an R&D study to refine the technical basis for
a future proposal to employ the BNL AGS as the source of a 1MW (or
possibly greater), ~1GeV neutrino beam for the continuing exploration
of neutrino physics, including CP-violation in the neutrino sector.
We also expect as the second element of this R&D study, to be key
organizers of an experimental physics and detector design effort that
will engage interested physicists in the U.S.  and other countries in
the preparation of the conceptual basis for a formal proposal to
design and build a neutrino detector system to exploit the BNL
neutrino beam and to carry out the associated neutrino physics
program.

To this end, the Laboratory will designate three R&D leaders for these
efforts: the Neutrino Team Leader; the Accelerator and Beam Systems
Team Leader; and the Physics Goals and Detector Design Team Leader.
These three leaders will, in turn, be responsible for organizing the
technical work that will enable a good scientific proposal to be
written to the funding agencies that are identified as potential
sponsors of this new U.S. particle physics effort.  The three team
leaders will serve until this R&D study is complete and documented in
a written report.  It is intended that the written R&D report should
be completed no later than June 1, 2002.

The specific roles of the three Team Leaders comprise:

Neutrino Team Leader: The Neutrino Team Leader (NTL) will have
responsibility for ensuring that the overall goals of a successful
neutrino physics program have been covered by appropriate R&D studies
in each of the important contributing technical systems and that there
is a coherent overall time evolution plan that is consistent with
preparing a compelling proposal that addresses the goals of neutrino
physics in a timely manner.  This role should be understood as
primarily a guidance and oversight role rather than a detailed
management role.  The balance and completeness of the study is the
primary responsibility of the NTL.

Accelerator and Beam Systems Team Leader: The Accelerator and Beam
Systems Team Leader (ABSTL) is the person primarily responsible for
planning, staffing, carrying out and reporting on the accelerator and
neutrino beam forming systems that are relevant for the preparation of
a credible proposal to construct and operate a 1MW or greater proton
target and associated useful neutrino beam(s) using the AGS (suitably
upgraded) as the proton driver.  To accomplish this mission, the ABSTL
will be helped by the relevant BNL department heads to identify
sufficient and appropriate technical staff to carry out the needed
studies.  The ABSTL is also expected to create an appropriate
discussion and reporting forum(s) where the ongoing progress in this
R&D effort can be reported and discussed for the general benefit of
interested parties and participants.  The ABSTL role is understood to
be the principal management role for accomplishing the desired R&D
studies in the accelerator and beam forming elements of the overall
R&D program.

Physics Goals and Detector Team Leader: The Physics Goals and Detector
Team Leader (PGDTL) is the person primarily responsible for planning,
staffing, carrying out and reporting on the physics goals and detector
strategies that are relevant for the preparation of a credible
proposal to construct and operate a detector array that can exploit
the 1MW or greater neutrino beams from the AGS proton driver.  To
accomplish this mission, the PGDTL will be helped by the BNL Physics
Department head and (hopefully) by neutrino community scientists and
engineers in other institutions to find sufficient and appropriate
scientific staff to carry out the needed studies.  The PGDTL is also
expected to create appropriate discussion and reporting forums where
the ongoing progress in this R&D effort can be reported and discussed
for the general benefit of interested parties and participants.  The
PGDTL role is understood to be the principal management role for
accomplishing the desired R&D physics and detector studies for the
overall neutrino R&D program.

\end{verbatim}

\newpage

\section{Appendix II Underground Detector Construction at Homestake}


Plans for the construction of a multiple module megaton \cerenkov{}
detector at the Homestake Mine have gone through a number of essential
evaluation and design stages consisting of rock strength and stability
evaluation, chamber design and layout, construction planning and
sequencing and development of budget and timetable.  Here is a summary
of these steps.

\subsection{Determination of Excavation Stability}

The Rock Stability Group at the Spokane Research Laboratory of NIOSH
(National Institute of Occupational Safety and Health) carried out an
evaluation of the stability of large excavations as a function of
depth in the Yates rock formation in the Homestake Mine.  This
involved a three-dimensional finite difference evaluation using the
FLAC3D program.  These results were compared with the empirical
prediction charts of Barton and Grimstad and Barton.  The conclusions
were that 50 meter diameter by 50 meter high chambers could be safely
excavated and would be stable for long term occupancy at 2150 meter
depth and probably somewhat deeper.

The Yates rock quality was determined by direct measurement of samples
taken from the accessible edges of this formation.  Before excavation
begins, it is essential that core samples from various internal
sections of the proposed rock formation are measured and the
excavation reevaluated.

\subsection{Construction of Multiple 100 kiloton Modules in the Homestake Mine}

Using the results of the stability evaluation a group of ex-Homestake
mining engineers, (Mark Laurenti - former Chief Mine Engineer, Mike
Stahl - former Mine Production Engineer and John Marks - former Chief
Ventilation Engineer) designed an array of ten 100 kiloton water
\cerenkov{} chambers.  The chambers are located along the circumference
of a 250 meter radius circle that is centered on the Winze 6 shaft.
The top of each chamber is connected to the 6950 ft station of the
shaft via a horizontal, radial tunnel.  A similar tunnel connects the
bottom of each chamber to the 7100 ft shaft station.  Fresh air will
be sent to each chamber via the top tunnel and exhaust air removed via
the bottom tunnel, thus providing independent air supplies to each
chamber.


During chamber construction, waste rock will be removed via the bottom
tunnel.  This will prevent rock dust from one chamber contaminating
the fresh air supply of another chamber.  Once construction is
completed, the bottom chamber to tunnel connection will be sealed.  A
spiral ramp that surrounds each chamber and is used for access during
construction will then complete the ventilation loop between top and
bottom tunnels.
 
Each chamber will have a concrete liner.  The inner surface of the
liner provides a smooth surface for the water tight plastic liner that
will separate the \cerenkov{} counter fill from the chamber walls.  The
liner also provides additional mechanical stability for the
excavation.  If necessary, drainage can be provided between the
concrete liner and the surrounding rock.

\subsection{Construction Timetable and Cost}

Marc Laurenti has worked out a detailed timetable and budget for the construction of these modules including initial rock evaluation coring, construction of both top and bottom access tunnels, removal of waste rock, maintenance of mining equipment, etc. 

The excavation process consists of continuous repetition of three
separate tasks, (1) drilling and blasting of rock, (2) removal of the
rock rubble, and (3) installation of rock and cable bolts to stabilize
the freshly exposed rock walls.  Each excavation cycle is about 10
weeks with 3 weeks for each of the above three steps.  There is a
considerable cost savings in excavating three chambers at the same
time, with a three week phase shift between steps in each module.
This arrangement permits each of the three specialized crews to move
from one excavation to the next every three weeks or so and continue
using the same equipment and carry out their specialized tasks.  In
contrast, using one crew to sequentially do three different tasks will
result in idle equipment for 2/3 of the time and inefficiency as they
switch from one task to another.

For the three module mode, the cost of excavating each chamber is
\$14.7M.  This includes \$3.25M for the concrete liner and a 15\%
contingency.  In contrast, the cost of excavating a single module is
\$16.9M including liner and contingency.  The total required equipment
cost is the same for both of these construction modes.

Assuming three shifts/day and 5 days/week operation, it will take 208
weeks or 4 years to excavate each 3 module group.  This time could be
reduced by going to a 6 or 7 day week.  The Homestake Company
frequently operated on 6 or 7 day per week basis.

\subsection{Rock Removal}

Each 100 kiloton module (105 m3) involves the removal of about 416,000
tons of rock including access tunnels, domed roof, etc.  For three
chambers this results in 1,248,000 tons of rock in 4 years or 312,000
tons of rock per year.  Since the hoisting capacity of the Winze 6 -
Ross shaft system is 750,000 tons per year, the simultaneous
construction of three modules utilizes only 40\% of the capacity of
this shaft system.

\subsection{Equipment Cost}

Since all mining equipment has now been removed, new mining equipment
will have to be purchased or leased.  The required equipment, one Face
Drill, two LHD loaders, 2 Bolters, 2 Underground Support Vehicles, 2
Lift Trucks, 1 LH Drill and 2 ITH Drills, costs about \$4.2M.  It may
be possible to arrange for leases instead of purchasing these items.
Normal equipment maintenance has been included in the construction
cost.  It is unclear whether the cost of this equipment should be
assigned to this specific task or should be part of the general
facility budget.

\subsection{Choice of Depth and Depth Dependent Cost}

There has been considerable discussion of depth necessary for very
large detectors and the costs associated with deep detector locations.
It is clear that the deeper the detector, the lower the cosmic ray
muon and associated particle background.  It is always preferable to
have lower background.  We can quantify the background limit by
specifying that there be less than one cosmic ray related event per
year within the megaton detector during the time that the accelerator
neutrino beam is on.  If we assume the accelerator beam is on for one
microsecond per second, this requirement specifies an upper limit of
$1.6 x 10^{-6} mu/m^2/sec$, essentially the cosmic ray flux at about
7000 ft depth.  The effect of this specification is that every event
observed in the detector during the beam-on time is due to a neutrino
from the accelerator without any cuts whatsoever.

The question then is one of access and rock strength, namely, does a
specific facility have ready access to a deep location and is the
local rock structure capable of supporting large chambers.  For
Homestake the answer to both of these questions is YES.  The present
mine extends to 8000 ft, about 1000 ft deeper than the proposed
detector location, and the rock seems strong enough to readily permit
the excavation of large chambers.

In the appendix we provide a comparison of costs of building the
megaton \cerenkov{} detector at 6950 ft depth vs at the 4850 ft depth.
As indicated there, the maximum additional cost for putting the
megaton \cerenkov{} array at 6950 ft versus at 4850 ft is 5-6\% of
excavation cost or less than 2\% of total detector cost.

\subsection{What Lessons About Depth Can be Learned from Previous Experience?}

Detectors are located underground to reduce background in the detector
due to cosmic rays.  The deeper the detector, the lower the cosmic ray
background.  We have yet to have a detector that claimed to be "too
deep".  The only issues are: (1) is there a substantial additional
cost associated with depth, and (2) are there technical limits
associated with rock strength, etc. that limit depth at a given
location?  For many existing laboratories, depth is specified by what
is available at that facility.  Only two locations, the Sudbury mine
(SNO), and the Homestake Mine (chlorine), have multiple levels available.
SNO chose to be at 6800 ft, essentially the same as the proposed
megaton detector.  Since chlorine was the first underground neutrino
detector, there were no precedents and so it might be instructive to
review the sequence of events that led to its location.

In 1962, Ray Davis tested a small perchloroethylene detector in a
limestone mine in Barberton, Ohio at a depth of 2200 ft.  The 37Ar
production was completely dominated by cosmic rays.  That started a
search for a much deeper site.  There were two possibilities in the
U.S., with Homestake the preferable one.  At that time, in 1965, 4850
ft was the deepest level that the Company would agree to.  At the time
the prediction for the solar neutrino signal was larger than now,
there was no thought about signal depression because of neutrino
flavor conversion and no one expected a final measurement with a 5\%
statistical precision.  By the early to mid-1970's it was already
clear to us that the cosmic ray induced background was too large,
given the observed signal, and that we needed a larger and deeper
detector.  Unfortunately, at the time, the Company was not willing to
consider a deeper and larger detector.

The final result was that the cosmic ray induced signal is 10\% of the
solar neutrino signal in the chlorine detector and the uncertainty in
that signal is the largest contributor to the systematic uncertainty.
The lesson is clear - locate detectors as deep as possible and be sure
that there is a roadmap to detector enlargement.

A detailed construction plan for the construction of three 100 kiloton
modules in four years at the 7000 ft depth in the Homestake Mine has
been developed.  The total construction cost of these three modules is
about \$44 M or \$11M/year.  In addition, there must be a one time
purchase of about \$4.2 M worth of mechanized mining equipment.  The
lead time in delivery of the mining equipment can be used to carry out
coring of the rock region in which the detector array is to be
constructed.

\subsection{Comparison of Costs at 4850 ft versus 6950 ft}

There are two depth dependent costs, the cost of hoisting rock and the
cost of rock and cable bolting.  To estimate this effect, we determine
the difference in costs between identical chambers built at the 4850
ft level (the bottom level of the Ross shaft, the upper hoist system,
and the beginning of the Winze 6, the lower hoist system) and the 6950
ft level.  The direct manpower costs for hoisting the extra 700 meters
in the Winze 6 are about \$0.30/ton.  The power costs add another
\$0.20/ton for a total of \$0.50/ton or \$208,000 per 100 kiloton
module, where shaft maintenance costs have not been included.

The incremental rock support costs are more difficult to determine.
The cable bolting planned and budgeted for these modules is far
greater than required.  This was done to insure that the chambers
would have a minimum 50 year occupancy.  A similar approach to
corresponding excavations at the 4850 ft level might result in exactly
the same bolting pattern and thus the same cost.  Another approach
would scale the bolting cost by the difference in rock stress between
the two levels.  The rock stress in the Homestake and Poorman
formations, the formations that have been extensively studied in mine,
are rather surprising.  The measured vertical stress $S_v= 28.3 \times
D kPa$, where D is the depth in meters, is exactly what is expected
for a fluid of density 2.9 (the rock density).  The horizontal stress
is very direction dependent.  Along the high stress axis $S_{h1} =
14,328 + 12.4 \times D kPa$, while along the low stress axis, $S_{h2}
= 834 + 12 \times D kPa$.  Presumably, the high horizontal stress
results from the rock folding that resulted in the upbringing of the
gold ore deposit to the surface and thus its discovery.

We assume that the effective stress at 6950 ft is about 35\% greater than the corresponding
 one at the 4850 ft level.  Since the total cost of the cable and rock bolts is 
\$910,000 and the related labor, including benefits, is about the same, 
we assign a depth dependent cost increase of \$630,000 for rock support.
  Combining this with the increase in hoisting costs gives a total of 
\$838,000 or 6\% of the total construction cost. 
 Note that this is less than 2\% of the complete detector cost.

However, there are three offsetting costs that reduce the cost of
constructing the \cerenkov{} detector array at 6950 ft vs. at 4850 ft.
The first of these is the water fill.  The total water fill for the
megaton detector is 250 million gallons.  Removing that much water
from the local streams would be quite significant, especially given
the present drought conditions in the area.  Instead, we plan to use
the water that is being pumped from the bottom of the mine at the 8000
ft level.  This water will be purified to remove any light scattering
or absorbing material and any radioactive contaminants.  Since the
mine now pumps out about 350 gallons per minute, we will require about
1.4 years worth of water distributed over the construction time of the
entire detector.  For a detector at the 6950 ft level, this water is
only pumped up 1000 ft while for a detector at 4850 ft, the water must
be pumped up about 3100 ft.  The cost savings here is about $1/4$ of
the increase in rock hoist cost or about \$50,000.

The second offsetting cost is that of cooling the \cerenkov{} detector.
Operating the detector at $10^oC$ gives $1/4$ the photomultiplier
noise of operation at $20^oC$.  Since the rock temperature at the 4850
ft level is over $35^oC$ and still higher at 6950 ft, cooling will be
necessary at either depth.  The mine has an enormous refrigeration
plant (2400 ton capacity) at the 6950 ft level, with a fairly short
path for the coolant from the refrigeration plant to the detector. A
detector at the 4850 ft level will either require a new refrigeration
plant at that level or the installation of 2000 ft of vertical coolant
piping in the mine shaft.  We have not estimated the cost of either of
these steps, but they are clearly very substantial.

The third offsetting cost is that the level structure at 4850 ft does not readily lend 
itself to the construction and ventilation system described above.  
If the upper detector access is at 4850 ft then the lower, rock 
removal tunnel is at 5000 ft.  Unfortunately, there is no ventilation 
exhaust system at that level and waste rock would have to be raised 
in order to get it into the hoist system.  The alternate approach, 
putting the top access at 4700 ft, would require additional excavation
 in order to provide the necessary tunnels for the upper access.

The material in this section was assembled and compiled by Kenneth
Lande based on work done by a number of senior mining engineers who
previously were in charge of mining operations at the Homestake Mine.

\clearpage
\addcontentsline{toc}{section}{\protect\numberline{}{List of Figures}}
\listoffigures
\clearpage
\addcontentsline{toc}{section}{\protect\numberline{}{List of Tables}}
\listoftables

\addcontentsline{toc}{section}{\protect\numberline{}{References}}

\end{document}